\let\ifarxiv\iftrue %\iftrue  \iffalse
\ifarxiv
\documentclass{vldb}%vldb_arxiv
\else
\documentclass{vldb}
\fi

\usepackage{graphicx}
\usepackage{balance}  % for  \balance command ON LAST PAGE  (only there!)

\usepackage{tikz}
\usepackage{times}
\usepackage{amsmath,epsfig}
\usepackage{amsfonts}
\usepackage{etoolbox}
\usepackage{booktabs} % For formal tables
\usepackage{graphicx,epsfig,color,endnotes,alltt}
\usepackage{caption}
\DeclareCaptionType{copyrightbox}
\usepackage{comment}
\usepackage{keyval}
\usepackage{subfig}
\usepackage[sort,nocompress]{cite} %automatic arrange references in increasing number
\usepackage{balance}
\usepackage [ruled,linesnumbered] {algorithm2e} % \usepackage{algorithm}
\usepackage[noend]{algorithmic}
\usepackage[noindent, nolineno]{lgrind}
\usepackage{listings}
\usepackage{epstopdf}
\usepackage{array}
\usepackage{multirow}
\usepackage{amsmath}
\usepackage{bm}
\usepackage{slashbox}
\usepackage{marginnote}
\usepackage{geometry}
\usepackage{soul}

\usepackage[labelfont=bf]{caption}%upper case: the title and caption. 

\SetKw{KwBy}{by}

\newcommand*\circled[1]{\tikz[baseline=(char.base)]{
		\node[shape=circle,fill,color=black,text=white,inner sep=0.05pt](char){#1};}}

\makeatletter %%%notsotiny: between tiny and scriptsize
\newcommand\notsotiny{\@setfontsize\notsotiny\@vipt\@viipt}
\makeatother

\graphicspath{{./figure/}}

\newcommand{\kaan}[1]{{\color{violet}#1}}

\ifarxiv

  \vldbTitle{Accelerating Generalized Linear Models with
  	MLWeaving: A One-Size-Fits-All System for Any-Precision Learning}
  \vldbAuthors{Zeke Wang, Kaan Kara, Hantian Zhang, Gustavo Alonso, Onur Mutlu, Ce Zhang}
  \vldbDOI{https://doi.org/10.14778/3317315.3317322}
  \vldbVolume{12}
  \vldbNumber{7}
  \vldbYear{2019}

\else

  \usepackage{nopageno}
\vldbTitle{Accelerating Generalized Linear Models with
MLWeaving: A One-Size-Fits-All System for Any-Precision Learning}
\vldbAuthors{Zeke Wang, Kaan Kara, Hantian Zhang, Gustavo Alonso, Onur Mutlu, Ce Zhang}
\vldbDOI{https://doi.org/10.14778/3317315.3317322}
\vldbVolume{12}
\vldbNumber{7}
\vldbYear{2019}

\fi

\begin{document}

\sloppy %%%%This is important to out-of-margin staff....

\ifarxiv
\title{ Accelerating Generalized Linear Models with
MLWeaving:\\
A One-Size-Fits-All System for Any-Precision Learning (Technical Report)} 
%\setcopyright{none} 
\else
\title{ Accelerating Generalized Linear Models with
MLWeaving:\\
A One-Size-Fits-All System for Any-Precision Learning} 
\fi
	
\numberofauthors{6} 
\author{
	Zeke Wang, Kaan Kara, Hantian Zhang, Gustavo Alonso, Onur Mutlu, Ce Zhang \\
	%\and
	\affaddr{\parbox{8cm}{\centering Systems Group, Department of Computer Science ETH Zurich, Switzerland \\ firstname.lastname@inf.ethz.ch}} 
}

	\maketitle

\begin{abstract}
%Decades of database research have left us with a 
%diverse range of data structures---sometimes
%called ``indices''---which can be used
%to accelerate relational queries. As modern database
%systems are increasingly being enriched by machine
%learning functionalities, we wonder: {\em 
%Can we build auxiliary data structures to
%accelerate the training of machine learning models over
%relational data?}
Learning from the data stored in a database is an important function increasingly available in relational engines. Methods using lower precision input data are of special interest given their overall higher efficiency. However, in databases, these methods have a hidden cost: the quantization of the real value into a smaller number is an expensive step. 
To address this issue, we present MLWeaving, a data structure and hardware acceleration technique intended to speed up learning of generalized linear models over low precision data. MLWeaving provides a compact in-memory representation that enables the retrieval of data at any level of precision. MLWeaving also provides a highly efficient implementation of stochastic gradient descent on FPGAs
%The solution adopted in MLWeaving is more efficient than existing designs in terms of space (since it can process any resolution on the same design) and resources (via the use of bit-serial multipliers). %\st{, and model convergence time (due to a low overhead synchronous approach used in the model calculation)}.
%\reversemarginpar
%\kaancomment{We should remove: , and model convergence time...} 
and enables the dynamic tuning of precision, instead of using a fixed precision level during learning. % on a \emph{per epoch} basis
%We illustrate this using a simple, dynamic precision schedule.
%As such, even a simple and fixed tuning policy can significantly speed up training while maintaining accuracy. }
%Moreover, in the paper we propose a method whereby the learning algorithm self-adjusts the level of precision, freeing up the user from determining first which resolution would work for the problem at hand. 
Experimental results show that MLWeaving converges up to $16 \times$ faster than low-precision implementations of first-order methods on CPUs. 
%over an FPGA implementation using full-precision. 
\end{abstract}

%that allows us to accelerate the training
%of machine learning models by taking advantage of
%the low-precision representation of the data.
%Given an input relation stored in full precision
%(e.g., 32-bits), MLWeaving, which is as large
%as a single copy of the data, provides a way to 
%transfer the input relation into multiple 
%precision levels (from 1 bit to 32 bits). Conceptually, MLWeaving is inspired
%by BitWeaving for relational queries, and our
%first contribution is the MLWeaving storage layout, which is specifically designed for machine learning.

%We then study how to use MLWeaving
%to accelerate the training of machine learning models.
%Our second contribution is an FPGA-based implementation
%that takes advantage of the MLWeaving data structure.
%There are two prominent features of our design.
%First, our design uses synchronous execution, which requires fewer epochs to converge than its 
%asynchronous counterpart, while maintaining roughly 
%the same computation speed on FPGA. 
%Second, as MLWeaving allows the precision to be changed dynamically per epoch, we propose
%an adaptive training system in which the user does not need to specify the precision level at all.

%Moreover, the adaptive precision
%tuning algorithm is robust across all datasets.
%Without any user intervention on the desired precision level, it matches, and sometimes
%even slightly outperforms, a well-tuned fixed-precision implementation.

\vspace{-2ex}
\section{Introduction}

Database engines have started to provide support for training machine learning (ML) models over relational data (e.g., MADlib~\cite{madlib_vldb12}). Generalized linear models, such as support vector machines and logistic regression solved using stochastic gradient descent (SGD), are among the most common approaches used within databases. Although useful in many applications, these models are expensive to compute and, thus, there is a great deal of activity exploring ways to reduce the overhead of training. One promising approach is quantization applied to either the values in the model~\cite{deep_compression_iclr16, brainwave_micro18, umuroglu2017finn} or, our focus in this paper, lower precision input data. Using lower precision on the input data reduces the amount of data accessed during training, thereby shortening training times~\cite{zipml_icml17}.

From a database perspective, these approaches are interesting as they alleviate the memory bottleneck. However, existing approaches quantizing the input data implicitly assume that either the data is always available in the correct precision
(e.g., for 32-bit fixed-point numbers, 32 copies of the original data with precision ranging from 1 to 32 bits are required) or the data is pre-quantized at its source (e.g., memory or disk)~\cite{asyn_lp_sgd_isca17, sgd_fpga_fccm17}.
A further challenge is that the correct precision needed by each application is not always known in advance and varies depending on the statistical characteristics of the data.
The overhead for machine learning that uses quantized datasets is even bigger when hardware accelerators are used, since existing solutions require 1) a different microarchitecture for each precision level and 2) a separate copy of the quantized data at the right level of precision \cite{sgd_fpga_fccm17}. 

To address these issues, we have designed MLWeaving, a novel \textit{end-to-end} system enabling \textit{any-precision} learning of generalized linear models in database engines. We implement a prototype of the MLWeaving system in DoppioDB, a column store database augmented with FPGA acceleration \cite{sidler2017doppiodb}, as a first step to explore how database storage formats can be combined with the requirements of machine learning algorithms. MLWeaving has two key innovations that make the learning process on FPGAs quantization-friendly ({\bf C1}) and synchronous ({\bf C2}).

%components:
%({\bf C1}) 
%We illustrate how to do this using a simple dynamic, per-epoch precision schedule (we leave the exploration of more efficient dynamic schedules to future work).

%({\bf C2}) A further contribution is an efficient synchronous computation scheme built into the hardware accelerator  , making training more appealing on FPGAs when compared with other devices such as CPUs or existing designs requiring different circuits for each level of precision.

% First, inspired by the database community, we propose a memory layout storing same-significant bits subsequently in the memory,
% leading to more efficient data access during quantized training.
% Second, we go beyond the von Neumann (vN) architectures and design a specialized FPGA-based architecture providing computation capability natively on our quantization-
% optimized memory layout.

\vspace{0.3em}
\noindent
{\bf C1: Flexible Memory Layout and Hardware Implementation. } 
%Instead, MLWeaving uses an auxiliary data structure inspired by BitWeaving~\cite{bitweaving_sigmod13} to represent all possible levels of precision in a compact form and facilitate their efficient retrieval in the same way an index works.Inspired by BitWeaving~\cite{bitweaving_sigmod13}, 
MLWeaving combines 1) a memory layout supporting the efficient retrieval of the input data at \emph{any level of precision} and 2) an FPGA-based design providing hardware acceleration to speed up SGD regardless of the precision used. 
%The approach has the additional advantage of enabling the dynamic adjustment of the precision used during training. 
The key idea behind MLWeaving is a transposed memory layout where different bits of a given value are stored separately. SGD is evaluated \emph{sample-at-a-time}, i.e., all the features of the sample are read before the gradient is computed. At full precision, reading a row corresponding to a data point accesses all the features for that sample. MLWeaving vertically partitions each data point (a row in a table) at the bit level so that the first bit of all features of a data point are stored consecutively, then the second bit, etc. This provides two benefits. First, the number of memory accesses needed to read a value is proportional to the precision used. Lower precision leads to fewer memory accesses. Second, the format allows the serialization of the data into a hardware accelerator in the form of a bit stream, improving the memory bandwidth utilization. 

We show that the bit stream format can be used to compute the gradient using bit-serial multipliers (i.e., process all the first bits in the first cycle, all the second bits in the second cycle, etc.~\cite{bit_serial_assp89, bismo_fpl18, filter_adaptive_precision_asicsoc99, dct_vlsi99, convnets_vlsic16, stripes_micro16, connection_machine_mit86}) a more efficient approach than that used in existing systems~\cite{asyn_lp_sgd_isca17, sgd_fpga_fccm17}. The resulting design employs only one third of the resources used in previous solutions and allows higher frequency, doubling the rate at which data can be processed on an FPGA accelerator. Furthermore, our approach supports dynamic selection of the level of precision by simply reading more or fewer bits when processing the data in each epoch of the SGD algorithm. %since these days most machine learning takes place on hardware accelerators

\vspace{0.3em}
\noindent
%\kaancomment{You are still presenting synchronous as a major contribution.}
{\bf C2:  Efficient Synchronous Execution. }
Due to the sequential nature of the SGD algorithm, a common approach to parallelizing it on modern CPUs/GPUs is to perform asynchronous updates to the model.
%to minimize the overhead of 
%\kaancomment{Why is vN related to asynchronous updates?}
By doing so, different cores do not need to acquire expensive locks for every gradient calculation~\cite{hogwild_NIPS11, modelAveraging_NIPS2010} (thus better hardware efficiency). This approach is guaranteed to converge under certain (often mild) conditions, but it could converge slower than the synchronous approach (thus worse statistical efficiency).
In MLWeaving, we show how synchronous SGD on an FPGA can be made almost as efficient, in terms of hardware efficiency, as asynchronous SGD on an FPGA. Meanwhile, MLWeaving often requires fewer epochs to converge than its asynchronous {\em CPU} counterpart. As a result, MLWeaving converges faster, in terms of end-to-end performance, than its asynchronous CPU counterpart.%For example, MLWeaving with a 3-bit precision requires only 40 epochs to converge for the dataset Epsilon, while the full-precision CPU counterpart needs at least 199 epochs to converge to the same loss (Figure~\ref{fig_loss_epoch_epsilon}). 

\vspace{0.3em}
%\noindent
We have implemented MLWeaving on an Intel Arria 10 FPGA using Intel's Xeon+FPGA platform (HARP)~\cite{harp2_fpl16} where the FPGA is integrated in the same package as a Xeon multicore CPU.  
%Experimental results show that MLWeaving achieves up to $27 \times$ speedup over state-of-the-art CPU implementations and up to $10.6\times$ over an FPGA implementation with full-precision inputs. 
Experimental results show that MLWeaving achieves up to $16 \times$ performance improvement over the state-of-the-art low-precision first-order CPU implementation (Table~\ref{table_of_per_epoch_tuning_precision}). % (or $25 \times$ memory traffic reduction)

\setlength{\textfloatsep}{1.5pt}

\vspace{-2ex}
\section{Background}
\label{sec_preliminaries}

MLWeaving combines ideas from several areas: databases, machine learning, and computer architecture. In this section we introduce the necessary background to understand the overall design. Table~\ref{t_parameters} summarizes the notation used throughout the paper.
\begin{table} [ht]
	\centering
	%\begin{spacing}{0.3}
	\begin{scriptsize}
    \vspace{-1ex}
	\caption{Notation used in the paper}
		\label{t_parameters}
		\vspace{-1.5ex}
		\begin{tabular}{|c||c|c|}
			\hline
			\textbf{Term} & \textbf{Definition} & \textbf{Range}\\
			\hline
			\hline
			$N$ & Number of samples for training &  Input\\
			\hline
			$M$ & Number of features in the sample &  Input\\
			\hline
			$B$ & Mini batch size & Hyper-param \\
			\hline			
			$\lambda$ & Learning rate & Hyper-param \\
			\hline			

			$s$ & Number of bits used at runtime & Input\\
			\hline
			
			$\vec{x}$ &  Model, i.e.,  a vector of parameters &  Output\\
			\hline
						
			$S$ & Maximum number of bits of the quantized value &  32\\
			\hline
			$T_{s}$ & $s$-bit fixed-point table (i.e., training dataset)  & $1 \le s\le S$\\
			\hline
			$a^{[i]}$ & $i$-th bit of fixed-point value $a$, $1 \le i \le S$  & 0 or 1\\			
			\hline
			%$T_{S}$
			$Q_s(A)$ & $s$-bit quantized value of full-precision A  & $1 \le s\le S$\\
			\hline			
			\hline			
			
			$\#CL$ & Number of bits of a cache line    & 512 bits\\
			\hline			
			$\#Freq$ & Frequency of the SGD hardware design  & 400 MHz\\
			\hline
			$\#M_{max}$ & Maximum dimensions of supported model & 32K\\
			\hline
			$\#Bank$ & Number of banks implemented in hardware  & 8\\
			\hline
		\end{tabular}
	\end{scriptsize}
		\vspace{-2ex}
	%\end{spacing}
\end{table}

%and preliminaries. We first discuss low precision
%data representations and mathematics, followed by 
%current theory and hardware-based implementation 
%of stochastic gradient descent
%over low-precision data representation.  

% of the pre-processing, low-precision stochastic gradient descent. %Then, we discuss the main challenges of the existing approach. % efficient storage layouts and the processing of complex predicates.
%

\vspace{-1ex}
\subsection{Normalization and Quantization } %  Machine Learning Arithmetic

For all ML algorithms, raw data must be converted into a suitable format before the learning process. In MLWeaving, data is stored in the format resulting from a normalization and a quantization step. 

%Most existing work on training machine learning
%models with low precision data representation 
%relies on very similar assumptions on quantization
%and low precision arithmetic. We now introduce 
%related background material.

%The necessary pre-processing steps of low-precision machine learning are normalization and quantization. 

%{\bf Normalization and Quantization. }
%\subsubsection{Normalization}
%Without normalization, the range of the values 
%of each feature can vary dramatically. 
\vspace{0.5ex}
\noindent
{\bf Normalization} reduces the range of values without affecting the overall result of the training. 
%Most work on \wzk{ML} assumes that data has been normalized before the learning process. 
Without loss of generality, MLWeaving uses the scaled range [0, 1]. Accordingly, for each column, the original value $f$ is normalized to the value $\tilde{f}$ (Equation~\ref{E_normalization}): %, as shown in Equation~\ref{E_normalization}
%In particular, we rescale the range of each feature to [0, 1].
\begin{equation} \begin{scriptsize}
\label{E_normalization}
	\vspace{-1ex}
\tilde{f} = \frac{f-f_{min}}{f_{max} - f_{min}},
	\vspace{-1ex}
\end{scriptsize} \end{equation}
where $f_{min}$ and $f_{max}$ are the maximum and minimum values in the column, respectively. An advantage of learning inside the database engine is that normalization can be accomplished using either the meta-data available on a relational table or computed using standard SQL (min, max). 

\vspace{0.5ex}
\noindent
{\bf Quantization} happens over the normalized dataset. It involves converting a full-precision floating-point value $\tilde{f}$ into a lower-precision fixed-point value. %\tilde{ }
Specifically, our goal is to construct a function 
$Q_s: \mathbb{R}^{d} \mapsto \mathbb{F}^{d}$, a \emph{deterministic quantization function} that maps a floating-point value to an $s$-bit fixed-point representation $Q_s(a)$.%\tilde{ } 32-bit  \footnote{We can prove that our deterministic quantization on the training dataset can converge with theoretical guarantee in Subsection~\ref{subsection_theorotiacal}.} 
The quantization process is conducted in two steps. 

%, represented in Equation~\ref{E_number_representation}.
%\begin{equation} \begin{scriptsize}
%\label{E_number_representation}
%	\vspace{-1ex}
%a =  $\tilde{f}$
%	\vspace{-1ex}
%\end{scriptsize} \end{equation}

\begin{figure}[ht]
	\centering
	\includegraphics[width=7.9cm]{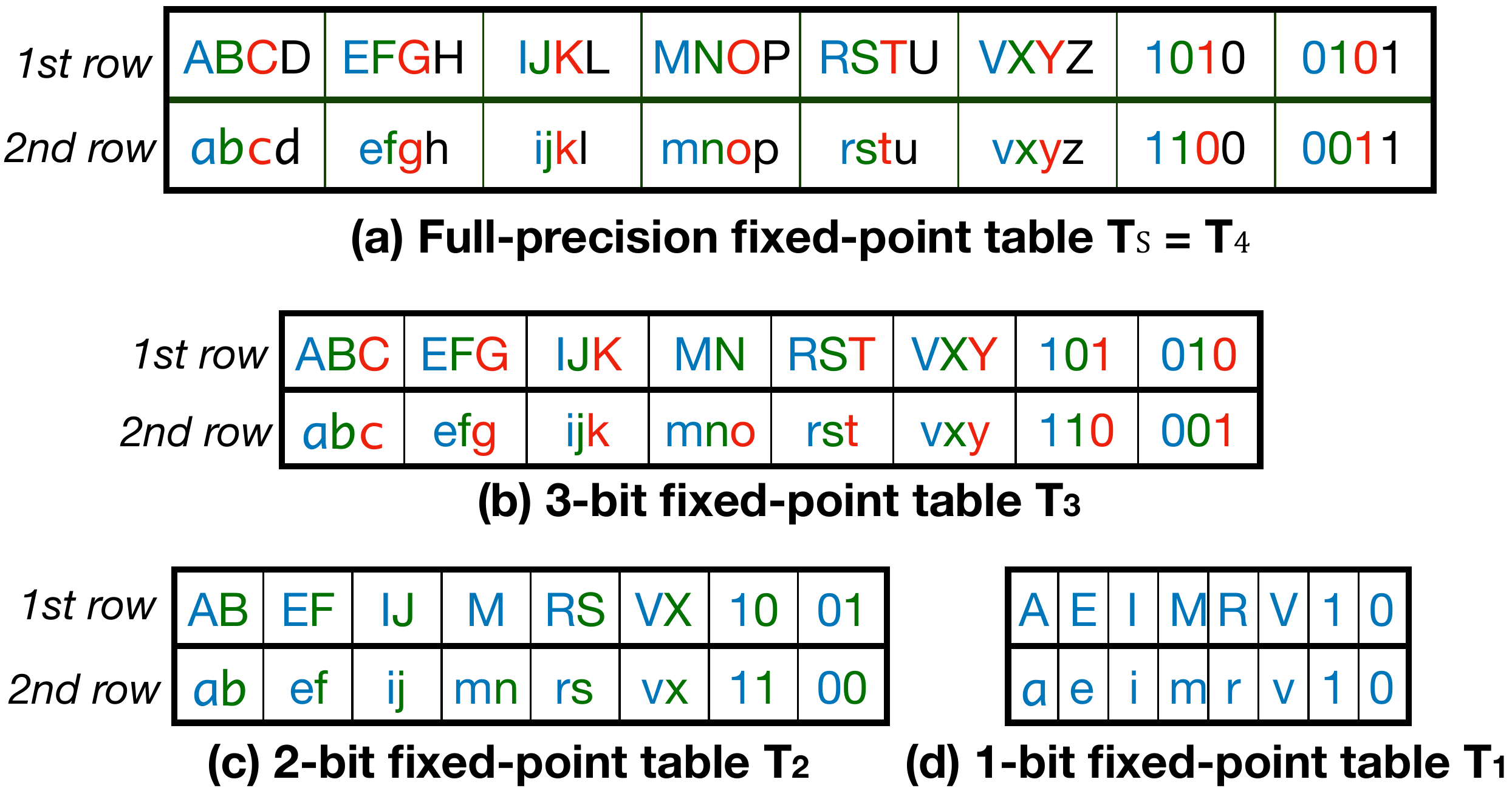} %
	\vspace{-1.4ex}
	\caption{Four fixed-point quantized tables ($T_4$, $T_3$, $T_2$, $T_1$) containing 2 data points with 8 features. Each table has two rows and eight columns, and each element has four bits, where the symbol (e.g., A-Z, a-z) is binary, 0 or 1. }
	\vspace{-2.5ex}
	\label{fig_quantization}
\end{figure}

{\em 1, Floating Point to Fixed Point Conversion.}
The first step is to generate a 
{\em full-precision fixed-point table $T_S$},
where $S$ is the maximum number of bits of the quantized value in the table. The new fixed-point value ($\tilde{a}$) in $T_S$ is calculated to be $\tilde{f}$ multiplied by $2^S-1$ (instead of $2^S$),\footnote{$\tilde{f} = 1.0$ leads to $\tilde{a} = 2^S-1$.} since the range of each floating-point value in the normalized table is between 0 and 1. Therefore, the larger the $S$ value, the higher the precision. 
In our FPGA design, the fixed-point value $\tilde{a}$ is interpreted as: 
\begin{equation} \begin{scriptsize}
	\vspace{-1ex}
%\tilde{f} = \frac{f-f_{min}}{f_{max} - f_{min}},
    \tilde{a} = \sum_{i=1}^{S} \tilde{a}^{[i]} \times 2^{-i}, 
	\vspace{-1ex}
\label{E_normalization_kk}
\end{scriptsize} \end{equation} %	\label{equ_bit_comp}
where $\tilde{a}^{[i]}$ represents the $i$-th bit of $\tilde{a}$, 0 or 1. The first bit ($i$ = 1) is the most significant bit.
%Without loss of any numerical accuracy, the original floating-point normalized value (a) is quantized via the deterministic quantization function $Q_{S}$, where $Q_{S}$ is the maximum number of bits of the quantized value. 
%From now on, we refer the training dataset (i.e., table) to .
Figure~\ref{fig_quantization}(a) shows an example of a full-precision fixed-point table $T_{S}$ where $S = 4$. The quantized value ($\tilde{a}=1010_2$) of the seventh column and the first row\footnote{We use three pairs of words (table, training dataset), (row, sample) and (column, feature) interchangeably within each pair.} represents the value of $\tilde{f}$:     %shows the full-precision fixed-point table , where.  
%\begin{equation} \begin{scriptsize}
$1 \times 2^{-1} + 0 \times 2^{-2} + 1 \times 2^{-3} + 0 \times 2^{-4} = 0.625$. 
%\end{scriptsize} \end{equation}

% and their corresponding custom arithmetics, where $T_S = T_4$. Each level of precision has a custom quantized table (software side) and the corresponding arithmetic (hardware side). 

{\em 2, Fixed Point Quantization.}
The second step of $Q_s$ is to quantize the fixed point table $T_S$ into the desired level of precision. This step is performed by keeping the $s$ most significant bits of $T_S$. 
Figures~\ref{fig_quantization}(b-d) illustrate the quantized 3-bit, 2-bit and 1-bit fixed-point tables for the original data shown in Figure~\ref{fig_quantization}(a).
For example, the full-precision fixed point representation $1010_2$ in $T_4$ is $101_2$, $10_2$, and $1_2$ for 3, 2, and 1 bit precision, respectively. 

%, as shown in 
%Figure~\ref{fig_quantization}.

%After performing the deterministic quantization function $Q_s$ ($s$ is 1, 2 or 3) on the full-precision fixed-point number, the new quantized value $Q_s(a)$ is computed, as shown in Equation~\ref{E_quantization_equation}. 
%\begin{equation} \begin{scriptsize}
%\label{E_quantization_equation}
%	\vspace{-1ex}
%Q_s(a) = \sum_{i=1}^{s} a^{[i]} \times 2^{-i}
%	\vspace{-1ex}
%\end{scriptsize} \end{equation}

%To take advantage of low-precision data representation, the full precision dataset can be quantized into low-precision format, e.g., 8-bit, to reduce the amount of data movement. Then, the low-precision multiplier (e.g., 8-bit) is employed to do the corresponding computation. Therefore, both the data movement and computational resources can be significantly reduced, compared with the original full precision format.

\vspace{-1ex}  
\subsection{Low-precision SGD}%Stochastic Gradient Descent
\label{subsection_lp_sgd}

SGD is a popular algorithm to train generalized linear models. Given a relation $A$, the full precision SGD solves the following optimization problem:
	\vspace{-1ex}
\[
\underset{\vec{x}}{\text{minimize}}: \frac{1}{N}  \sum_{i=1}^{N} f(\vec{x} \cdot \vec{a_i}, b_i),
\]
%	\vspace{-1ex}
where $\vec{a_i}$ is one row in the input relation, $b_i$ is the corresponding training label, $\vec{x}$ is the model, and $f(-)$ is a loss function. SGD solves this problem by iteratively scanning the input relation $A$ --
for each row $\vec{a_i}$, it calculates the gradient with respect to $\vec{x}$, and updates the model. Each pass over the input data is called an {\em epoch}. SGD usually runs for multiple epochs until convergence. 

Training SGD over low precision
representation of the input data using a generalized linear model targets the following function: 
%\begin{equation} \begin{scriptsize}
%\label{E_sgd_objective}
\[
	\vspace{-1ex}
\underset{\vec{x}}{\text{minimize}}: \frac{1}{N} \sum_{i=1}^{N} f(\vec{x}\cdot Q_s(\vec{a_i}), b_i),
    \vspace{-1ex}
\]
%\end{scriptsize} \end{equation}
	
where the $i$-th sample consists of a vector of quantized $s$-bit values ($Q_s(\vec{a_i}) \in \mathbb{F}^{1 \times M}$). % and the true label value ($b_i \in \mathbb{R}^{1}$)

%SGD over deterministically quantized input data is known to be biased~\cite{zipml_icml17} and converges to a slightly different solution than the full precision counterpart.
%Moreover, it is difficult to automatically map an error tolerance of the solution to the precision level in the input. This motivates the adaptive precision scheduler used in MLWeaving, which increases the precision level during execution to provably converge to the same solution as full precision SGD, while taking advantage of the speedups introduced by low precision data representation.

\vspace{1ex}
\noindent
{\bf Low-Precision Mini-batch SGD on FPGAs.} One variant of SGD that is popularly implemented in many systems is {\em mini-batch SGD}~\cite{mini_batch_sgd_kdd14} -- instead of calculating the gradient using a single sample, mini-batch SGD uses multiple samples, called a ``mini-batch'', to calculate the \emph{average} gradient. Mini-batch SGD can be easier to accelerate because all samples in a mini-batch share the same model and thus can be processed independently in parallel. For applications such as distributed deep learning for image classification, mini-batch SGD, instead of the standard SGD, is the {\em de facto} training algorithm. When deploying SGDs on FPGAs, we also use mini-batch SGD~\cite{sgd_fpga_fccm17}.\footnote{We explain more in Section~\ref{sec_mlweaving}.} 

Algorithm~\ref{alg_sgd_flow_bank} illustrates the flow of low-precision mini-batch SGD, which is iteratively evaluated in $E$ epochs (Line 1). 
In each epoch, the entire low-precision training dataset is scanned, one mini-batch of $B$ samples per iteration (Line 2). Inside a mini-batch, we initialize the average gradient ($\vec{g}$) to zero at the beginning (Line 3), and compute the average gradient of this mini-batch (Lines 4-13).\footnote{In this context, we assume that $Bank$ is 1 for ease of understanding such that one sample is processed at a time. $Bank$ is 8 in MLWeaving that addresses the constraints encountered when deploying any-precision SGD on an FPGA (Subsection~\ref{subsec_mlweaving}). } 
Each sample is processed as follows.
First, we compute the dot product of two vectors: the $i$-th low-precision sample $Q_s(\vec{a_i})$ and the full-precision model\footnote{In this paper, we do not consider the quantization of the model.} $\vec{x}$ (Line 8). Its output $a\_dot\_x$ is a full-precision scalar value.
Second, we compute the scaling value $scale$, based on the derivative (i.e., $df$) of the given loss function (Line 9). For different learning algorithms, we only need to modify this function to compute the scaling value, while keeping the other parts unchanged. 
Third, the gradient ($\vec{g_{j+k}}$) of this sample is computed (Line 10).
Fourth, $\vec{g_{j+k}}$ is accumulated into the gradient of this mini-batch (Line 11). 
Fifth, the model ($\vec{x}$) is updated with the average gradient $\vec{g}$ (Line 14).
\begin{algorithm} [t]
	\SetAlFnt{\tiny} \linespread{1.0} \selectfont \caption{\sc Low-Precision Mini-Batch SGD}% on FPGAs
	\label{alg_sgd_flow_bank}
	\SetKwInOut{Input}{Input}
	\SetKwInOut{Output}{Output}	
    \newcommand\mycommfont[1]{\notsotiny\ttfamily\textcolor{blue}{#1}}
    \SetCommentSty{mycommfont}
	
	\begin{scriptsize}
		%	\DontPrintSemicolon
		\Input{$N$: number of samples, \\
		    $E$: number of epochs, \\
			$\gamma$: learning rate, \\
			$Q_s(\vec{a_i})$: $s$-bit quantized data set of the $i$-th sample, $\in \mathbb{F}^{1 \times M}$, \\ %Q_s(\vec{a_i})
			$b_i$: label value of the $n$-th sample, $b_i \in \mathbb{R}^{1}$. }
		
		\Output{$\vec{x}$: model with a set of parameters.}
		
		%\tcc{Add the ($\vec{a_i}, b_i$)'s gradient to $\vec{x}$.}
\tcc{Evaluate the $e$-th epoch.} %//Evaluate the $e$-th epoch.%
\For{$e = 1$ \KwTo $E$}{ 
            \tcc{Mini-batch processing}
	\For{($i = 0$; $i < N$; $i += B$)}{
	 \tcc{Zero the gradient of this mini-batch}
		$\vec{g}$ = 0;	\\
		\For{($j = 0$; $j < B$; $j += \#Bank$)}{
            \tcc{Multi-bank processing}
			$\#$pragma parallel in hardware \\
			\For{($k = 0$; $k < \#Bank$; $k++$)}{
				int32 $t = i + j +k$; \\
				\tcc{Dot product}
			int32 $a\_dot\_x = Q_s(\vec{a_{t})} \boldsymbol{\cdot} \vec{x}$;
	
			\tcc{Serial part}%\tcc{$df$ varies with ML algorithm.} %, scalar computation
			int32 $scale = \gamma \times df(a\_dot\_x, b_{i+j+k})$;
			
			\tcc{Gradient computation}%\tcc{Compute the gradient for each bank.} % of $i$-th sample
			$\vec{g_{j+k}} = scale \times Q_s(\vec{a_{t}})$;
			
			\tcc{Gradient accumulation}%\tcc{Accumulate the gradient.} % of $i$-th sample
			$\vec{g} = \vec{g} + \vec{g_{j+k}}$;
			}
		}
		\tcc{Model update}%%\tcc{Accumulate the gradient to $\vec{x}$.} %($\vec{a_i}, b_i$)'s
		$\vec{x} = \vec{x} - \vec{g}/B$;
	}
}	
	\end{scriptsize}
\end{algorithm}

% \vspace{-1ex}
% \begin{algorithm} [ht]
% 	\SetAlFnt{\tiny} \linespread{1.0} \selectfont \caption{\sc Low-precision SGD}
% 	\SetKwInOut{Input}{Input}
% 	\SetKwInOut{Output}{Output}	
%     \newcommand\mycommfont[1]{\footnotesize\ttfamily\textcolor{blue}{#1}}
%     \SetCommentSty{mycommfont}
% 	\label{alg_sgd_flow}
% 	\begin{scriptsize}
% 		%	\DontPrintSemicolon
% 		\Input{$N$: number of samples, \\
% 			$E$: number of epochs, \\
% 			$df$: derivative of the given loss function, \\
% 			$\gamma$: learning rate, \\
% 			$Q_s(\vec{a_i})$: $s$-bit quantized data set of the $i$-th sample, $\in \mathbb{F}^{1 \times M}$, \\ %Q_s(\vec{a_i})
% 			$b_i$: label value of the $n$-th sample, $b_i \in \mathbb{R}^{1}$. }
		
% 		\Output{$\vec{x}$: model with a set of parameters.}
		
% 		%\tcc{Evaluate the $e$-th epoch.} %//Evaluate the $e$-th epoch.%
% 		\For{$e = 1$ \KwTo $E$}{ 
			
% 			%\tcc{Evaluate each sample}%\tcc{Add the ($\vec{a_i}, b_i$)'s gradient to $\vec{x}$.}
% 			\For{$i = 1$ \KwTo $N$}{	
% 				\tcc{Dot product}%\tcc{Compute the dot product of $\vec{a_i}$ and $\vec{x}$.}
% 				int32 $a\_dot\_x = Q_s(\vec{a_i}) \boldsymbol{\cdot} \vec{x}$;
				
% 				\tcc{Serial part}%\tcc{$df$ varies with ML algorithm.} %, scalar computation
% 				int32 $scale = \gamma \times df(a\_dot\_x, b_i)$;
				
% 				\tcc{Gradient computation}%\tcc{Compute the gradient.} % of $i$-th sample
% 				$\vec{g} = scale \times Q_s(\vec{a_i})$;
				
% 				\tcc{Model update }%\tcc{Add the gradient to $\vec{x}$.} %($\vec{a_i}, b_i$)'s
% 				$\vec{x} = \vec{x} - \vec{g}$;
% 			}
% 		}
% 	\end{scriptsize}
% \end{algorithm}
% \vspace{-1ex}

\vspace{0.5ex}
\noindent
{\bf Performance Metrics.} The end-to-end performance of SGD can be measured
as the time that it takes to achieve the target
loss. This can be further decomposed into
two metrics~\cite{dimmwitted_vldb14} that we will use throughout this paper. \emph{Hardware efficiency} measures 
the time that SGD requires to finish one epoch. \emph{Statistical efficiency} represents the number of epochs that SGD requires to converge.

\vspace{-1ex}
\subsection{Hardware Acceleration}
Most ML algorithms are known to be compute- and data-intensive.
%\kaan{\st{ML is a computationally demanding process both as a result of the actual computation (SGD) as well as because of the number of iterations needed to produce a model and the number of versions of the same model needed to reach a solution suitable for inference.}}f
Not surprisingly, in recent years, we have seen a significant increase in the number of specialized hardware solutions for ML, from GPUs to FPGAs to specialized processors such as TPUs~\cite{tpu_isca17}. In this paper, we focus on FPGAs since they provide a higher degree of versatility in exploring possible algorithms and designs, something important in an area that is evolving as quickly as ML. Since the early pioneer work exploring the use of FPGAs on databases \cite{Glacier_sigmod10, streams_vldb09, data_processing_vldb09}, a growing number of database operations accelerated with FPGAs have been proposed \cite{pattern_matching_sigmod17, reg_fccm16, partitioning_opencl_fpl15, multi_kernel_tvlsi17, kara2017fpga, query_opencl_fpga_fpl16, bionicdb_cidr13,Ibex_vldb14,histogram_sigmod14}. FPGAs are also increasingly available, specially in cloud platforms such as Microsoft's Catapult \cite{capult_micro16} and Brainwave~\cite{configurable_dnn_isca18, brainwave_micro18} projects, or Amazon F1 instances, making them a suitable target for hardware acceleration. 

%\st{For reasons of space we cannot go into many details of how FPGAs function. We refer the interested reader to, e.g.,} \cite{fpga_book_mc13}. \st{For the purposes of this paper, it suffices to know that FPGAs provide reconfigurable hardware in the form of a number of logic elements that can be use to build different digital circuits.} 
The degree of microarchitectural freedom in designing an algorithm is much higher on an FPGA than in software in general as it is possible to design specialized compute units from scratch and tailor the hardware to the application at hand\kaan{~\cite{fpga_book_mc13}}. 
%As such, we can go beyond the von Neumann (vN) architecture with FPGAs and implement a computing task in a more elegant way.
The limiting constraints in FPGA designs are {\em meeting timing} (how fast the hardware design can be clocked) and {\em resource usage} as the FPGA is a physical device and, at a certain point, there are no more gates available to implement additional function. Often, designs have to balance one against the other and strive for an efficient trade-off between parallelism on the one hand and FPGA resource usage on the other hand.

\vspace{-2ex}
%\input{system_overview.tex}
%\vspace{-2ex}
%\vspace{-1ex}
%\section{System Overview} %: Pitfalls of Existing Low-precision Designs

%\vspace{-0.5ex}
\section{System Overview}
%\kaancomment{Should we introduce Xeon+FPGA also around here with a figure?}

\begin{figure}[b]
\vspace{-0.5ex}
\includegraphics[width=\columnwidth]{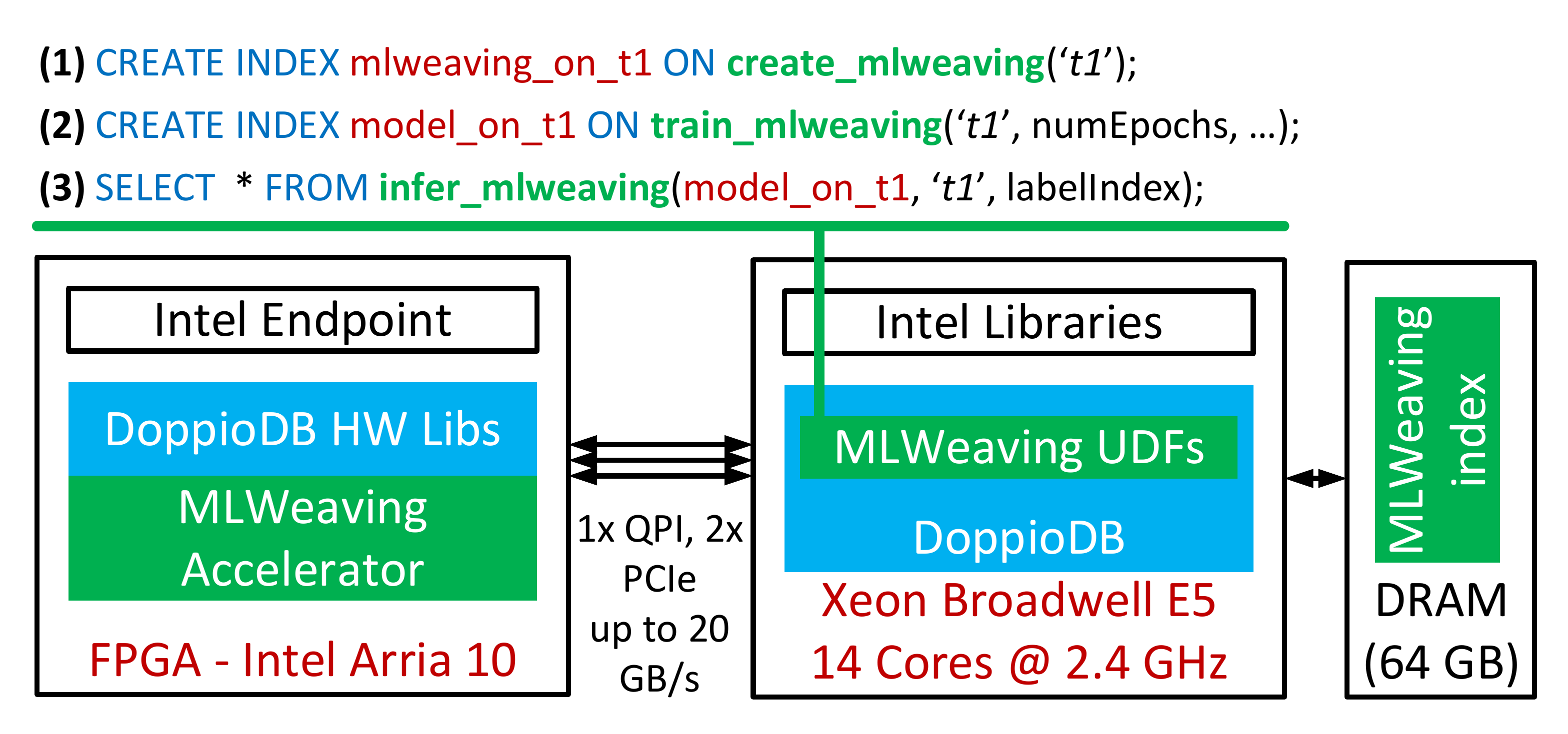}
\vspace{-5ex}
\caption{Target platform (Intel Xeon+FPGA Gen2), overview of DoppioDB on this platform and sample queries with MLWeaving UDFs in DoppioDB.}
\vspace{-0.5ex}
\label{fig_doppiodb}
\end{figure}

\noindent
\textbf{Target Platform.} The target platform (Figure~\ref{fig_doppiodb}) for this work is the 2nd generation Intel Xeon+FPGA~\cite{harp2_fpl16}, combining a Broadwell 14-core CPU E5-2680v4 with an Arria 10 FPGA. The FPGA has cache-coherent access to the main memory (DDR4: 64GB) of the CPU via 1 QPI link and 2 PCIe links, resulting in around 20 GB/s combined read and write bandwidth. We perform all our experiments on this machine.

\noindent
\textbf{Database Integration.} We integrate our FPGA-accelerated training designs into DoppioDB~\cite{sidler2017doppiodb}, an open source solution for FPGA-accelerated databases, based on MonetDB~\cite{idreos2012monetdb}. DoppioDB enables easy integration of FPGA-based accelerators through a set of software and hardware libraries~\cite{centaur_fccm17}. The hardware libraries expose a native memory interface, via which the FPGA accelerators use to access the main memory of the host CPU. On the software side, FPGA accelerators can be started and monitored as separate threads. These FPGA threads run in parallel to software threads.

We implement three User Defined Functions (UDFs) in DoppioDB, as shown in Figure~\ref{fig_doppiodb}. 
%(1) The first UDF is to initiate the creation of the MLWeaving data structure. This data structure is kept internal in the database, similar to an index, associated with a certain table.
(1) The first UDF is to initiate the creation of the \emph{MLWeaving index} that is associated with a certain table and kept internal in the database. 
(2) The second UDF is to initiate training using MLWeaving. Under the hood, DoppioDB will look for the MLWeaving index that belongs to the table given by this UDF. If found, the FPGA thread is started. It uses the MLWeaving index to train a model which then gets transferred to the main memory. (3) The third UDF is to perform inference on tuples using the model that has been trained before. Similar to the MLWeaving index, the trained model is also associated with a certain table and will check during query execution if inference can be performed, conditioned on a model having been trained before.

MLWeaving combined with DoppioDB offers users a SQL front-end for learning models from relational data in a seamless manner as the data selection and transformation needed for the learning can be done using SQL and without incurring expensive data transfers in and out of the database.

\vspace{-1ex}
\section{MLWeaving Design}
\label{sec_mlweaving}
%\vspace{-1ex}
MLWeaving involves many aspects that interact in a tight manner. To make it easier to understand how it works, we develop MLWeaving in three stages: ``Quantized'' (Subsection~~\ref{subsec_quantized}), ``BWeaving" (Subsection~\ref{sec_bit_serial_method}) and ``MLWeaving" (Subsection~\ref{subsec_mlweaving}). Table~\ref{t_comparison} summarizes the comparison results. %, each one with a design addressing a particular issue

\begin{table}[ht]
	\centering
	%\begin{spacing}{0.3}
	\begin{scriptsize}
		\vspace{-2ex}
	\caption{Comparison of three approaches. $\#CL$ is 512. ``n" represents any positive integer. }
	\label{t_comparison}	
	\vspace{-2ex}
		\begin{tabular}{|c||c|c|c|}
			\hline
			\textbf{Hardware Metrics}& \textbf{Quantized} & \textbf{BWeaving} & \textbf{MLWeaving} \\%~\cite{sgd_fpga_fccm17}
			\hline
			\hline
			Supported precision levels & 1  &  32 & 32 \\
			\hline
			Required memory layouts &  Per any $s$-bit  &  1 & 1 \\%One for
			\hline
			Bitwidth (bits) of model $\vec{x}$ &   $\lfloor\tfrac{512}{s}\rfloor \times 32$  &  16K & 2K  \\
			\hline
			Number of banks, $\#Bank$ &  1  &  1 & 8 \\
			\hline			
			Mini-batch size, $B$ &  n  &  n & 8*n \\
			\hline			
		\end{tabular}
		\vspace{-3ex}
	\end{scriptsize}
	%\end{spacing}
\end{table}

\subsection{Quantized}
\label{subsec_quantized}
Understanding of MLWeaving requires understanding the interplay between the data representation and the processing required by the accelerator. In the case of SGD, the most important components in the design are the \emph{computing circuits}, \emph{how the input data is transferred from the external memory}, and \emph{how the model is accessed on the FPGA's local memory}. In the following, we discuss the hardware properties of Quantized~\cite{sgd_fpga_fccm17}.

\vspace{0.5ex}
\noindent
{\bf Computing Circuits. }Quantized employs a fixed circuit for each precision level. It, thus, requires the data to be available at the corresponding precision.  `Quantized'' processes features using fixed-point bit-parallel multipliers and can only support one precision level per hardware design. %The other two designs, ``BWeaving" and ``MLWeaving" use bit-serial multipliers~\cite{bit_serial_assp89}, enabling on-the-fly precision selection. Quantized can support one precision level while both ``BWeaving" and ``MLWeaving" can support 32 precision levels. 

\vspace{0.5ex}
\noindent
{\bf Memory Layouts. }In the Xeon+FPGA platform~\cite{centaur_fccm17} used to implement MLWeaving, the FPGA has cache coherent access to the main memory of the CPU. The data arrives to the FPGA  in the form of cache lines, {\em \#CL} (512 bits per cache line). Efficiency is achieved by processing in parallel as many elements within that cache line as possible. How many elements are within a cache line depends on how the data is stored in memory.  Quantized stores each data point at a given precision in a consecutive manner.
If the precision is 16 bits, every cache line brings in 512/16 data points (i.e., features or columns) to be processed. Because the value for each data point arrives as a whole, the multipliers needed for the gradient computation are based on the 16-bits used to represent a value. Such 16-bit wide multipliers are complex and take up significant hardware space. 
%In ``BWeaving" and ``MLWeaving", data points are represented in memory not as a sequence of values but as an interwoven sequence of bits from different data points. Reading a cache line brings in the first bit of 512 data points. Thus, the gradient for a data point cannot be computed right away. Instead, we use a bit-serial multiplier, where the required multiplication is performed one bit at a time with the result accumulated and added to the next one-bit multiplication until the whole multiplication for a data point is completed. With this design, ``BWeaving" and ``MLWeaving" use one memory layout to support arbitrary-precision data retrieval from memory, while 
Besides, Quantized requires one memory layout for each precision level.

\vspace{0.5ex}
\noindent
{\bf Accessing the Model on FPGAs. }The SGD hardware reads not only the input data from the external memory but also the corresponding entry in the model on the FPGA. In particular, the SGD hardware that operates on each dimension needs to read a value from the model as a whole. Since the SGD hardware processes a great number of dimensions concurrently, an additional design factor we should consider is how to efficiently access the corresponding values of the model. This is referred to the {\em bitwidth of the model}. The bitwidth is a very important parameter in an FPGA design because it affects the complexity of the interconnects and the way data has to move from the model storage to the computing units. The higher the bitwidth, the more complex the design and, thus, the higher the probability that it will not meet timing at higher clock rates.
Quantized processes data points as a whole, so its model bitwidth is $\lfloor\tfrac{512}{s}\rfloor \times 32$ bits, where the precision of the model is 32 bits and $s$ is the precision level of input data. 
%``BWeaving" has a high model bitwidth of $16K$ (32*512), making it unsuitable in real settings.\footnote{In Intel Arria 10 FPGA~\cite{Intel_arria_10_device}, the model is implemented with 20-Kb memory blocks (i.e., M20Ks). Each M20K can provide a 32-bitwidth with ECC (or 40 without ECC) for the model, so a 16K-bitwidth requires 512 M20Ks with ECC (or 410 without ECC). It is extremely difficult to access 410 M20Ks in a lock-step manner (e.g., sharing the same read/write address) while maintaining high frequency, as M20Ks are uniformly distributed inside an FPGA.} 
%``MLWeaving" reduces the bitwidth of the model by using a mini-batch SGD algorithm (Subsection~\ref{subsection_lp_sgd}) where each mini-batch is processed without modifying the model. This allows MLWeaving to adopt a multiple-bank hardware design (similar to data partitioning but in hardware) such that each bank can accommodate a sample. The hardware design of MLWeaving has 8 banks, each bank only needs to consume 64 bits to achieve the necessary throughput, 512 bits per cycle in the interwoven representation. Since all the banks share the same model, the model bitwidth becomes 2K, significantly smaller than 16K of ``BWeaving" which literally uses only one bank. MLWeaving can only support a mini-batch size, $B$, that must be a multiple of 8. The other two designs work with any positive integer size.

\vspace{-1ex}
\subsection{BWeaving} %: Flexibility \& Performance
\label{sec_bit_serial_method}
BWeaving extends Quantized~\cite{sgd_fpga_fccm17} such that both memory access and computation time linearly decrease with a lower number of bits used for  ML training. %BWeaving contains its own memory layout and arithmetic. %
To do so, we propose a bit-serial memory layout (Subsection~\ref{bweaving_software}) and a specialized hardware design powered by bit-serial multipliers (Subsection~\ref{bweaving_hardware}).

\vspace{-1ex}
\subsubsection{BWeaving Memory Layout (Software)}% (Software Component)
\label{bweaving_software}
%{\bf Objective. }
%Inspired by BitWeaving, 
The BWeaving memory layout is based on ideas proposed by BitWeaving~\cite{bitweaving_sigmod13, widetable_vldb14}. In a nutshell, we transpose the training dataset to allow the runtime selection of the precision level and to reduce memory traffic for lower precision levels. 
Contrary to BitWeaving that performs the weaving on one column, BWeaving transposes each row (a sample) to preserve access locality, since the training dataset is accessed by SGD in a \emph{sample-at-a-time} manner. %the bit-level columnar organization

\begin{figure}[ht]
\vspace{-1ex} 
	\centering
	\includegraphics[width=8cm]{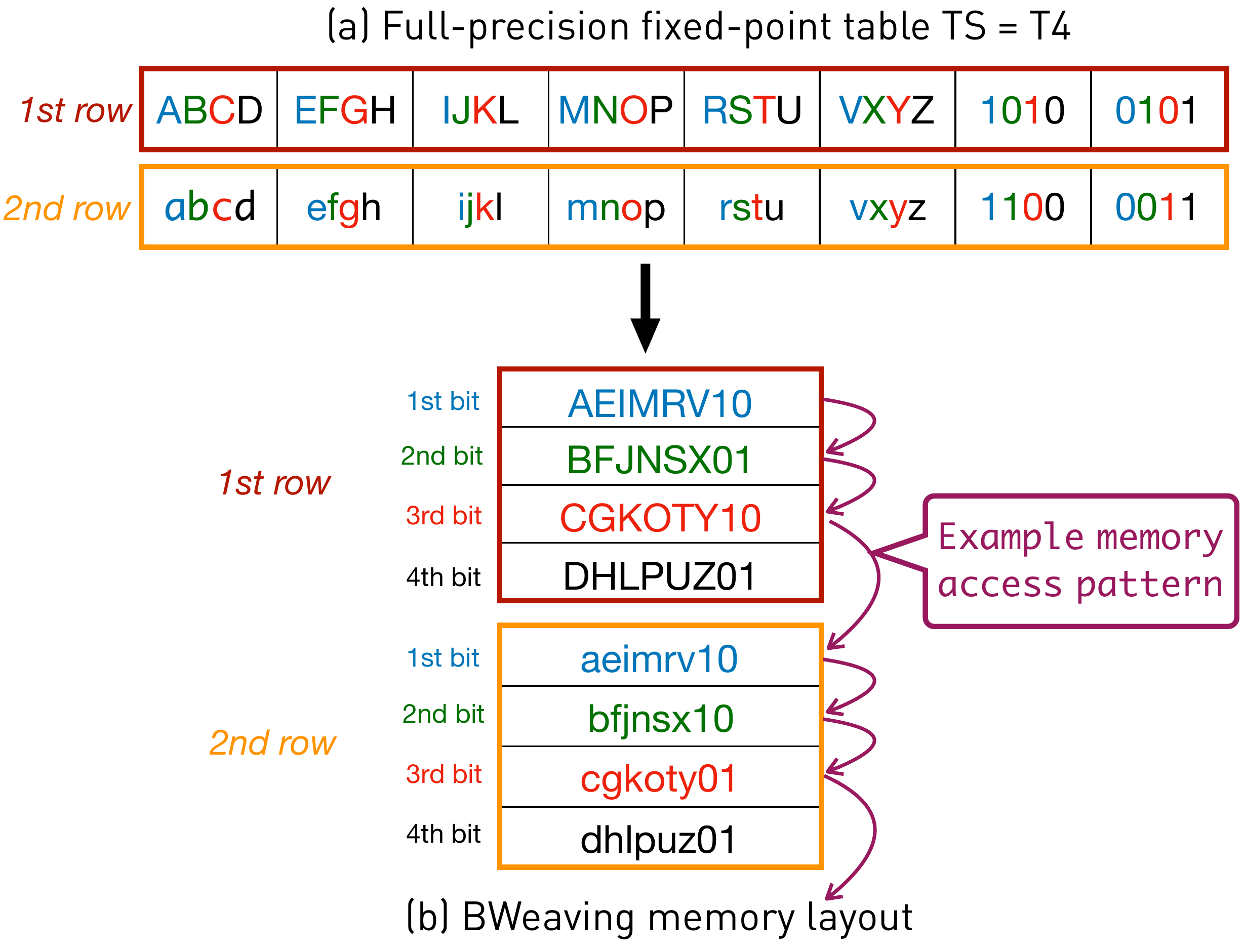}
	\vspace{-2.0ex} %bit_serial_layout
	\caption{Full-precision fixed-point memory layout (a) converted into the BWeaving memory layout (b). Each symbol (e.g., A-Z, a-z) in the table is binary, 0 or 1. The BWeaving memory layout enables the flexible selection of precision in memory. As an example, we show the ``memory access pattern'' with a 3-bit precision ($s$ = 3). }
	\vspace{-1ex} 
	\label{fig_bit_serial_layout}
\end{figure}

The transposition of the data used in BWeaving is shown in Figure~\ref{fig_bit_serial_layout}. $M$ $S$-bit features of a sample are transposed into $S$ $M$-bit words (where $M$ is 8 and $S$ is 4 in the example shown).
Inside the first row, eight 4-bit features are transposed into four 8-bit words. The first bits (i.e., {\color{blue} AEIMRV10}) of the first sample are stored in an 8-bit word.
The second word {\color{red} BFJNSX01}, which contains the second bits, is stored next to the first word, and so on.  
Between rows, the second row is stored consecutively to the first row. 

Under the BWeaving memory layout, data at any level of precision can be retrieved by following a different access pattern over the same data structure. 
In Figure~\ref{fig_bit_serial_layout}(b), we show the accesses needed to retrieve the data set at a precision of 3 bits. The first bits in the first row are accessed, followed by the second and third bits in the first row. Then, the access jumps to the first bits of the second row, skipping the fourth bits of each row. 

\vspace{0.5ex}
\noindent
{\bf Instantiation of Memory Layout. } Table~\ref{t_bit_serial_instance} shows an example of a BWeaving memory layout with $\#CL$ = 512 and $M$ (number of features) = 2048. For instance, the first memory slot (with index = 0) is populated with the first bits of the first 512 features of the first sample, while the second memory slot (with index = 1) is populated with the second bits of the first 512 features of the first sample. If $M$ is not a multiple of 512, we use padding to fit a 512-bit boundary such that we can easily retrieve bits. %\emph{x\_y:z\_w} denotes the \emph{w}-th bits of 512 features (from $z$-th to $y$-th) in the $x$-th sample and $y - z = 511$.
%\newline
%\newline

\begin{table} [t]
	\centering
	%\begin{spacing}{0.3}
	%	\vspace{-1ex} 
	\begin{scriptsize}
	\caption{BWeaving memory layout: \emph{x\_y:z\_w} denotes the \emph{w}-th bits of 512 features (from $z$-th to $y$-th) in the $x$-th sample.}% and $y - z = 511$
	\label{t_bit_serial_instance}	
	\vspace{-1ex}
		\begin{tabular}{|p{2cm}|c||c|}
			\hline
			\textbf{Description} &\textbf{Index} & \textbf{Content ($\#CL$=512 bits)} \\%
			\hline
			\hline
			\hline
			\multirow{6}{*}{\parbox{2cm}{First 512 features of the first sample}} & $0$ & 0\_511:0\_0\\
			\cline{2-3}
			& $1$ & 0\_511:0\_1\\
			\cline{2-3}
			& $2$ & 0\_511:0\_2\\
			\cline{2-3}
			&$3$ & 0\_511:0\_3\\
			\cline{2-3}
			& $...$ & ... \\
			\cline{2-3}
			& $31$ & 0\_511:0\_31\\
			\hline
			\hline
			\multirow{4}{*}{\parbox{2cm}{Second 512 features of the first sample}} & $32$ & 0\_1023:512\_0\\
			\cline{2-3}
			&$33$ & 0\_1023:512\_1\\
			\cline{2-3}
			&$34$ & 0\_1023:512\_2\\
			\cline{2-3}
			&$...$ & ... \\
			\hline
			\hline
			$...$&$...$ & ... \\
			\hline
			\hline			\multirow{3}{*}{\parbox{2cm}{First 512 features of the second sample}} & $128$ & 1\_511:0\_0\\
			\cline{2-3}
			&$129$ & 1\_511:0\_1\\
			\cline{2-3}
			& $...$ & ... \\
			\hline
		\end{tabular}
	\end{scriptsize}
		\vspace{2ex}
\end{table}

%\vspace{-1ex}
\subsubsection{BWeaving Arithmetic (Hardware)} % (Hardware Component)
\label{bweaving_hardware}
Using a bit-serial multiplier operating on the data one bit per cycle~\cite{stripes_micro16, convnets_vlsic16, filter_adaptive_precision_asicsoc99, bismo_fpl18, bit_serial_assp89, dct_vlsi99, connection_machine_mit86}, we design the BWeaving arithmetic that provides bit-level flexibility (i.e., supporting any precision with a single hardware design) while maintaining a processing rate of a cache line per clock cycle.
We now present the difference between bit-serial and bit-parallel multipliers, followed by the SGD hardware design powered by a bit-serial multiplier.  

\begin{figure}[t]
	\vspace{-1.0ex}
	\centering
	\includegraphics[width=8.3cm]{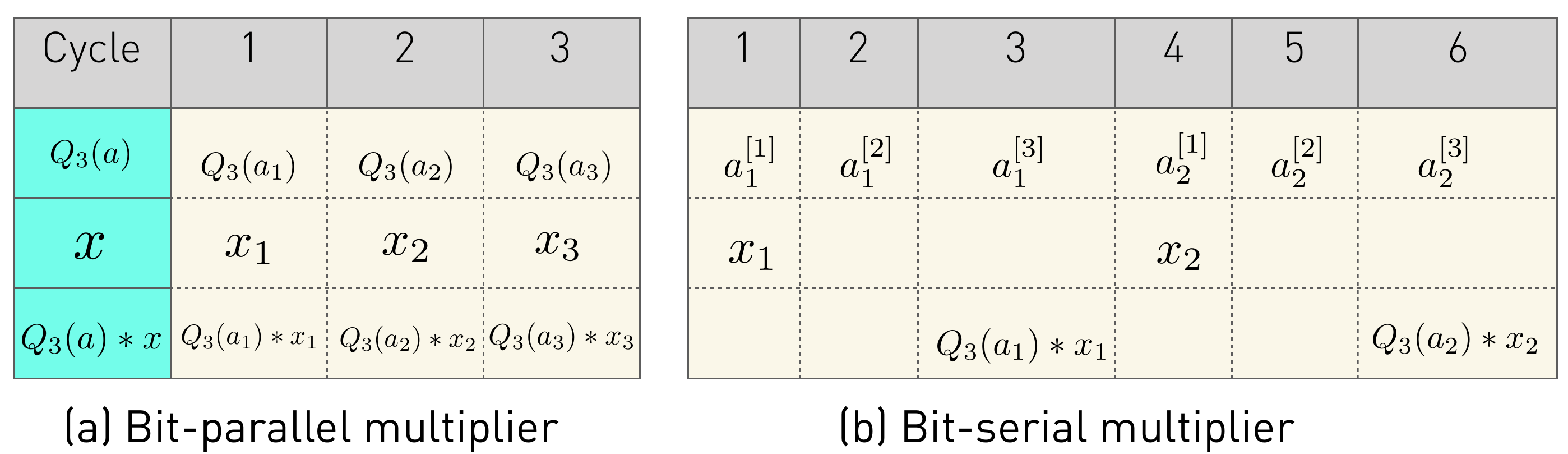}
	\vspace{-1.5ex}
	\caption{Multipliers: bit-parallel (a) vs. bit-serial (b). Bit-parallel multiplier produces one multiplication result per cycle, while bit-serial multiplier produces every three cycles, e.g., on cycle 3 or 6. }
	%\vspace{-3.0ex}
	\label{fig_bit_serial_vs_paralle}
\end{figure}

%the computation time scales linearly with the number of bits used. 
\vspace{0.5ex}
\noindent
{\bf Bit-serial Multiplier vs. Bit-parallel Multiplier}.  
Figure~\ref{fig_bit_serial_vs_paralle} illustrates the difference between a bit-parallel multiplier and a bit-serial multiplier with one example multiplying $3$-bit $Q_3(a)$ (low-precision) by $4$-bit $x$ (full-precision). %Here, 

%Suppose we want to , and w
In a bit-parallel multiplier, each clock cycle can enable the multiplication of two numbers, in this case the quantized input $Q_{3}(a)$ and the corresponding value from the model $x$ (Figure~\ref{fig_bit_serial_vs_paralle}a). This is the type of multiplier used in conventional CPUs and also previous specialized hardware solutions~\cite{sgd_fpga_fccm17}. %However, it cannot efficiently support another level of precision, except $s$-bit.

In a bit-serial multiplier, one multiplication result is produced every three cycles, a bit of $Q_{3}(a)$ per cycle. After $Q_3(a)$ is replaced with $\sum_{i=1}^{3} a^{[i]} \times 2^{-i}$ (Equation~\ref{E_normalization_kk}), the product $Q_3(a) \times x$ (Equation~\ref{E_bit_serialmultiplication}) is computed to be the sum of the product of $a^{[i]}$ and ($x \ggg i$), where the binary value $a^{[i]}$ represents the $i$-th bit of $a$ (0 or 1), $\ggg$ means signed right shift, and $i$ is from 1 to 3. 
%The first bit is the most significant bit.  within three cycles
In terms of cycles, the product $Q_3(a) \times x$ is set to be $a^{[1]} \times (x \ggg 1)$ in the first cycle, $a^{[2]} \times (x \ggg 2)$ is added to the product in the second cycle, and $a^{[3]} \times (x \ggg 3)$ is added in the third cycle. The advantage of bit-serial multiplier is that shift-and-add operations are enough for the calculation, considerably simplifying the design. The disadvantage is that multiple cycles are needed to complete an operation due to its inherent bit-serial nature. 
\begin{equation} \begin{scriptsize}
\label{E_bit_serialmultiplication}
	\vspace{-1ex}
Q_3(a) \times x = x \times \sum_{i=1}^{3} a^{[i]} \times 2^{-i} = \sum_{i=1}^{3} a^{[i]} \times (x \ggg i)
%	\vspace{-1ex}
\end{scriptsize} \end{equation}

\begin{figure}[t]
	\centering
	\includegraphics[width=8cm]{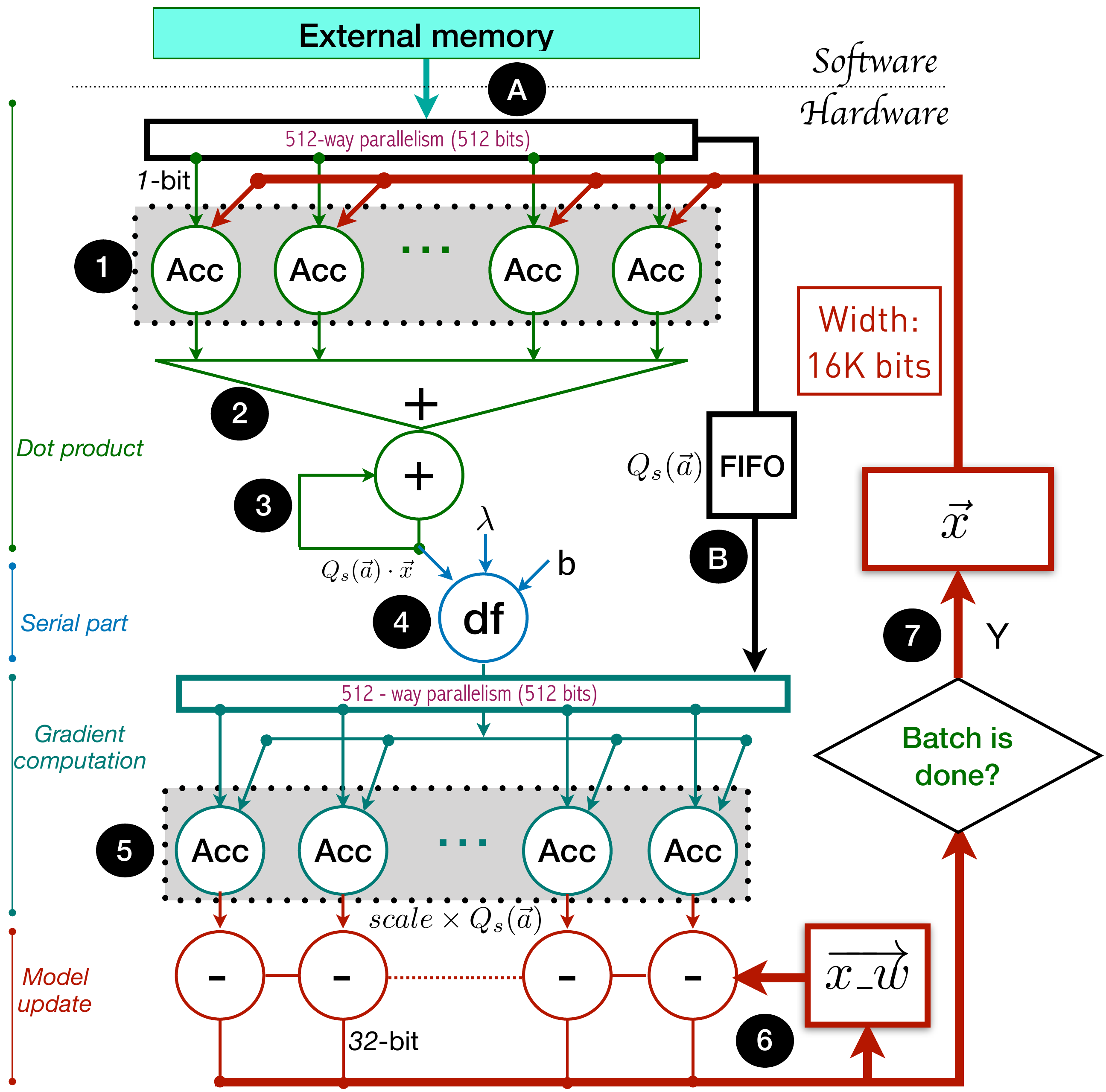}
	%\vspace{-1.5ex}
	\caption{Fully pipelined BWeaving hardware design with 1 bank ($\#Bank$ = 1), according to Algorithm~\ref{alg_sgd_flow_bank}.}% For the instantiation, $\#CL$ is 512.
	\vspace{1.5ex}
	\label{fig_bit_serial_hardware}
\end{figure}

\vspace{0.5ex}
\noindent
{\bf Hardware Design of BWeaving}. The goal of BWeaving arithmetic is to consume 512-bit data per cycle from the BWeaving memory layout described above. We implement the fully-pipelined hardware design for the BWeaving arithmetic in Figure~\ref{fig_bit_serial_hardware}, according to Algorithm~\ref{alg_sgd_flow_bank} with $\#Bank$ = 1. The design has four stages, each of which occupies unique hardware resources of an FPGA. In the following, we discuss the detailed implementation for each stage. 
%\vspace{-1ex}

In the ``dot product" stage, 512 bit-serial multipliers are instantiated (\circled{1}) to consume the 512-bit data stream (\circled{A}) from the BWeaving memory layout per cycle, where each bit-serial multiplier can handle a bit from a feature per cycle. At the same time, the data stream is also fed to the ``FIFO" (\circled{B}). In the first cycle, the first bit of each of the first 512 features from a sample enter the pipeline in a lock-step manner, and 512 values are read from the \emph{architectural model} (labelled $\vec x$ in Figure~\ref{fig_bit_serial_hardware}), where the architectural model stores the committed state of the model $\vec x$ to preserve the semantics of mini-batch SGD. After processing $s$ bits of the first 512 features (where $s$ is the precision level used in the execution), 512 multiplication results are passed to the fully-pipelined adder tree (\circled{2}), whose depth is $log_2(512)$. The output of the adder tree is connected to an accumulator (\circled{3}), which aggregates $\lceil\tfrac{M}{512}\rceil$ valid results from the adder tree and then computes the final result ($Q_s(\vec{a}) \boldsymbol{\cdot} \vec{x}$) of the dot product,\footnote{In the actual implementation, we adopt the distributed arithmetic computation approach~\cite{stripes_micro16, bit_serial_assp89} to compute the dot product, since it produces the same result with fewer hardware resources. Here, we use the basic bit-serial multiplier for ease of understanding. } where $M$ is number of features in the sample. %, so the degree of intra-sample parallelism is $\#CL$ 

In the ``serial part" stage, the scaling value \emph{scale} (\circled{4}) is computed to be $\lambda \times df(Q_s(\vec{a}) \boldsymbol{\cdot} \vec{x}, b)$, where $\lambda$ is the learning rate (parameterizable at runtime), $df$ is the derivative of the given loss function, and $b$ is the label\footnote{In our actual implementation, we also load the label $b$ from memory. However, we omit the related data path in Figure~\ref{fig_bit_serial_hardware} for clarity.} of the sample.   

In the ``gradient computation" stage, we again instantiate 512 bit-serial multipliers (\circled{5}) to compute the full-precision gradient with comparable throughput to the ``dot product" stage. The scaling value \emph{scale} is broadcast to each bit-serial multiplier as its bit-parallel input, while the bit-serial input $Q_s(\vec{a})$ can be read from the ``FIFO" (\circled{B}). The bit-serial multipliers in this stage require exactly the same bit stream order of $Q_s(\vec{a})$ as that required by the ``dot product" stage.    

In the ``model update" stage, the part of the gradient from the first 512 features is computed after $s$ cycles, and then used to update the \emph{working model} ($\overrightarrow{x\_w}$). The working model keeps the temporary model updated after each data sample is processed (\circled{6}).\footnote{The pair (architectural model, working model) is only used in the hardware design (Figures~\ref{fig_bit_serial_hardware}, \ref{fig_mlweaving_hw}), analogous to the pair (architectural register, physical register) in computer architecture. The architectural register indicates the register specified by instruction set architecture (ISA), visible to the programmer. The physical register is used to store temporary results, invisible to the programmer. } Later, $\overrightarrow{x\_w}$ is updated with the part of gradient from the second 512 features after next $s$ cycles, and so on. $\overrightarrow{x\_w}$ is updated at the rate of every sample, while the architectural model ($\vec x$) is updated only after a mini batch ($B$) of samples (\circled{7}), where $B$ is the mini batch size (parameterizable at runtime). Therefore, the semantics of mini-batch SGD is preserved. %samples in this mini batch are able to read the expected model without any pollution from the gradient of this mini batch; in other words, %It mean that the width of the model 

\vspace{0.5ex}
\noindent
{\bf Instantiation of Hardware Design. }%~\cite{Intel_arria_10_device}
``BWeaving" has a really high model bitwidth of $16K$ (32*512), making it unsuitable in real FPGA implementations.\footnote{In Intel Arria 10 FPGA, the model is implemented with 20-Kb memory blocks (i.e., M20Ks). Each M20K can provide a 32-bitwidth with ECC (or 40 without ECC) for the model, so a 16K-bitwidth requires 512 M20Ks with ECC (or 410 without ECC). It is extremely difficult to access 410 M20Ks in a lock-step manner (e.g., sharing the same read/write address) while maintaining high frequency, as M20Ks are uniformly distributed inside an FPGA.}

%\vspace{-1ex}
\subsection{MLWeaving} %: Flexibility \& Performance \& Deployment
\label{subsec_mlweaving}

MLWeaving develops the ideas behind BWeaving so as to make them implementable. In particular, MLWeaving changes both the memory layout and the design on the FPGA to dramatically reduce the required bitwidth of the model. 
Inspired by the mini-batch SGD that uses the same model to process one mini batch of $B$ samples, we can process multiple samples in the same mini-batch simultaneously such that we can reduce the model bitwidth without compromising on throughput: 512 bits per cycle. As shown in Algorithm~\ref{alg_sgd_flow_bank}, MLWeaving instantiates 8 physical banks to accommodate 8 samples ($\#Bank$ = 8) in the same mini-batch simultaneously (Lines 5-6) such that 8 samples reading the same portion of the model are processed in a lock-step manner. This is why $B$ must be a multiple of 8.\footnote{The larger $\#Bank$ leads to less complexity of hardware design but more limitation on mini-batch size. } At the same time, we adjust the related memory layout such that the data stream from the memory flows into the MLWeaving hardware without any transposition overhead. 
%Since $B$ samples in the same mini batch read the same model (Line 7),
%Next, we show the intuition behind MLWeaving, followed by the related memory layout and hardware design.
%With this approach, MLWeaving is more amenable to hardware implementation, while maintaining precision flexibility and high throughput (i.e., $\#CL$ bits per cycle), as shown in Table~\ref{t_comparison}.

%In the BWeaving design in Figure~\ref{fig_bit_serial_hardware}, a cache line contains the x-bit of 512 features for a given sample. In MLWeaving, a cache line contains the x-bit of 256 features of sample $n$ and 256 features of sample $n$+1 (for the case with 2 banks). So while bit-serial processes, in principle, data form a single sample (row) in a cycle, MLWeaving processes data from 64 rows (samples) per cycle but less bits for each one of them. Thus, each cycle in MLWeaving processes less data from a sample (using a bank) but it processes several samples, while bit-serial processes one sample at a time. Since all the samples processed in all the banks correspond to the same part of the model, MLWeaving only needs to read that part of the model for all banks. Therefore, the bitwidth becomes 2K bits under MLWeaving, compared with 16K bits under bit-serial. 

%\newline
%\newpage
%\newline

%	\vspace{1ex}
\subsubsection{MLWeaving Memory Layout (Software)} 
\label{subsec_mlweaving_layout}
\begin{figure}[t]
	\centering
	\includegraphics[width=6cm]{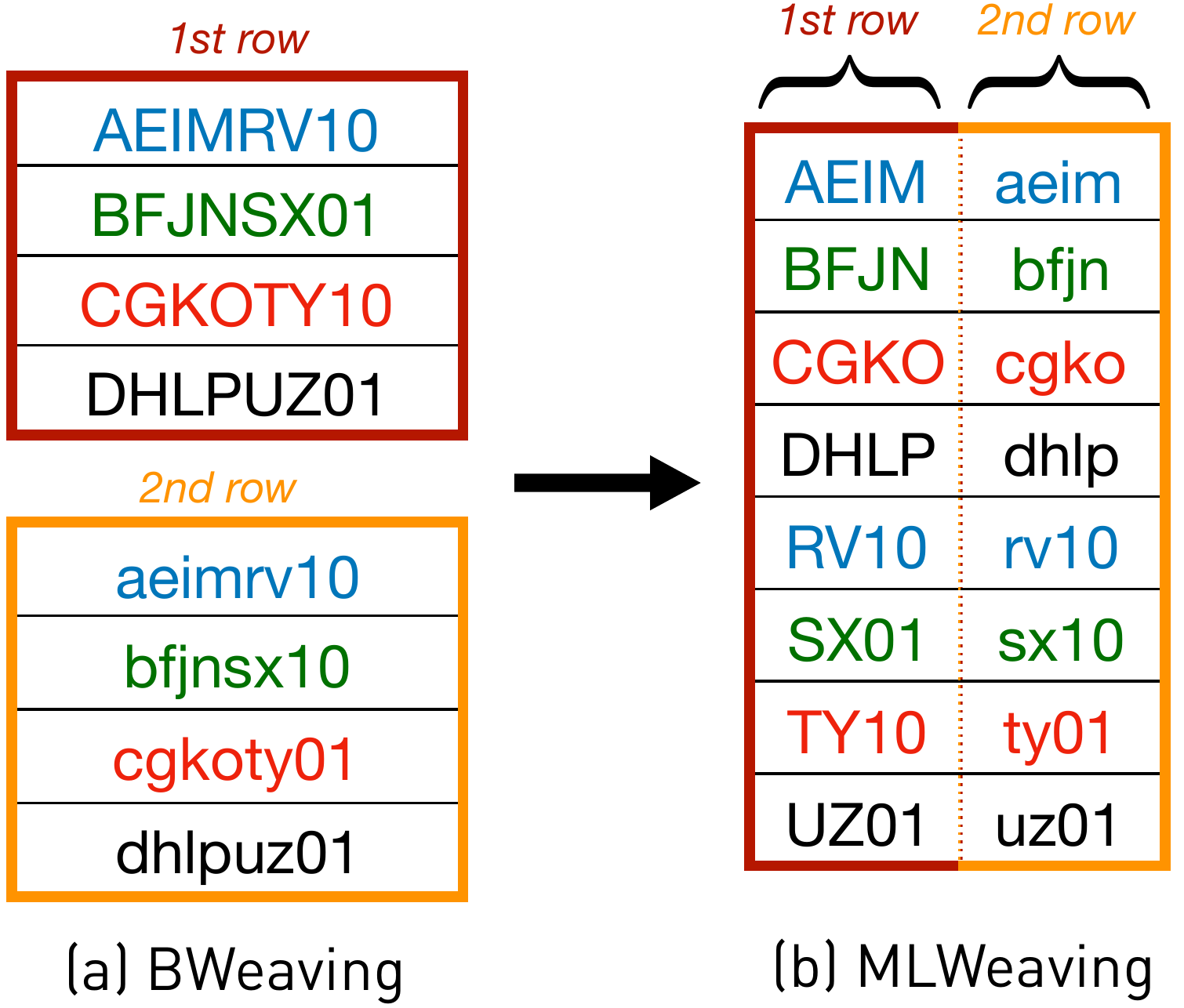}
	\vspace{-1.5ex}
	\caption{BWeaving memory layout (a) converted to MLWeaving memory layout (b): $M$ = 8, $\#CL$ = 8 and $\#Bank$ = 2. } %The model only needs to provide four weights per cycle under MLWeaving, compared with eight under bit-serial.
	\vspace{-1ex}
	\label{fig_mlweaving_bank_layout}
\end{figure}

	\vspace{-0.5ex}
% partitioning each sample at the bit level, 
%The objective of MLWeaving memory layout can provide the exact bit-serial stream to the corresponding hardware design, such that the performance and energy efficiency is guaranteed.  
%No resource overhead is due to the fact that the data flows into the computing pipeline smoothly, without any 
%The bit-serial memory layout saturates the entire $CL$-bit memory transaction with the bits from a single sample (i.e., only exploring intra-sample parallelism). In contrast, each memory transaction assembles the data from $\#Bank$ samples, each of which contributes $\frac{\#CL}{\#Bank}$ bits under MLWeaving. 
Starting from the BWeaving memory layout in Figure~\ref{fig_bit_serial_layout}, we show the memory layout transition to MLWeaving in Figure~\ref{fig_mlweaving_bank_layout}. 
BWeaving populates each memory transaction with eight bits from the same row, e.g., {\color{blue} AEIMRV10} of the first row. In contrast, the first/second row contributes four bits for each memory transaction under MLWeaving. For instance, the first bits of the first four features of two rows assemble into the first memory transaction, e.g., {\color{blue} AEIMaeim}. %It can perf
Since the bits {\color{blue} AEIM} and {\color{blue} aeim} share the same weights, the model only needs to provide four weights within a cycle. %contains four bits from the first row and four bits from the second row 

\vspace{0.5ex}
\noindent
{\bf Instantiation of Memory Layout.} We instantiate the MLWeaving memory layout by setting $\#CL$ (or $\#Bank$) to be 512 (or 8). Now we use the case with $M$ = 2048 as an example in Table~\ref{t_mlweaving_instance}. For instance, the first memory transaction is populated with the first bits of the first 64 features of the first eight samples. If $M$ is not a multiple of 64, we use padding to align it to 64 bits. Thus, the memory traffic $MT$ (in terms of bits) for each sample consists of two parts, as shown in Equation~\ref{E_memory_traffic}. The first equality shows the memory traffic from all the features, evaluated to be the precision level $s$ multiplied by the value that rounds up $M$ to the nearest multiple of 64. The second equality (32 bits) comes from the label of each sample. 
\begin{equation} \begin{scriptsize}
\label{E_memory_traffic}
	\vspace{-1ex}
MT = s \times \lceil \frac{M}{64} \rceil \times 64 + 32
	\vspace{-1ex}
\end{scriptsize} \end{equation}

%Suppose we want to achieve 8-way parallelism ($\#CL$ = 8).   
 
%MLWeaving also exploit not only intra-sample $\frac{CL}{\#Bank}$-way parallelism, but also inter-sample $\#Bank$-way parallelism. 
%In particular, we 
%to share the memroy transaction with multiple samples. 

%Each $\#CL$-bit memory transaction contains features from $\#Bank$ rows. 
%Serial part: 

\begin{table} [t]
	\centering
	%\begin{spacing}{0.3}
	\begin{scriptsize}
		\caption{Instantiation of MLWeaving memory layout. \emph{x\_y:z\_w} denotes the \emph{w}-th bits of 64 features (from $z$-th to $y$-th) in the $x$-th sample and $y - z = 63$. A row contains a cache line (512 bits). }
		\label{t_mlweaving_instance}
		\vspace{-1ex}	
        \begin{tabular}{|p{1.6cm}|c||c|c|c|c|}
			\hline
			\textbf{Description} &\textbf{Index} & \textbf{Bank 7} & ... & \textbf{Bank 1} & \textbf{Bank 0} \\
			\hline
			\hline
			\hline
			\multirow{6}{*}{\parbox{1.6cm}{First 64 features of the first 8 samples}}&$0$ & 7\_63:0\_0 & ... & 1\_63:0\_0 & 0\_63:0\_0  \\
			\cline{2-6}
			&$1$ & 7\_63:0\_1 & ... &  1\_63:0\_1 & 0\_63:0\_1  \\
			\cline{2-6}
			&$2$ & 7\_63:0\_2 & ... &  1\_63:0\_2 & 0\_63:0\_2  \\
			\cline{2-6}
			&$...$ & ... & ... & ... & ...  \\
			\cline{2-6}
			&$31$ & 7\_63:0\_31 & ... & 1\_63:0\_31 & 0\_63:0\_31  \\
			\hline
			\hline
			\multirow{3}{*}{\parbox{1.6cm}{Second 64 features of the first 8 samples}} & $32$ & 7\_127:64\_0 & ... & 1\_127:64\_0 & 0\_127:64\_0  \\
			\cline{2-6}
			&$33$ & 7\_127:64\_1 & ...  & 1\_127:64\_1 & 0\_127:64\_1  \\
			\cline{2-6}
			& $...$ & ... & ... & ...  & ...  \\
			\hline
			\hline
			$...$ & $...$ & ... & ... & ...  & ...  \\
			\hline
			\hline
			\multirow{3}{*}{\parbox{1.6cm}{First 64 features of the second 8 samples}}&$1024$ & 15\_63:0\_0 & ... & 9\_63:0\_0 & 8\_63:0\_0  \\
			\cline{2-6}
			&$1025$ & 15\_63:0\_1 & ... & 9\_63:0\_1 & 8\_63:0\_1  \\
			\cline{2-6}
			&$...$ & ... & ... & ... & ...  \\
			\hline
		\end{tabular}
		\vspace{3ex}
	\end{scriptsize}
	%\end{spacing}
\end{table}

\begin{figure*}[t]
	\centering
	\subfloat[Overall architecture]{\includegraphics[width=3in]{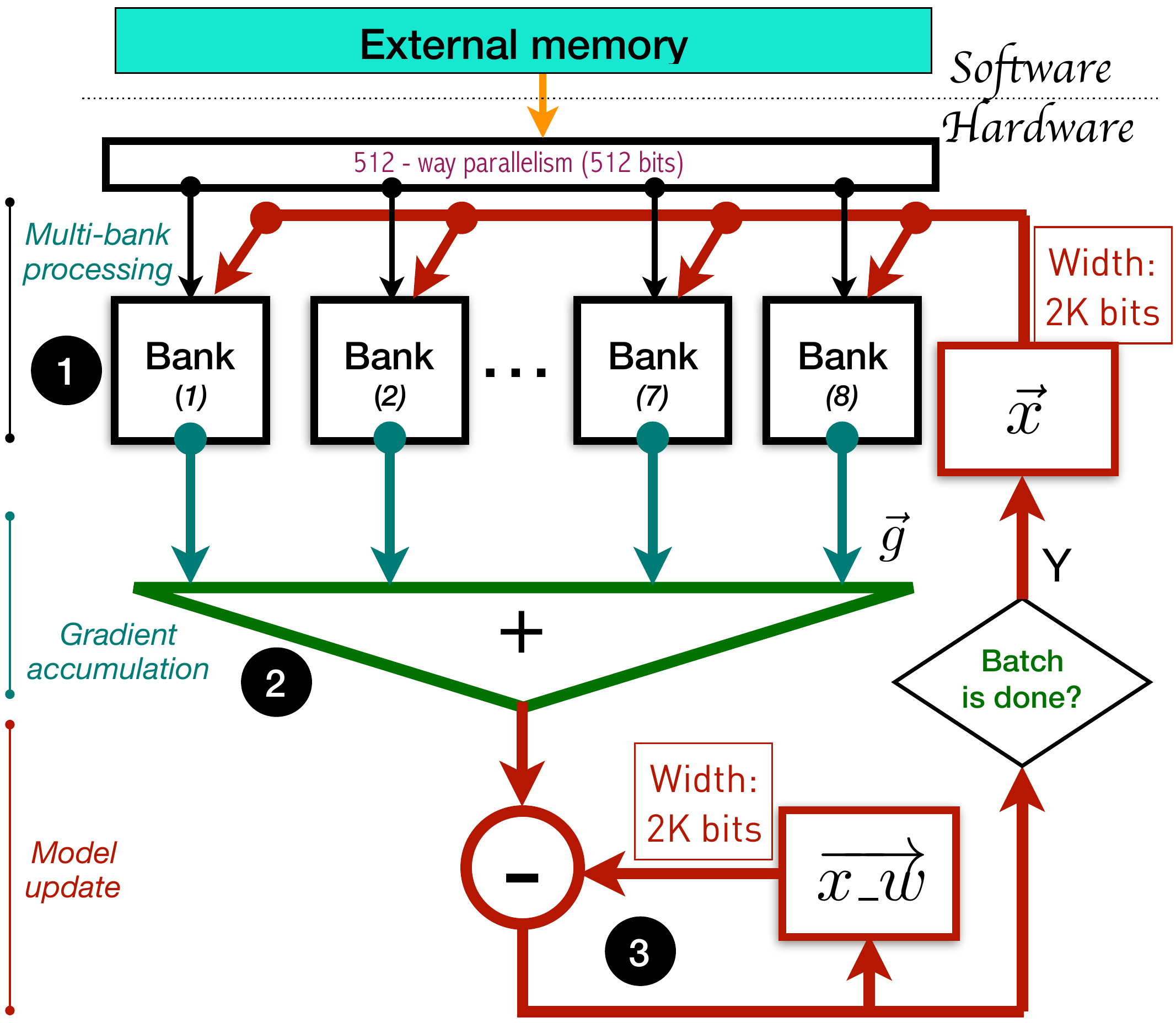} 
		\label{fig_mlweaving_overall}} % \caption{}
	%\hfill
	\subfloat[Detailed design of each bank]{\includegraphics[width=2.4in]{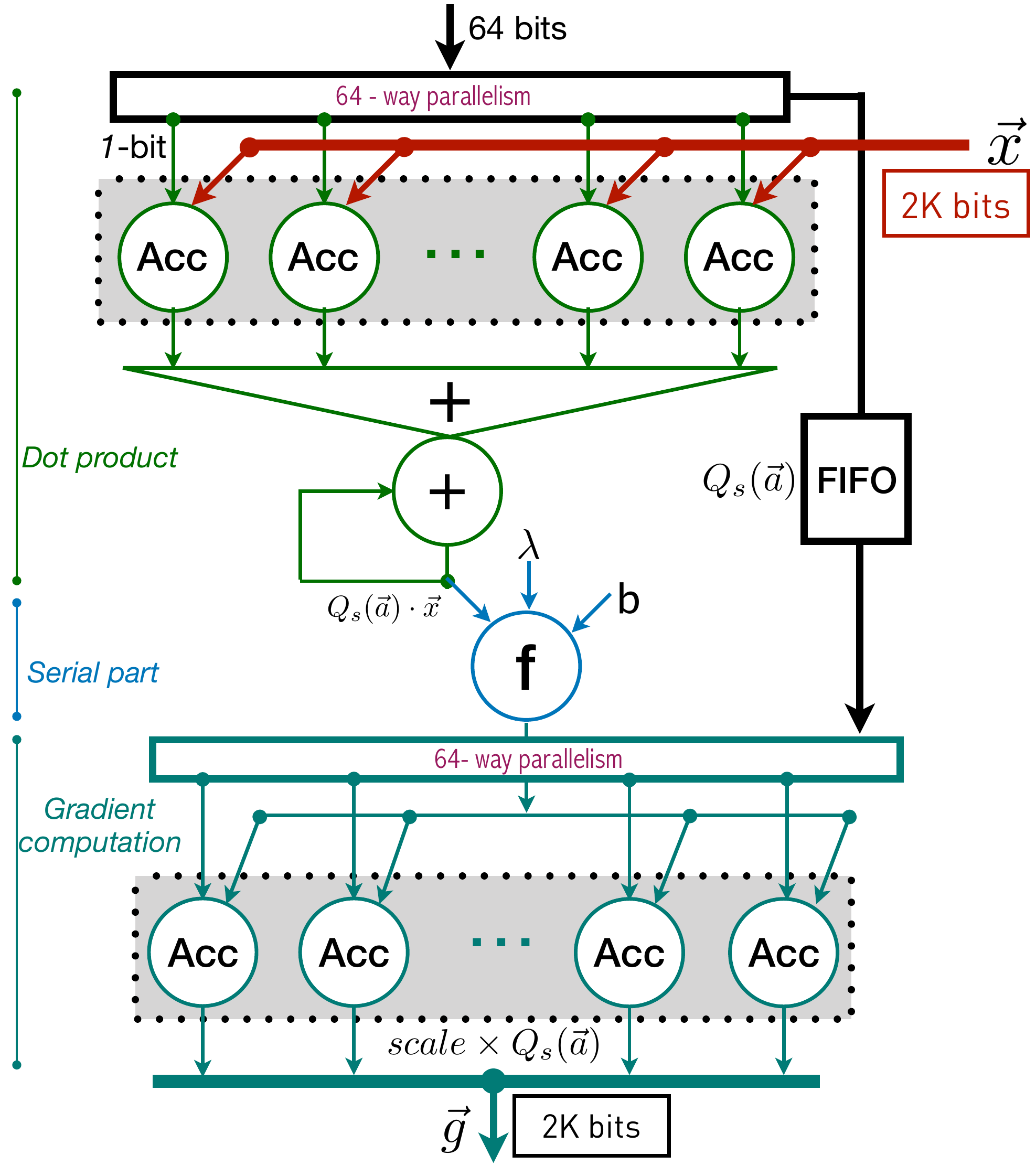} \label{fig_mlweaving_bank}} % \caption{}
	\vspace{-1.0ex}
	\caption{Fully pipelined MLWeaving hardware with 8 banks ($\#Bank$ = 8), according to Algorithm~\ref{alg_sgd_flow_bank}. Its throughput is 512 bits per cycle. } 
	\label{fig_mlweaving_hw} 
	\vspace{-3ex}
\end{figure*}  

\vspace{-1ex}
\subsubsection{MLWeaving Arithmetic (Hardware)} %
\label{subsec_mlweaving_arithmetic}
Following Algorithm~\ref{alg_sgd_flow_bank}, we present the fully-pipelined hardware design of MLWeaving arithmetic in Figure~\ref{fig_mlweaving_hw}. The targeted throughput of MLWeaving arithmetic is 512 bits per cycle. It consists of three main pipeline stages. In the following, we explain the design details of each pipeline stage. %overall architecture, followed by the details inside a bank. 
% Each bank provides $\#CL/\#Bank$-way intra-sample parallelism. 

In the ``multi-bank processing" stage, the 512-bit computing pipeline is divided into 8 banks, each of which consumes 64 bits from one sample (\circled{1}), as shown in Figure~\ref{fig_mlweaving_overall}. The most important property of this stage is that 8 banks read the same portion of the model at a given time. %Now, the model ($\vec{x}$) only needs to provide $32 * \#CL/\#Bank$ bits per cycle and broadcast them to all the banks. 
Each bank behaves the same as that in the BWeaving arithmetic in Figure~\ref{fig_bit_serial_hardware}, except that each bank instantiates 64 bit-serial multipliers in the ``dot product" and ``gradient computation" stages (Figure~\ref{fig_mlweaving_bank}). Since we instantiate 8 banks, the total throughput of the MLWeaving arithmetic is still 512 bits per cycle. 

In the ``gradient accumulation" stage, a portion of the gradient (64 32-bit elements) from 8 banks is fed to 64 element-wise adder trees, each of which generates one element of the average gradient within a cycle (\circled{2}). %In other words, $\#Bank$-way inter-sample parallelism is exploited. 
%We can see that the adder trees and model are fully pipelined.  (e.g., s)

In the ``model update" stage, the average gradient is used to update the working model ($\overrightarrow{x\_w}$) (\circled{3}) for every 8 samples. The architectural model ($\vec x$) is updated only after a mini batch ($B$) of samples is finished. %It mean that the width of the model 

\vspace{0.5ex}
\noindent
{\bf Instantiation of Hardware Design. }
%Figure~\ref{fig_mlweaving_hw} illustrates the instantiation of MLWeaving with $\#CL = 512$ and $\#Bank = 8$. 
Table~\ref{t_resource_conparison} shows the resource consumption in our FPGA when we implement the computing pipeline of MLWeaving. MLWeaving achieves high clock frequency (400MHz) while requiring a reasonable amount of FPGA resources. This is because 1) the proposed multi-bank architecture of MLWeaving leads to fewer BRAMs (on-chip memory blocks on FPGAs) for the architectural and working models in Figure~\ref{fig_mlweaving_hw}, and 2) the memory layout allows  MLWeaving to directly consume data from memory, without auxiliary hardware modules for transposition. The theoretical throughput of MLWeaving's hardware design is roughly  25.6GB/s (400M * 512 bits per cycle), much larger than the available memory read bandwidth: 15GB/s. 

\begin{table} [ht]
	\centering
	%\begin{spacing}{0.3}
	\begin{scriptsize}
	\vspace{-1ex}
 	\caption{FPGA resource consumption }
	\label{t_resource_conparison}	
	\vspace{-1.5ex}
	\begin{tabular}{|c||c|c|c|c|}
		\hline
		\textbf{Name} & \textbf{Logic (ALMs)} & \textbf{DSPs} & \textbf{BRAMs} & \textbf{Frequency} \\ 
		\hline
		\hline
		\textbf{BWeaving} & N.A & N.A & N.A & N.A\\
		\hline	
		\textbf{MLWeaving} & 35670 (8.4\%) & 0 (0\%) & 3.25Mb  (6.1\%) & 400 MHz\\
		\hline
	\end{tabular}
	\vspace{-3ex}
	\end{scriptsize}
	%\end{spacing}
\end{table}

%\input{bit_serial.tex}
%\vspace{-2ex}
%\input{mlweaving.tex}
%\vspace{-2ex}
%\vspace{-1ex}
%\subsection{Data Hazard Resolution \& Chaining Technique}
\section{Preservation of Precedence}%Scheme for  for Synchronous SGD
\label{subsection_chaining}
We propose a simple yet efficient scheme to keep SGD hardware design synchronous, without compromising processing speed. Our scheme is orthogonal to the BWeaving and MLWeaving designs, so it can be applied to both.\footnote{In the following, we describe our scheme for MLWeaving. The scheme can easily generalize to BWeaving. } Our aim is three-fold. {\bf G1}: it lets SGD read the up-to-date model. {\bf G2}: it supports various batch sizes. {\bf G3}: it exploits the greatest possible overlap between computation (i.e., dot product) and communication (i.e., model update). Next, we identify the performance issue of synchronous SGD, followed by our mechanisms to achieve three goals.

\begin{figure}[t]
	\centering
	\subfloat[Basic mechanism: without chaining]{\includegraphics[width=3.2in]{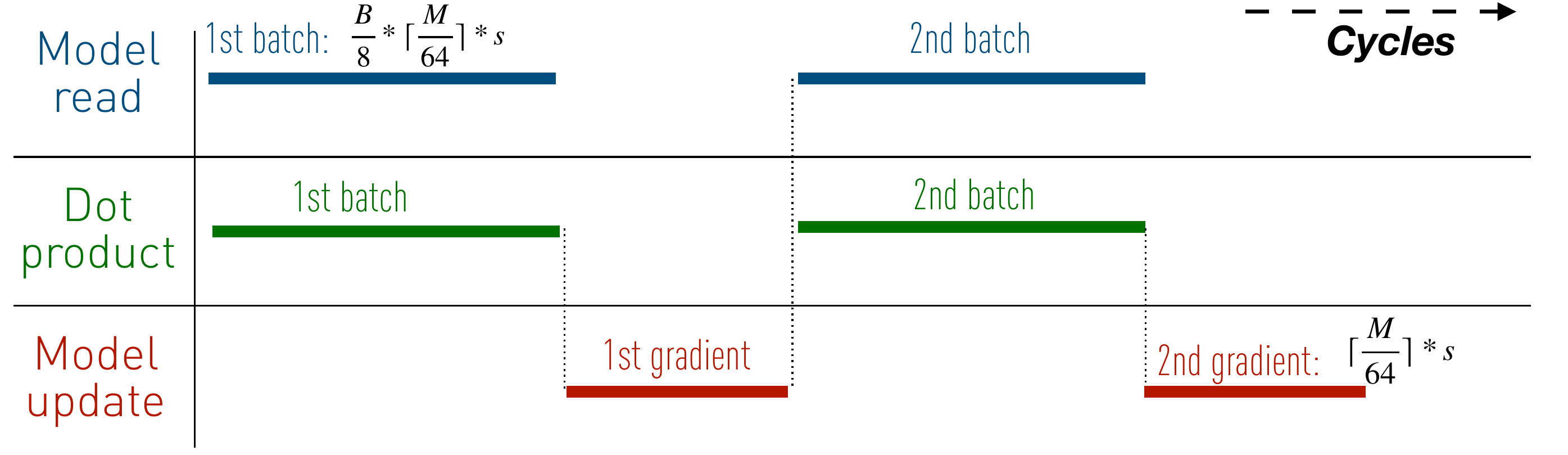} 
		\label{fig_bit_serial_raw}} % \caption{}
	\hfill
	\subfloat[Chaining-enhanced mechanism]{\includegraphics[width=3.0in]{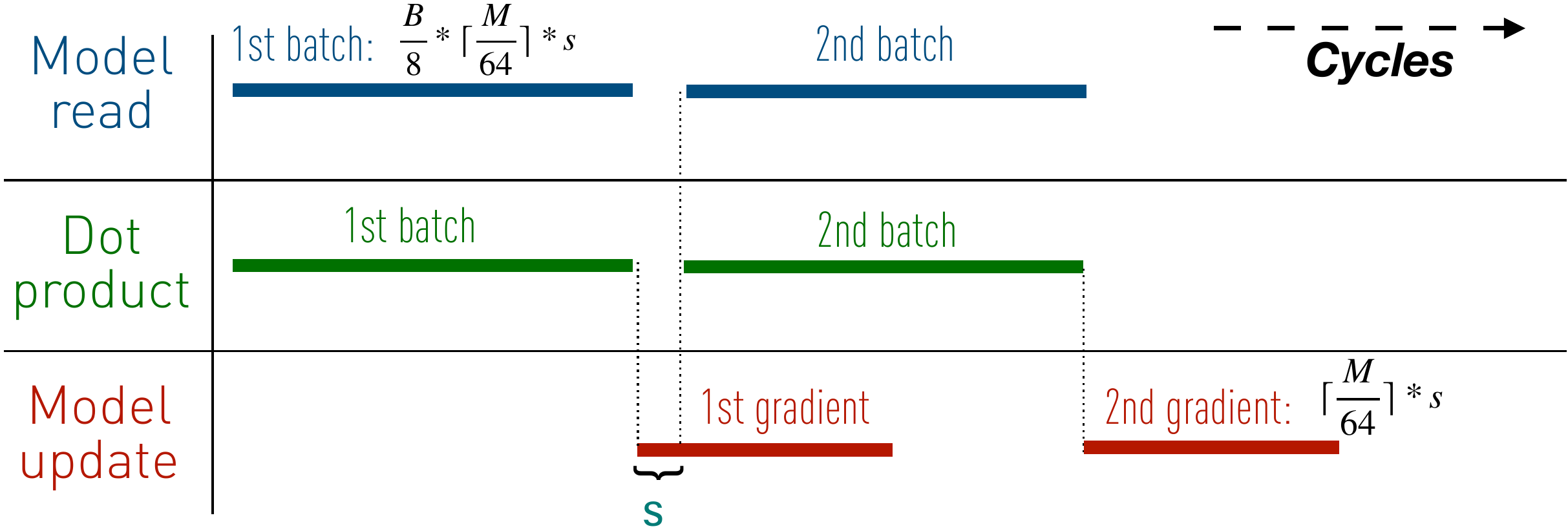} 
		\label{fig_mlweaving_raw}} 
	\vspace{-1.5ex}
	\caption{MLWeaving with and without chaining. } 
	\label{fig_raw} 	
	\vspace{-1ex}
\end{figure}

%\vspace{-1ex}
%\subsection{Performance Issue of Synchronous SGD} 
%Long pipeline latency and {\bf Issue 2: Long pipeline delay. } 
\vspace{0.5ex}
\noindent
{\bf Performance Issue of Synchronous SGD. }According to Algorithm~\ref{alg_sgd_flow_bank}, model reading (Line 8) and model update (Line 14) has a Read After Write (RAW) dependency, due to the inherently sequential nature of synchronous SGD. For example, the model read by the second batch ($B$ samples) should be up-to-date such that the gradient from the first batch has already been accumulated into the model, as illustrated in Figure~\ref{fig_bit_serial_raw}. In other words, the second batch has to wait until the model is updated. The performance issue is not trivial, as it takes $\lceil M/64 \rceil*s$ cycles to update the model for each batch, where $M$ is the number of features, 64 is the number of elements written to the model within a cycle, and $s$ is the number of cycles to do one multiplication with a bit-serial multiplier. 

%In particular, it takes multiple cycles to finish one multiplication (i.e., round), compared with one cycle for bit-parallel multiplier. %Therefore, long pipeline delaywhich can degrade the computing throughput.is its long pipeline delay 
%When the RAW dependency is violated, the execution does not read the latest model, which causes the \emph{staleness} in the context of machine learning. The staleness has the potential to affect the statistical efficiency, e.g., more epochs. %for the training.  

%\subsection{Details of Scheme} 
%In what follows we present the details of the mechanism which is dedicated to the synchronous SGD design. 
%We aim to preserve the dependency and improve the performance of the synchronous SGD design. 
%The first two requirements are to preserve the precedence for synchronous SGD, so we propose a basic mechanism to meet them. The third requirement is to maximize performance, and we address it separately using a chaining-enhanced mechanism.
\vspace{0.5ex}
\noindent
{\bf Basic Mechanism.}  
	\label{subsubsection_basic} 
The goal of the basic mechanism is to preserve the precedence for synchronous SGD ({\bf G1} and {\bf G2}). In particular, our mini-batch SGD reads the up-to-date model. The key idea of the basic mechanism is to record the read/write operations performed on the model $\vec x$ such that MLWeaving will not read the out-of-date model and wait until the model becomes up-to-date. To do so, we introduce two 16-bit registers: \emph{wr\_counter} and \emph{rd\_counter}. Table~\ref{t_hazard_resolution} illustrate how to manipulate such two counters to preserve the dependency. %The intuition is that the ``dot product" stage can read the model only after the model is up-to-date. maintenance
The \emph{wr\_counter} records the times of writing operations performed on the model $\vec x$ in the ``model update" stage. Its initialization value is $B$, where $B$ is the input batch size. It means that $B$ credits are provided at the beginning for $B$ samples to read $\vec x$ in the ``dot product" stage. \emph{wr\_counter} is incremented by $B$ when the model $\vec x$ is updated, indicating that $\vec x$ is updated by the average gradient from every $B$ samples in Algorithm~\ref{alg_sgd_flow_bank}. We can update the \emph{wr\_counter} once the average gradient is ready, since there is no WAR or WAW dependency.  
The \emph{rd\_counter} records the times of reading operations performed on the model $\vec x$ in the ``dot product" stage. Its initialization value is $0$, indicating that $\vec x$ has not been read yet. It is incremented by 8 when the model is read, since MLWeaving processes 8 samples concurrently. Its updating condition is that \emph{rd\_counter} is not equal to \emph{wr\_counter}, indicating there are still enough credits for samples to read $\vec x$. This is the critical step to preserve the dependency. %such that MLWeaving reads the up-to-date model. 

\begin{table} [t]
	\centering
	%\vspace{-1.5ex}
	%\begin{spacing}{0.3}
	\caption{Basic mechanism to preserve RAW dependency}
		\vspace{-1.5ex}
		\label{t_hazard_resolution}
	\begin{scriptsize}
		\begin{tabular}{|c||c|c|c|c|}
			\hline
			& \textbf{Init Value}  & \textbf{Step Size} & \textbf{Updating Condition} \\
			\hline
			\hline
			\emph{wr\_counter} & $B$ & $B$ & Always ok \\
			\hline
			\emph{rd\_counter} & 0 & 8 &  \emph{rd\_counter} != \emph{wr\_counter}\\
			\hline
			
		\end{tabular}
	\end{scriptsize}
	%\vspace{-1ex}
	%\end{spacing}
\end{table}
% and maximize performance of synchronous SGD  In particular, we try to resolve the data hazard and relax unnecessary dependencies regarding the model. 

\vspace{0.5ex}
\noindent
{\bf Chaining-Enhanced Mechanism.} % for Synchronous SGD
	\label{subsubsection_chaining} 
To increase the overlap between computation and model update ({\bf G3}), we propose the chaining enhanced mechanism for synchronous SGD. The key idea is to allow computation when updating the model, while preserving the dependency. To do so, we treat the model $\vec{x}$ as a vector register that requires multiple cycles to read/write and follows a sequential access pattern. Then, we use the chaining technique~\cite{Tarantula_isca02, vector_processor_micro02, vector_magize_ieee_micro03, chaining_cray_78} of vector processing to maximally overlap computation and model update. 
%we implement enhanced chaining, which can relax unnecessary dependencies to a certain extent between block registers, as long as data dependencies between elements are preserved.
In particular, the model updated with the gradient from the first batch can be forwarded to the second batch before the entire updating operation completes, as shown in Figure~\ref{fig_mlweaving_raw}. The second batch can begin to read after $s$ cycles, where $s$ is the precision level. We observe that it is safe for the second batch to read the model after the first part of the model (e.g., 64 values) is updated into the model. The data dependency is preserved since the model updating speed is not slower than the model reading speed. 

\vspace{-2ex}

\section{Per-epoch Tuning of Precision}
\label{sec_per_epoch_tuning}

MLWeaving introduces one tuning knob for the user: {\em the precision level to be used for training}. Most existing work assumes that it is the user's responsibility to set the right precision level. This is understandable, as the right level depends on both the data and the error tolerance and as no tight theory can map error tolerance back to the right precision level.

In this paper, we do not address the problem of determining the right precision level for the user, as it goes well beyond the scope of the work. Instead, we provide a simple, dynamic schedule of precision that harvests the potential of MLweaving. Such a schedule works robustly on all data sets we have. %for supporting dynamic precision scheduling. 

\vspace{0.5ex}
\noindent
{\bf Dynamic Precision Schedule.}
Our schedule is based on a very simple observation:
at the beginning of the training, the system is 
less sensitive to the error introduced by low precision
data representation; at the end of the training,
the system often requires more bits to converge.
MLWeaving allows us to dynamically change
the number of bits to use for each epoch. %We explore this flexibility and build probably the simplest dynamic precision schedule: use 1-bit for the 1st epoch, 2 bits for the 2nd and 3rd epochs, 3 bits for the 4th-7th epochs, 4 bits for the 8th-15th epochs, and so on. 
We exploit this flexibility and build a simple dynamic precision schedule: use 2 bits for the 1st-4th epochs, 3 bits for the 5th-8th epochs, 4 bits for the 9th-16th epochs, 5 bits for the 17th-32nd epochs, and so on. That is, the number of bits grows over time until we reach the targeted loss.

This simple schedule is inspired by the following theoretical observation: for SGD to converge with $O(1/\sqrt{K})$ rate, at each iteration $K$, it only requires that the bias introduced by the low precision representation decreases faster than $O(1/K)$. The above schedule is one example that satisfies this property, as the bias introduced by low precision is halved when one more bit is used.

\vspace{0.5ex}
\noindent
{\bf Remarks.} Note that the above schedule is far from optimal and perfect. One can design more adaptive schedules by, for example, monitoring the speed of the decrease of loss and dynamically choosing when to switch to the next level of precision. MLWeaving allows this possibility by providing an end-to-end solution that enables dynamic precision schedule. We will explore more sophisticated precision schedule schemes as well as the problem of mapping precision and error levels as part of future work.

\vspace{-2ex}
\section{Experimental Evaluation}
\label{sec_evaluation}
%In this section, we describe the experimental setup and evaluate the performance of MLWeaving.% on synthesized workloads. the performance and energy efficiency of 
\vspace{-1ex}
\subsection{Experimental Setup}
%\vspace{-1ex}
\label{sec_experiment_stup}
%{\bf Hardware Configuration.}
%We conduct our experiments on the Intel Xeon+FPGA platform, which contains an Intel Broadwell CPU and an Intel Arria 10 FPGA. The CPU features 14 cores/35MB last level cache and its read memory bandwidth can reach 60GB/s. The FPGA can directly access the CPU memory (bandwidth: $\approx$ 15GB/s) via one QPI and two PCIe links. All the CPU programs are compiled using ICC 18.0.0 with the highest optimization effort -O3. We use the Intel Performance Counter Monitor~\cite{Intel_PCM} to collect the performance counters (e.g., external memory traffic) on the program of interest.
%\reversemarginpar
%\kaancomment{This is the first time we ever mention Centaur. I think it should at least be mentioned briefly much earlier. Also, why do we mention \textit{with and without} DoppioDB? We don't have to mention this at all.} \wzk{We employ the built-in time-measuring tool of the Centaur~\cite{centaur_fccm17} to obtain the elapsed time in our experiment.} % \st{The elapsed time difference between with and without DoppioDB is quite trivial, since the system noise from DoppioDB is relatively small, compared with the actual training time.}
% in Figure~\ref{fig_fpga_platform}
%\begin{figure}[t]
%	\centering
%	\includegraphics[width=5.1cm]{experiment_platform.pdf}
%	\vspace{-0.5ex}
%	\caption{Experimental platform: Intel Xeon+FPGA v2. Hogwild and ModelAverage runs in 14 cores, while MLWeaving runs on the FPGA.}% We focus on the FPGA and main memory.
%	\vspace{-2ex}
%	\label{fig_fpga_platform}
%\end{figure}
\noindent
{\bf Workloads. }We carry out our experiments with the five data sets shown in Table~\ref{t_dataset}. 
For the multi-class dataset TL~\cite{columnml_vldb19}, we use 10 classes of ImageNet~\cite{imagenet_nips12}. Instead of directly using the images in ImageNet, we use 2048 features per sample extracted by a neural network (InceptionV3~\cite{inception_v3_cvpr16}) that can be used for transfer learning, and we train binary classifiers using the one-vs-one strategy~\cite{one_vs_all_jmlr04}. For the dataset Madelon, the training samples are duplicated 10 times so that the size exceeds the capacity of the last level cache in the CPU. For the dataset KDD, since its original dataset is highly skewed, i.e., 93\% samples are labelled 0, we uniformly delete the samples labelled 0 so that the final dataset is balanced. %\kaan{This sounds very artificial. Why don't we use some synthetic data instead of duplication?}
\begin{table} [ht]
	\centering
    \vspace{-1ex}
	%\begin{spacing}{0.3}
	\begin{scriptsize}
	\caption{Evaluated datasets. }
	\label{t_dataset}
		\vspace{-1.5ex}
		\begin{tabular}{|c||c|c|c|c|}
			\hline
			\textbf{Dataset} & \textbf{Features} & \textbf{Training samples} &  \textbf{Testing samples} & \textbf{Classes} \\%testing samples\\
			\hline
			\hline
			Gisette~\cite{libsvm_tist_11} & 5000 &  6000 & 1000 & 2\\
			\hline
			TL~\cite{columnml_vldb19} & 2048 & 26,000 & 5200 &10\\
			\hline
			Epsilon~\cite{libsvm_tist_11} & 2000 & 40,000 & 10,000 & 2\\		
			\hline
			KDD~\cite{mlbench_vldb18} & 2399 &  40,000 & 44772  & 2\\
			\hline
			Madelon~\cite{libsvm_tist_11} & 500 & 20,000 & 600 & 2 \\		
			\hline
		\end{tabular}
	\end{scriptsize}
	\vspace{-2ex}
	%\end{spacing}\begin{tabular}{@{}c@{}}ImageNet \\ (Transfer Learning)\end{tabular} $\#$  $\#$ $\#$ $\#$ 
\end{table}

\vspace{0.5ex}
\noindent
{\bf FPGA Implementations. }There are two FPGA implementations. First, MLWeaving (sync) represents the MLWeaving hardware that satisfies the RAW dependency. Second, MLWeaving (async) represents the MLWeaving hardware that violates the RAW dependency such that it directly reads the model even when the model is still out-of-date.\footnote{In the following experiments, by default, MLWeaving means MLWeaving (sync) on FPGAs and with chaining enabled.} %Intuitively, MLWeaving (async) could achieve more throughput than MLWeaving (sync). However, since both approaches are memory-bound on the targeted FPGA, their relative performance difference, i.e., (MLWeaving (sync)- MLWeaving (async))/MLWeaving (sync), is small, 

\vspace{0.5ex}
\noindent
{\bf CPU Baselines. }Since SGD is inherently sequential, keeping consistency when running SGD leads to no parallelism among cores. Two first-order variants of SGD (``Hogwild" and ``ModelAverage") have been proposed to parallelize SGD on modern CPUs and we use both as baselines, employing existing optimization methods on CPUs: multi-core (14 cores), low-precision (8-bit) and AVX2 instruction (256-bit). %In this paper, all the CPU implementations use all the 14 cores to maximize the performance on CPU.

``Hogwild"~\cite{hogwild_NIPS11, dimmwitted_vldb14} allows each core to compute the gradient from its own portion of dataset and then to perform asynchronous update on a single copy of the model without any synchronization. Therefore, the parallelism among cores is exploited at the cost of low statistical efficiency due to asynchronous updates. Even though no synchronization is required, Hogwild still suffers from cache coherence overhead, i.e., invalidating the model copies in the private caches of other cores before the real write operation. The cache coherence overhead is so severe on a multi-core CPU that Hogwild cannot benefit from using low-precision (Figure~\ref{fig_loss_time_epsilon}). %Since multiple cores can simultaneously  

``ModelAverage"~\cite{modelAveraging_NIPS2010} allows each core to have its own copy of the model so that no costly invalidation among cores occurs. It averages the models at the end of each epoch and then broadcasts the aggregated model to each core at the beginning of the next epoch. Therefore, its multi-core implementation can saturate the maximum memory bandwidth of the CPU (leading to high hardware efficiency). However, ModelAverage has relatively lower statistical efficiency since each worker uses its local (not global) model to compute the gradient. Since its 32-bit floating-point implementation is memory-bound, ModelAverage can significantly benefit from a low-precision dataset that requires less memory traffic. In our experiment, we choose 8-bit precision because 1) the smallest bank width of a SIMD register is 8 bits and 2) ModelAverage becomes compute-bound when the dataset is quantized to 8 bits, indicating lower performance for a lower-than-8-bit precision.  %its inherently sequential nature. \footnote{Deploying linear model training on a cluster of GPUs is not that beneficial for the same reason: the more computing power (e.g., cores) causes more communication overhead. So, we skip the comparison with GPU.}

\vspace{0.5ex}
\noindent
{\bf Learning Rate Schedules. }During training, the learning rate decays based on a pre-defined schedule to achieve a higher convergence rate. We describe our concrete learning rate schedules on FPGAs and CPUs. 
On CPUs, the schedule determines the learning rate ($\lambda_e$) of the $e$-th epoch to be $\lambda \times \beta^{\sqrt{e}}$, where $\lambda$ is the initial learning rate while $\beta$ is the decay factor. In our experiment, a more aggressive decay policy, e.g., $\lambda/\sqrt{e}$ or $\lambda/e$, slows down the convergence rate. We find that a constant learning rate leads to the best performance on several datasets.  
On FPGAs, we employ a relatively simple schedule, as shown in Equation~\ref{E_learning_rate_fpga}, where $\alpha$ is the threshold to decay the learning rate, because 1) on FPGAs, we apply the learning rate ($\lambda_e$) as a right-shift operator and thus $\lambda_e = 2^{-j}$, where $j$ is an integer; and 2) MLWeaving needs a lower number of epochs to converge to the same training loss versus its CPU rivals that are not synchronous.   
\begin{equation}
		\vspace{-1ex}
\begin{scriptsize}
\label{E_learning_rate_fpga}
\lambda_e=
\left\{
\begin{array}{cc}
\lambda, & e \le \alpha \\
\lambda*0.5, & e > \alpha \\
\end{array}
\right.
\end{scriptsize}
\end{equation}

%\begin{equation}
%		\vspace{-1ex}
%\begin{scriptsize}
%\label{E_learning_rate_schedule}
%\lambda_e= \lambda*drop^{\lfloor e/\alpha \rfloor}, 
%\end{scriptsize}
%\end{equation}
%where $drop$ is set to its typical value: 0.5.
\vspace{0.5ex}
\noindent
{\bf Comparison Methodology. }
Our evaluations mainly validate three hypotheses. First, MLWeaving can achieve linear speedup when a smaller number of bits is used in the training (Subsection~\ref{subsec_experiment_hardware_efficiency}). %the performance of MLWeaving increases linearly as the number ($s$) of bits decreases (hardware efficiency). 
Second, MLWeaving converges faster than its first-order counterparts on CPUs (Subsection~\ref{subsec_experiment_loss_time}). Third, using a dynamic precision schedule further accelerates the convergence process (Subsection~\ref{subsec_experiment_precision_schedule}).
%to show the statistical efficiency, we mainly 
%show the performance advantage of the proposed per-epoch tuning of precision under MLWeaving. 
%Third, we compare the hardware resource consumption of MLWeaving with the existing work.  

%In order to demonstrate the advantages of MLWeaving,  quantitatively analyze its linear scalability 
%and per-epoch selection of precision.  
	
	%The 
%The intuition behind flexible selection is that lower level of precision (higher hardware efficiency) can reduce the training loss to a moderate extent, and then the higher level of precision is used to reach the given loss target. In the end, it can converge to the same loss with less computation time.    
%%In order to demonstrate the performance advantages of the proposed memory-efficient execution model, we use three implementations. 
%Three implementations are used for performance comparison. The first one is \NameOfModel (denoted as ``Hebe"). The other two implementations come from the state-of-the-art column-first execution model~\cite{ByteSlice_sigmod_15} under ByteSlice memory layout. ``BS\_best" indicates the implementation with an optimal evaluation order while ``BS\_worst" indicates the implementation with the worst evaluation order. 

%ImageNet (Transfer Learning)
\vspace{-1ex}
\subsection{Hardware Efficiency: Throughput}
\label{subsec_experiment_hardware_efficiency}
In this subsection, we demonstrate the hardware efficiency of MLWeaving, i.e., elapsed time for each epoch.\footnote{The elapsed time for each epoch is inversely proportional to the throughput that each implementation can achieve. In the following, we use both terms interchangeably to illustrate hardware efficiency. } Our objective is two-fold. First, we analyze the performance characteristics of MLWeaving on FPGAs. Second, we compare MLWeaving with the state-of-art implementations on CPUs. %Since the hyper-parameter, e.g., learning rate, would not affect the hardware efficiency, we omit the real value in this subsection. %the performance scales linearly with the number of bits used. 
\vspace{-1ex}
\subsubsection{Hardware Characteristics of MLWeaving}
We analyze five different hardware properties of MLWeaving. In our analysis, we typically run 50 epochs and get the average time for each epoch. 
%performance difference vs bits (applications).Its impact varies with the number of features, the number of bits and batch size.the following 

\begin{figure}[t]
	\centering
    \vspace{-2ex}
	%\hfill
	\subfloat[Effect of chaining]{\includegraphics[width=1.65in]{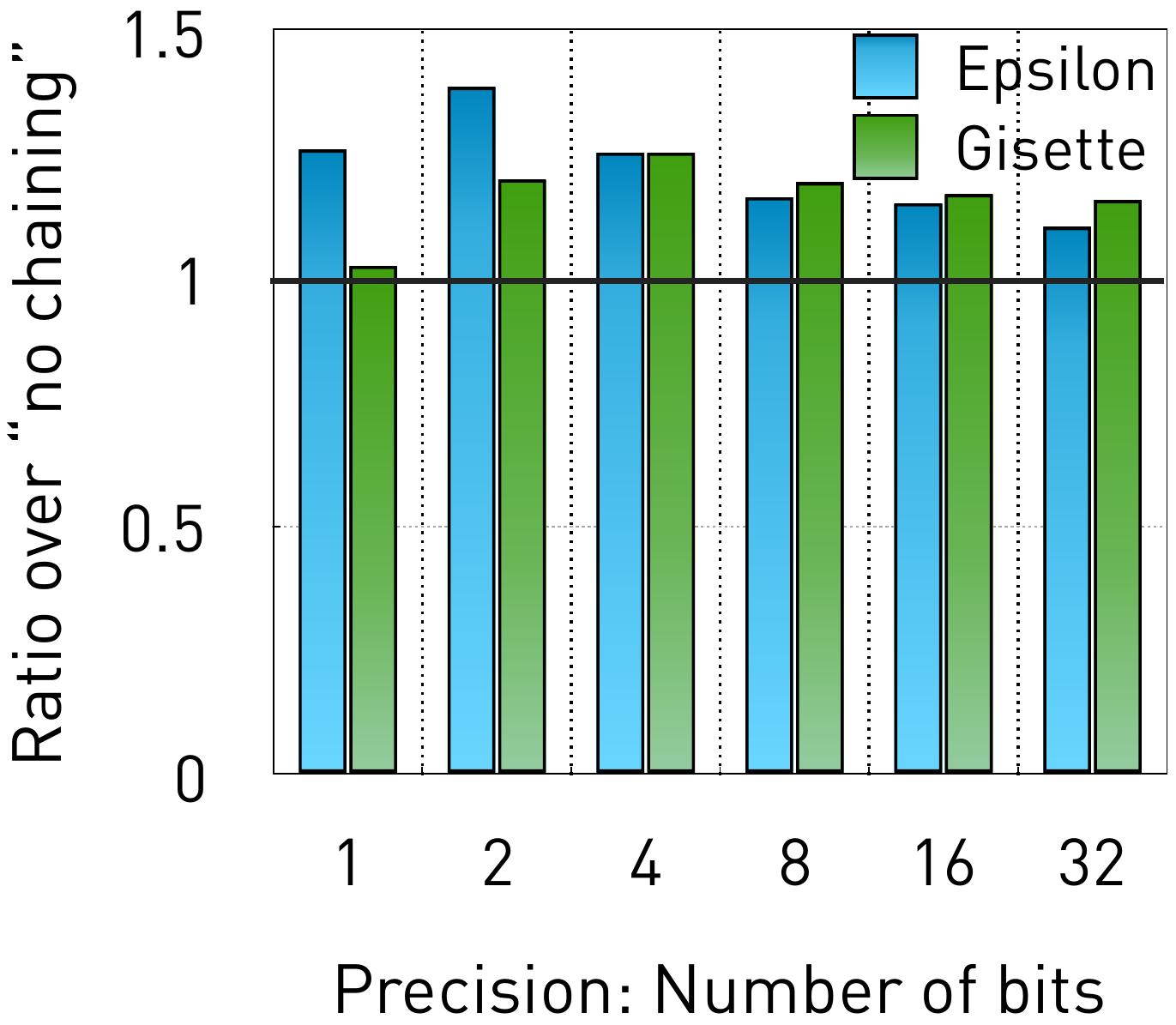} %Speedup of ``chaining" over ``no chaining
		\label{fig_imapct_of_chaining}} 
	\subfloat[Effect of batch size on Gisette]{\includegraphics[width=1.65in]{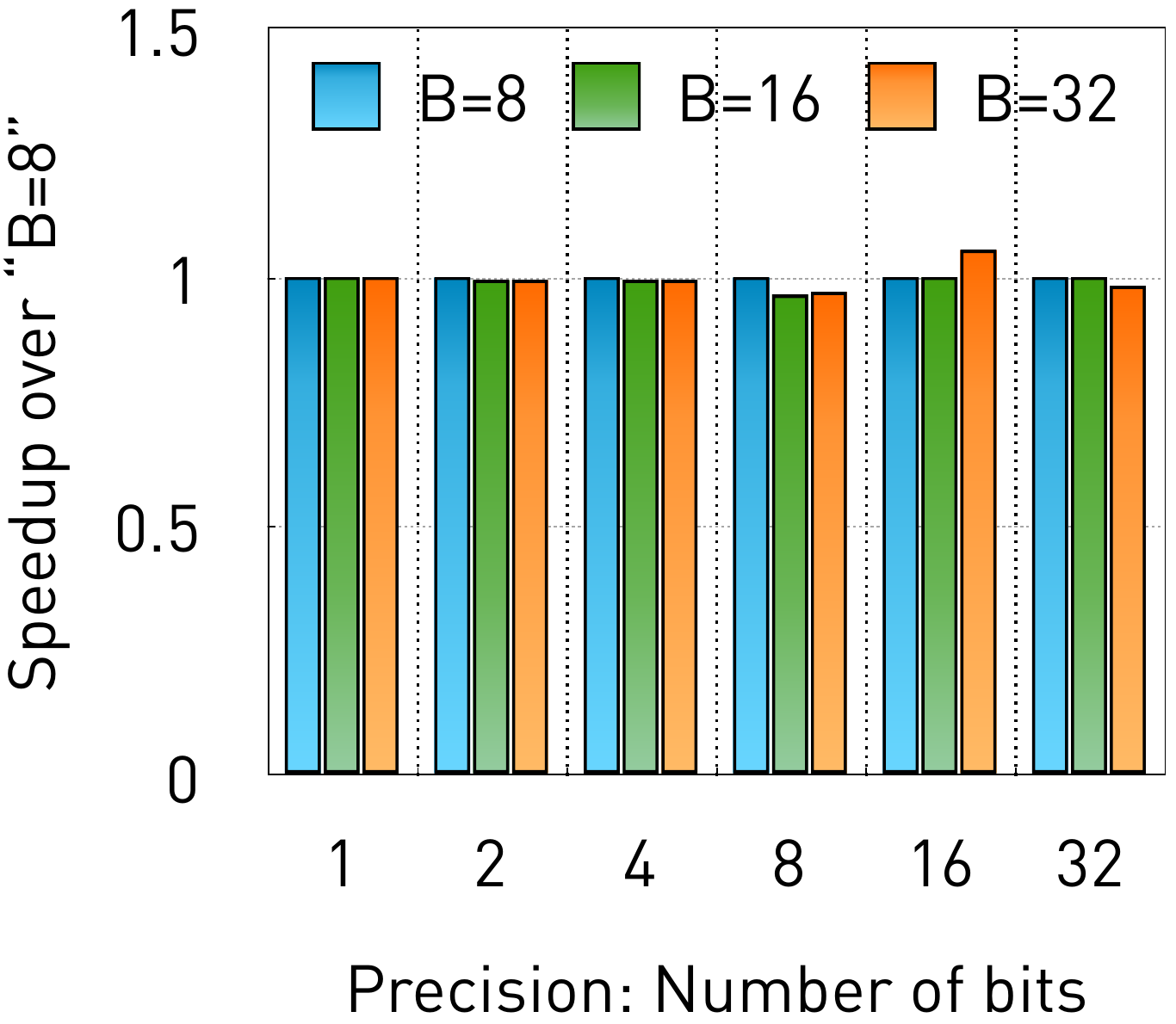} %Speedup of batch size (B) over ``B=8" for Gisette
		\label{fig_imapct_of_batch_size_hw}} % \caption{}
	\vspace{-1.5ex}	
	\caption{Hardware characteristics on hardware efficiency.} %: Hogwild and ModelAverage$\lambda$ is $1/2^{10}$ (``with RAW"), or $1/2^{13}$ (``without RAW"). 
	\label{fig_he_optimizations} 
	%\vspace{-1.5ex}
\end{figure}

\vspace{0.5ex}
\noindent
{\bf Effect of Chaining. }We examine the effect of our chaining technique that relaxes the unnecessary RAW dependency for MLWeaving (Section~\ref{subsection_chaining}). Figure~\ref{fig_imapct_of_chaining} decipts the speedup of ``chaining" over ``no chaining" for two datasets. The batch size is 8. The x-axis depicts the precision $s$. We observe that with chaining, we can achieve up to 1.4X speedup over no chaining for different combinations of precision level ($s$) and number of features, since chaining can fully overlap computation and memory access.  We conclude that chaining significantly increases the hardware efficiency. %``chaining" is vital to achieve high performance while the extra hardware cost (i.e., $<$ 100 slices in FPGA) is trivial.  
\ifarxiv
In order to fully understand the trend, we develop an analytical cost model to predict the performance for both chaining and no chaining settings, as shown in Appendix.
\else
In order to fully understand the trend, we develop an analytical cost model to predict the performance for both chaining and no chaining settings, as shown in Appendix of our technical report~\cite{mlweaving_tr}.
\fi
% \begin{figure}[t]
% 	\centering
% 	\includegraphics[width=6.0cm]{impact_of_chaining.pdf}
% 	\vspace{-1ex}
% 	\caption{Speedup of ``chaining" over ``wo chaining.  }
% 	\vspace{-2.5ex}
% 	\label{fig_imapct_of_chaining}
% \end{figure}

%\normalmarginpar
%\kaancomment{Figure 9 does not provide much information, no? Should we actually explain why larger batch size does not increase performance?}
\vspace{0.5ex}
\noindent
{\bf Effect of Mini Batch Size. }We examine the effect of mini batch size ($B$) on hardware efficiency. Figure~\ref{fig_imapct_of_batch_size_hw} illustrates the speedup of various batch sizes over ``B=8" on the dataset Gisette. We observe that performance is roughly stable for different batch sizes, since MLWeaving is able to maximally overlap computation and memory access, regardless of mini batch size. We conclude that the batch size has a negligible effect on hardware efficiency. % is vital to achieve high performance while the extra hardware cost (i.e., $<$ 100 slices in FPGA) is trivial.  

\begin{figure}[ht]
	\centering
	\vspace{-2ex}
	%\hfill
	\subfloat[Gisette]{\includegraphics[width=1.56in]{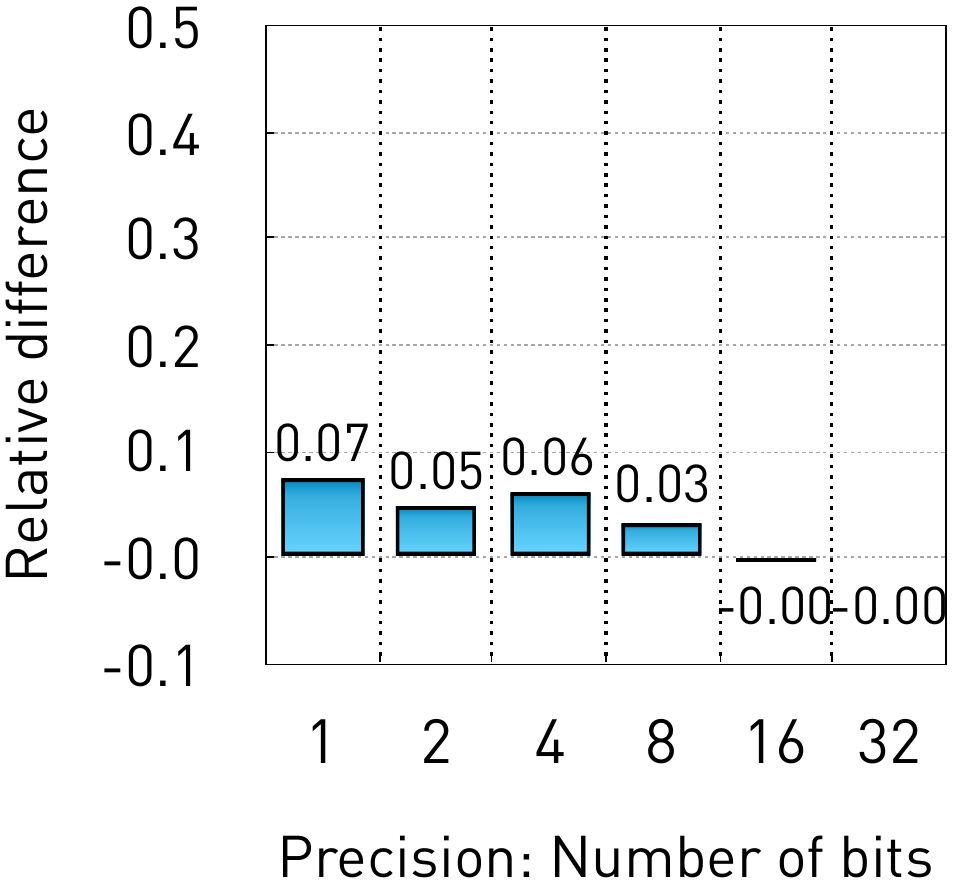} 
		\label{fig_cmp_RAW_time_gisette}} 
	\subfloat[Epsilon]{\includegraphics[width=1.56in]{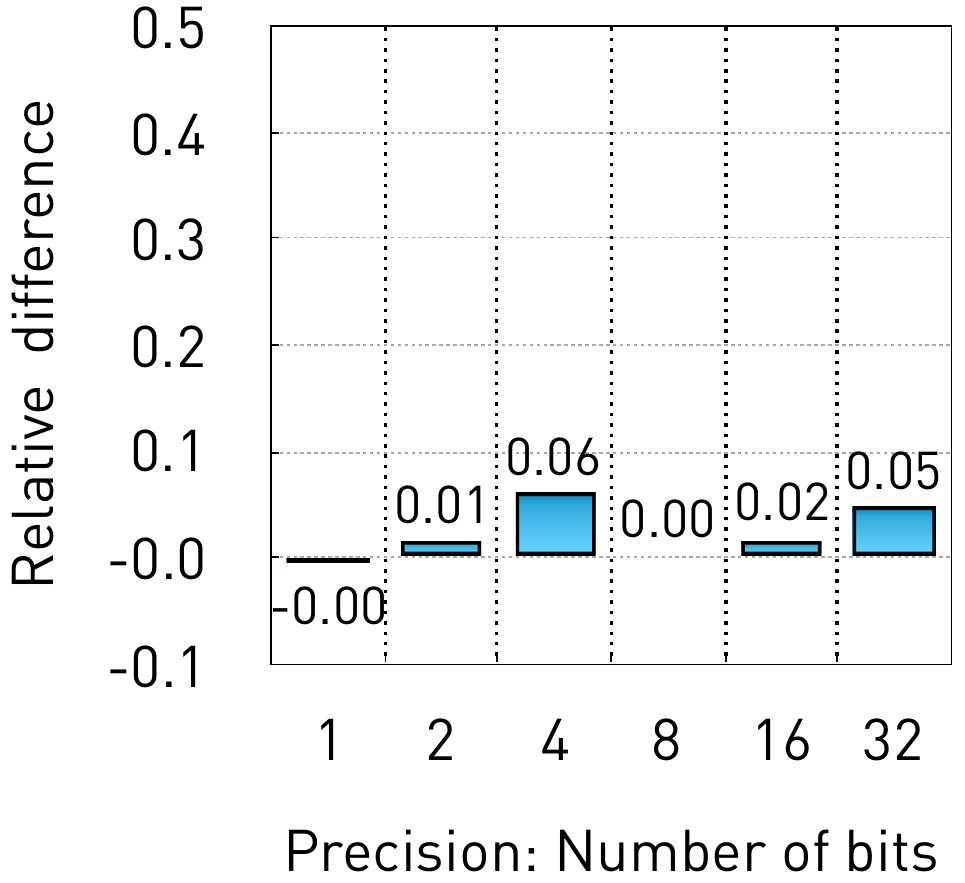} 
		\label{fig_cmp_RAW_time_epsilon}} % \caption{}
	\vspace{-1ex}	
	\caption{Relative performance difference between MLWeaving (sync) and MLWeaving (async). $B$ is 16. $s$ is 8.} %: Hogwild and ModelAverage$\lambda$ is $1/2^{10}$ (``with RAW"), or $1/2^{13}$ (``without RAW"). 
	\label{fig_cmp_RAW_time} 
	\vspace{-2ex}
\end{figure}

% \begin{figure}[t]
% 	\centering
% 	\includegraphics[width=8.0cm]{impact_of_batch_size_hw_gisette.pdf}
% 	\vspace{-2ex}
% 	\caption{Speedup of various batch size (B) over ``B=8".  }
% 	\vspace{-3.5ex}
% 	\label{fig_imapct_of_batch_size_hw}
% \end{figure}
%performance difference vs bits (application)  
\vspace{0.5ex}
\noindent
{\bf MLWeaving (sync) vs. MLWeaving (async). }We examine the effect of the RAW dependency on hardware efficiency. Intuitively, MLWeaving (async) would achieve more throughput than MLWeaving (sync), since MLWeaving (async) does not need to wait the RAW dependency to be resolved. However, since both approaches are memory-bound on the targeted FPGA, their relative performance difference, i.e., (MLWeaving (sync)- MLWeaving (async))/MLWeaving (sync), is small, as shown in Figure~\ref{fig_cmp_RAW_time}. We conclude that MLWeaving (sync), which preserves the RAW dependency, has little effect on hardware efficiency, with the help of our chaining technique.%The batch size is 16. %shows the relative performance difference between ``with RAW" and ``Without RAW". 
%\kaancomment{In Figure 10, how is the difference calculated? Is it async-sync or the other way around?}

\begin{figure}[t]
	\centering
	%\hfill
	\subfloat[Gisette (5000 features)]{\includegraphics[width=1.6in]{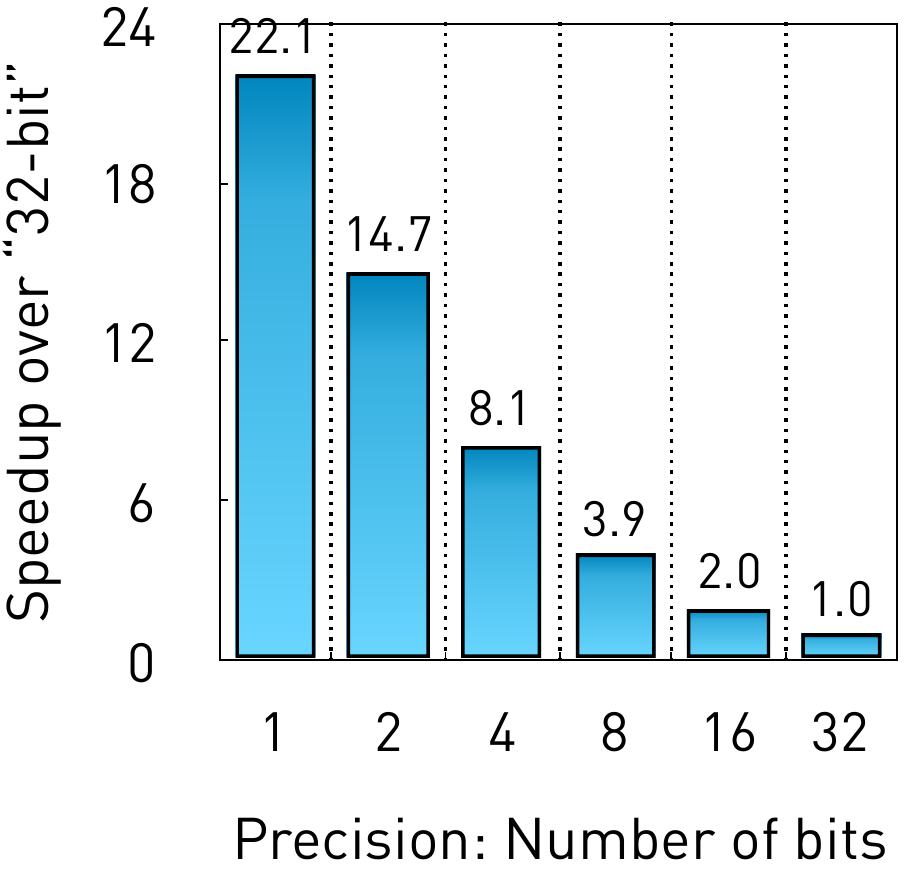} 
		\label{fig_time_bits_gisette}} 
	\subfloat[Epsilon (2000 features)]{\includegraphics[width=1.6in]{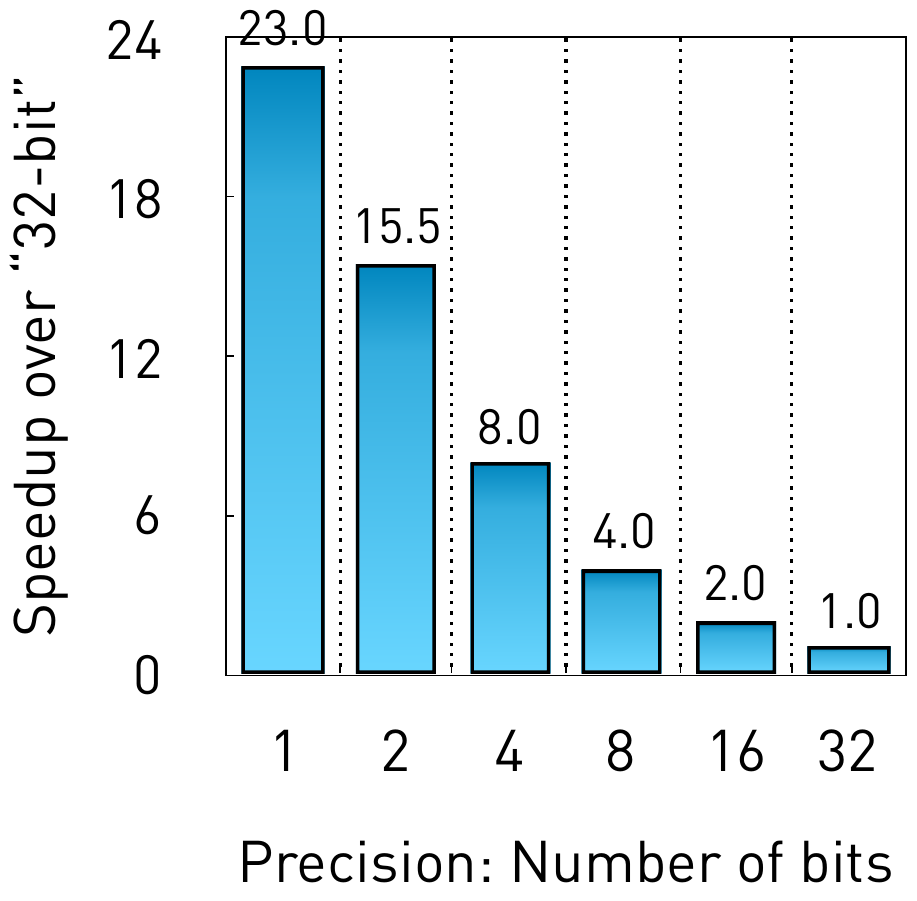} 
		\label{fig_time_bits_epsilon}} % \caption{}
	\vspace{-1ex}
	\caption{Relative performance improved with various precision levels over ``32-bit", where ``32-bit" is the case with $s$ = 32. } 
	\vspace{1ex}
	\label{fig_time_bits} 
	%\vspace{-3ex}
\end{figure}

%Time vs bits (one application + CPU: one line), three applications.  
\vspace{0.5ex}
\noindent
{\bf Effect of Precision Level on Execution Time. }Figure~\ref{fig_time_bits} provides the performance improvement of various precision levels over the full-precision implementation ``32-bit" under MLWeaving for two datasets. We make two observations. First, the performance of MLWeaving improves roughly linearly as bit precision reduces, especially when $s$ is larger than 4. Second, when $s$ is less than 4, the speedup is sub-linear since the benefit from our chaining technique cannot fully amortize the negative impact of the inherent pipeline latency. We conclude that MLWeaving significantly reduces the elapsed time when using a smaller number of bits. %MLWeaving can roughly achieve linear speedup when the product of precision level $s$ and number of features $M$ is relatively large.
%\kaancomment{Maybe we need to be less vague here?}.     
%we can conclude that CPU is 
%\vspace{-2ex}

\begin{figure}[t]
	\centering
		%\vspace{-2ex}
	%\hfill
	\subfloat[Gisette (5000 features)]{\includegraphics[width=1.63in]{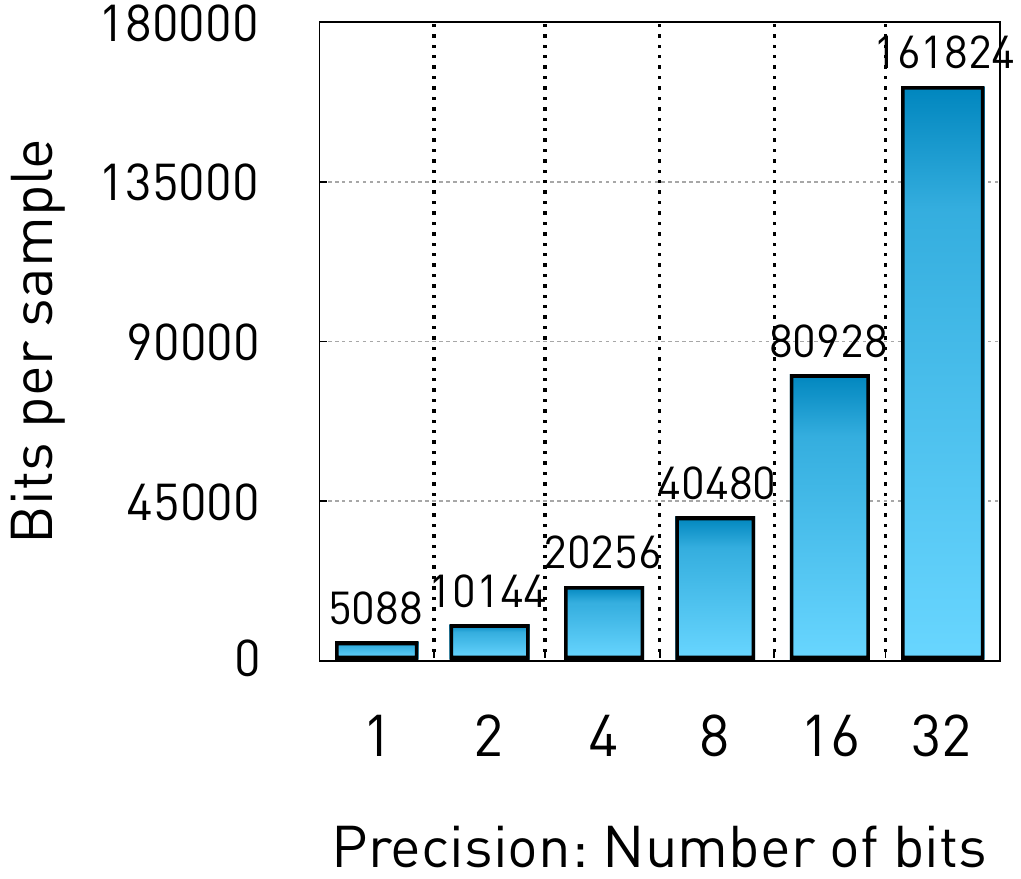} 
		\label{fig_mem_bits_gisette}} 
	\subfloat[KDD (2399 features)]{\includegraphics[width=1.63in]{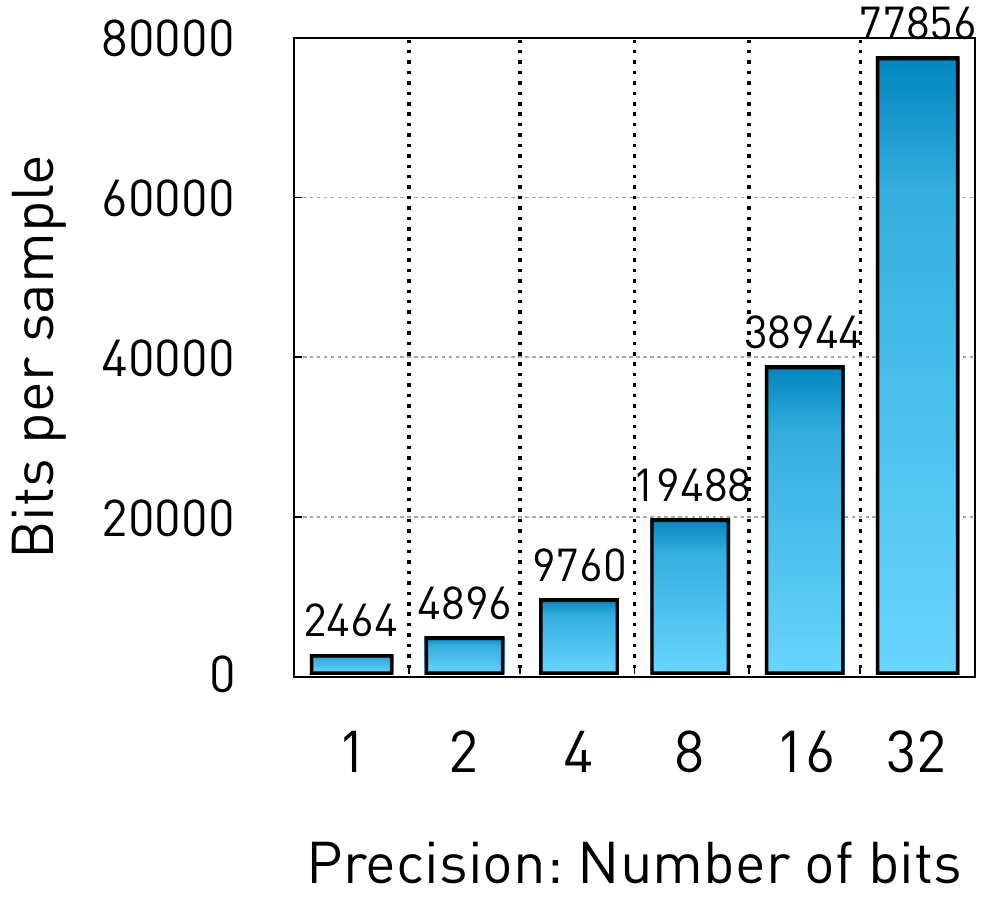} 
		\label{fig_mem_bits_imagenet}} % \caption{}
	\vspace{-1.5ex}	
	\caption{Memory traffic (bits) per sample as the precision varies. } 
	\label{fig_mem_bits} 
	%\vspace{-1ex}
\end{figure}

%Memory traffic vs bits (one application + CPU: one line), three applications.  
%\vspace{-1.5ex}
\noindent
{\bf Effect of Precision Level on Memory Traffic. }Figure~\ref{fig_mem_bits} illustrates the memory traffic required by each sample for two datasets, as $s$ is varied. We observe that the required memory traffic almost increases linearly as the precision level increases, consistent with the trend demonstrated in Equation~\ref{E_memory_traffic}. We conclude that the MLWeaving memory layout enables efficient data retrieval from external memory for any precision level at runtime.  %

%Based on the above analysis, we can validate the hypothesis that MLWeaving can use one index and one hardware design to efficiently support arbitrary-precision machine learning training. 

\vspace{-1.5ex}
\subsubsection{Comparison with CPU Implementations}
\begin{figure}
	\centering
	%\hfill
	\subfloat[Normalized throughput]{\includegraphics[width=2.36in]{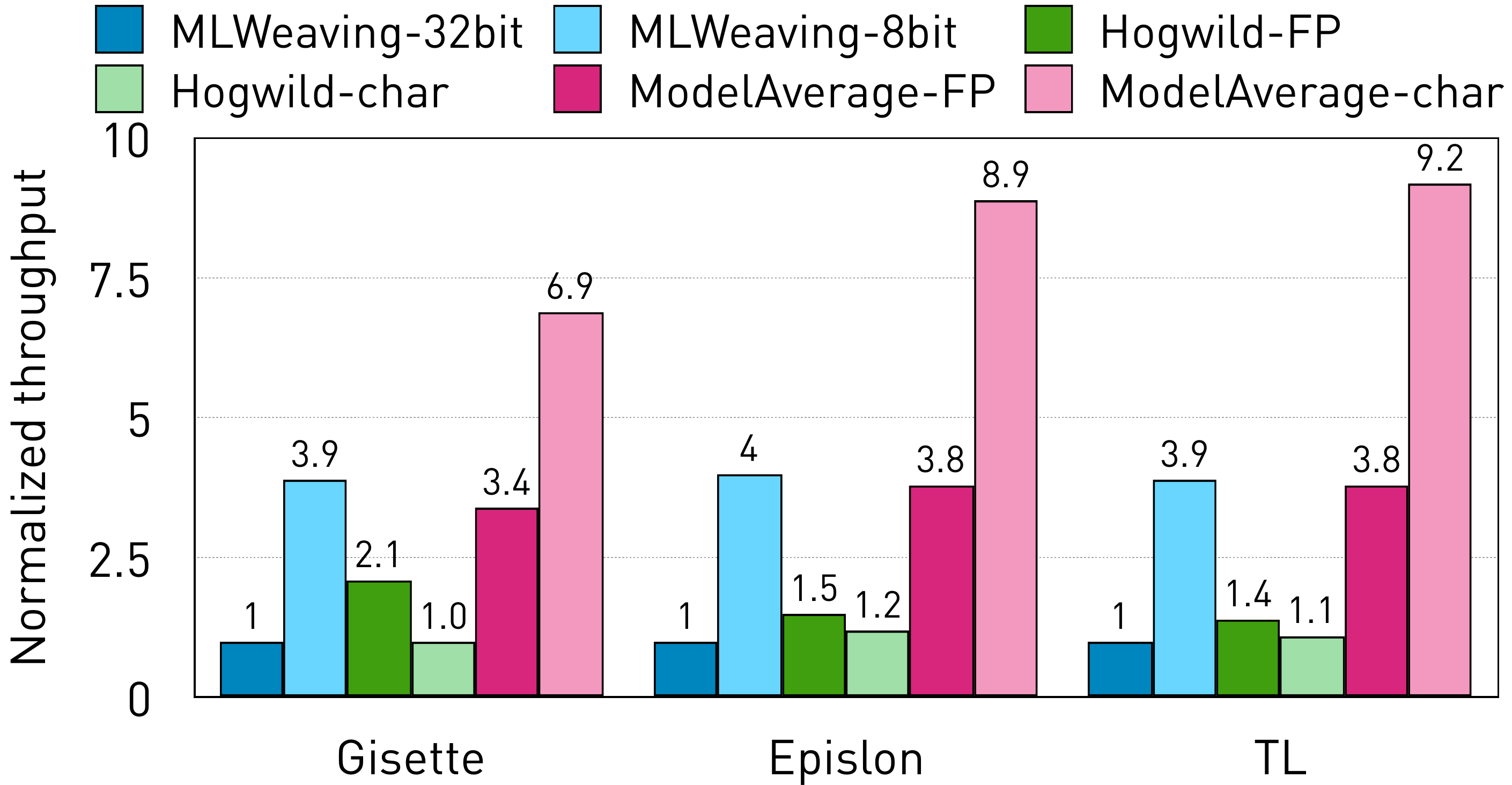} 
		\label{fig_cmp_cpu_time}} 
		\hfill
	\subfloat[Normalized memory traffic]{\includegraphics[width=2.25in]{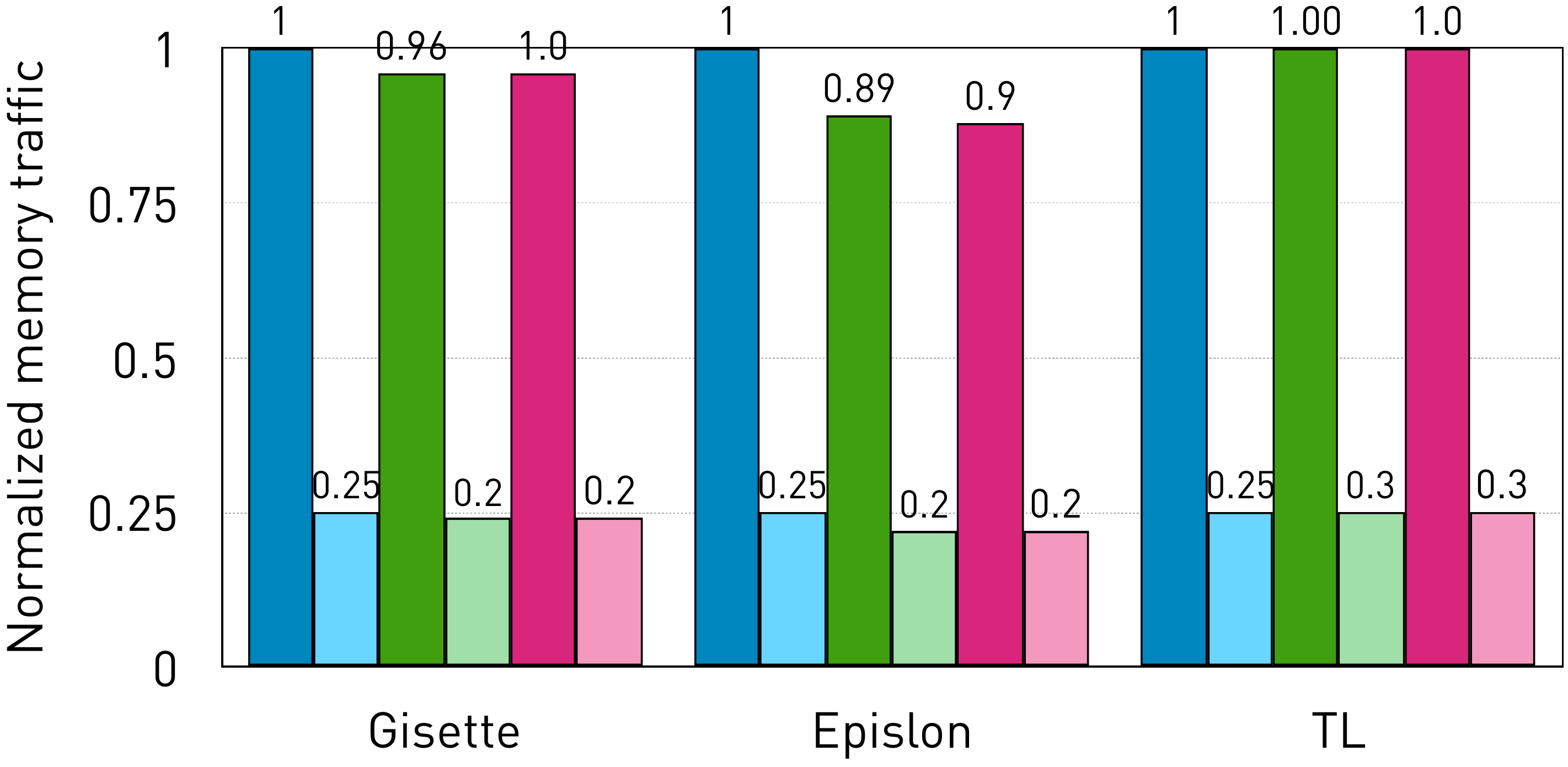} 
		\label{fig_cmp_cpu_bytes}} % \caption{}
	\vspace{-1.5ex}	
	\caption{Hardware efficiency: MLWeaving vs. CPU rivals. } %: Hogwild and ModelAverage
	\label{fig_cmp_cpu} 
	\vspace{1.5ex}
\end{figure}

We compare the throughput of MLWeaving with two state-of-art CPU algorithms: Hogwild and ModelAverage. Figure~\ref{fig_cmp_cpu} shows the comparison result. Both CPU algorithms are fully optimized. They use a low-precision dataset, employ all the 14 cores and are AVX2-enhanced. ``x-FP" (or ``x-char") indicates ``x'' with floating-point (or char) dataset, where ``x'' is Hogwild or ModelAverage. We use two metrics for comparison: time and memory traffic. ``MLWeaving-32bit" (or ``MLWeaving-8bit") means MLWeaving with $s$ = 32 (or 8) on FPGAs.

\vspace{0.5ex}
\noindent
{\bf Time. }Figure~\ref{fig_cmp_cpu_time} illustrates the normalized throughput of two CPU approaches and MLWeaving.
We make two major observations. First, ModelAverage-FP achieves roughly 4 times more throughput than MLWeaving-32bit, since both are memory-bound and the achievable memory read bandwidth of the CPU (i.e., 60GB/s) is roughly four times as much as that of the FPGA (i.e., 15GB/s). ModelAverage-FP and MLWeaving-8bit roughly have the same throughput, while ModelAverage-char (which is compute-bound) achieves obviously more throughput than MLWeaving-8bit. Second, even though Hogwild-FP achieves more throughput than MLWeaving-32bit, it is still compute-bound since it suffers from severe cache coherence overhead due to the fact that multiple cores try to update the same memory address. When the model dimension becomes smaller, the overhead becomes larger as cache invalidation occurs more frequently among cores.
\ifarxiv
\footnote{In Appendix, we add the experiments about MLWeaving on CPUs, where the lack of hardware instructions makes it hard to exploit MLWeaving on the CPU.  }
\else
\footnote{In Appendix of our technical report~\cite{mlweaving_tr}, we add the experiments about MLWeaving on CPUs, where the lack of hardware instructions makes it hard to exploit MLWeaving on the CPU.  }
\fi
%\kaancomment{Maybe we need to dive a bit deeper here in terms of explanation? First, we need to say for which data sets Hogwild is a good match (Gisette) and for which ones it is not (MNIST), and then why: because it performs asynchronous updates to the same memory address, leading to the same cache locations to be invalidated very frequently.}
%Generally, CPU is faster, FPGA can beat CPU with less bits. 
% \begin{figure}
% 	\centering
% 	%\hfill
% 	\subfloat[Gisette (5000 features)]{\includegraphics[width=1.6in]{time_bits_gisette.pdf} 
% 		\label{fig_time_bits_gisette}} 
% 	\subfloat[IM (2048 features)]{\includegraphics[width=1.6in]{time_bits_imagenet.pdf} 
% 		\label{fig_time_bits_imagenet}} % \caption{}
% 	\hfill
% 	\vspace{-1ex}	
% 	\subfloat[Epsilon (2000 features)]{\includegraphics[width=1.6in]{time_bits_epsilon.pdf} 
% 		\label{fig_time_bits_epsilon}} % \caption{}
% 	%\hfill
% 	\subfloat[MNIST (784 features)]{\includegraphics[width=1.6in]{time_bits_mnist.pdf} 
% 		\label{fig_time_bits_mnist}} 
% 	\vspace{-1ex}
% 	\caption{Performance increase of various precision level $s$ over ``32-bit", where ``32-bit" means the case with $s$ = 32. } 
% 	\vspace{-1ex}
% 	\label{fig_time_bits} 
% 	\vspace{-3ex}
% \end{figure}  

\vspace{0.5ex}
\noindent
{\bf Memory Traffic. } Figure~\ref{fig_cmp_cpu_bytes} illustrates the normalized memory traffic, where the memory traffic on the CPU is collected using the Intel Performance Counter Monitor~\cite{Intel_PCM}. We observe that both Hogwild-FP and ModelAverage-FP require roughly the same memory traffic as MLWeaving-32bit, since their datasets are all full-precision (32-bit), while Hogwild-char and ModelAverage-char require roughly the same amount of memory traffic as MLWeaving-8bit. We conclude that low-precision datasets causes less memory traffic on both CPUs and FPGAs.  

\vspace{0.5ex}
To sum up, due to the small amount of memory bandwidth available on FPGAs, MLWeaving has raw throughput advantage over CPU approaches only when the chosen precision level is relatively low, e.g., less than 4.

	\vspace{1ex}	
\subsection{Statistical Efficiency: Loss vs. Epochs}
\label{subsec_experiment_statistical_efficiency}
We analyze the statistical efficiency of MLWeaving. We mainly validate that MLWeaving requires significantly fewer number of epochs to converge than its state-of-the-art CPU rivals.%First, we validate that MLWeaving (sync) converges faster than its asynchronous counterpart MLWeaving (async). Second, we validate that MLWeaving can converge much faster than its state-of-the-art CPU rivals.

\begin{figure*}[t!]
	\centering
	\subfloat[Epsilon (loss vs. epoch)]{\includegraphics[width=2.05in]{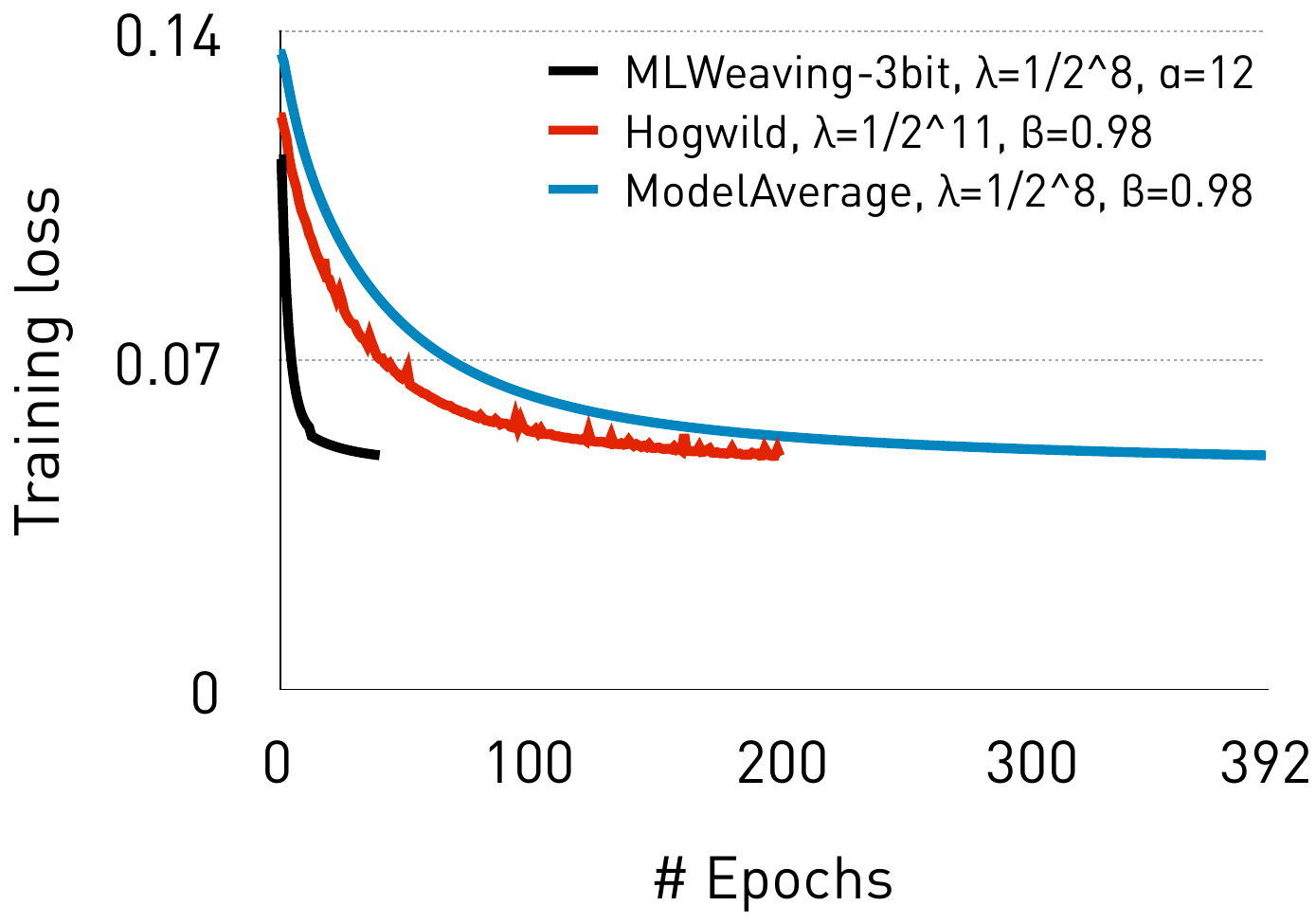} 
		\label{fig_loss_epoch_epsilon}} % \caption{}
	\subfloat[Epsilon (loss vs. time)]{\includegraphics[width=2.25in]{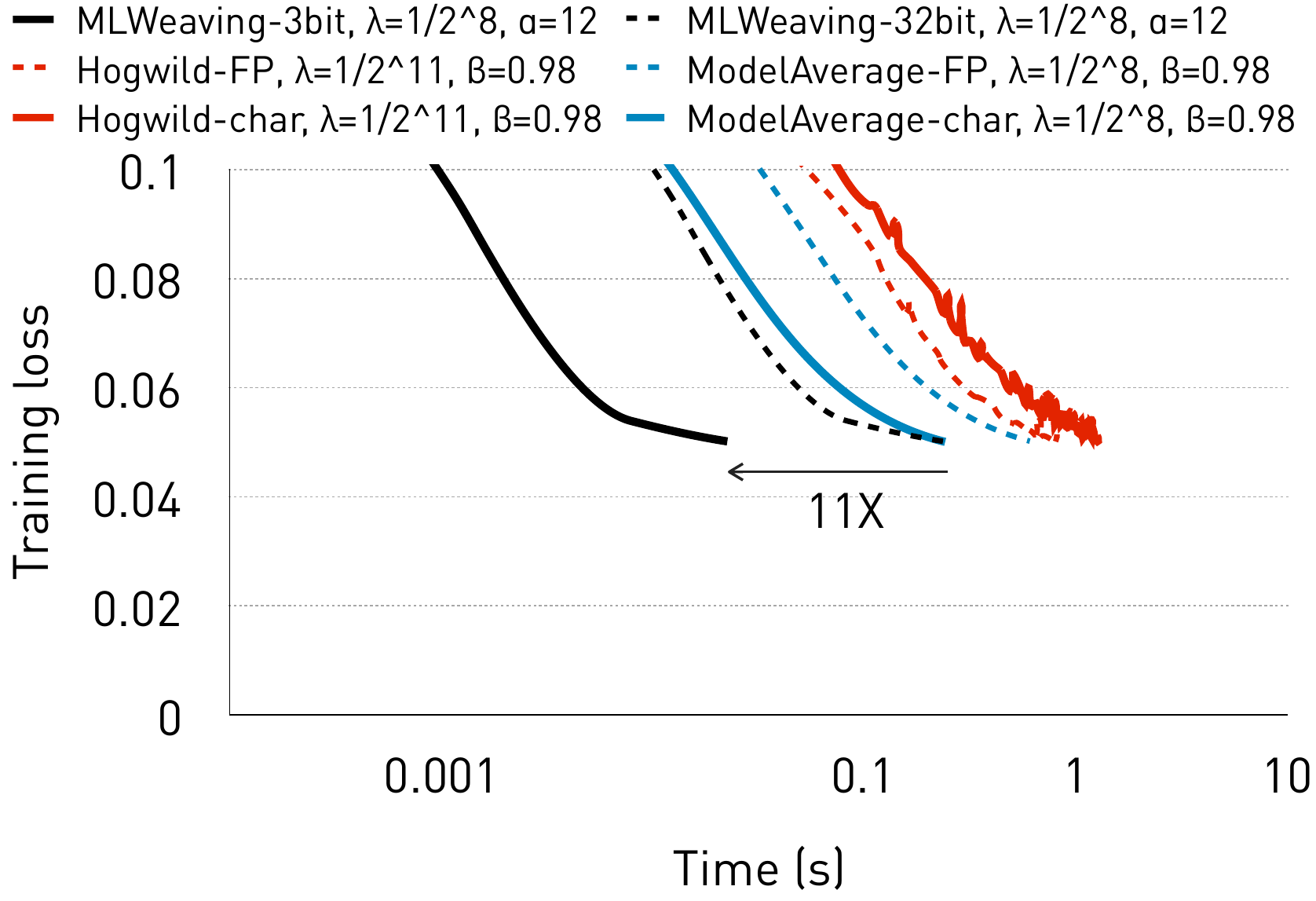} 
	\label{fig_loss_time_epsilon}} % \caption{}
	\subfloat[Epsilon (loss vs. memory traffic)]{\includegraphics[width=2.25in]{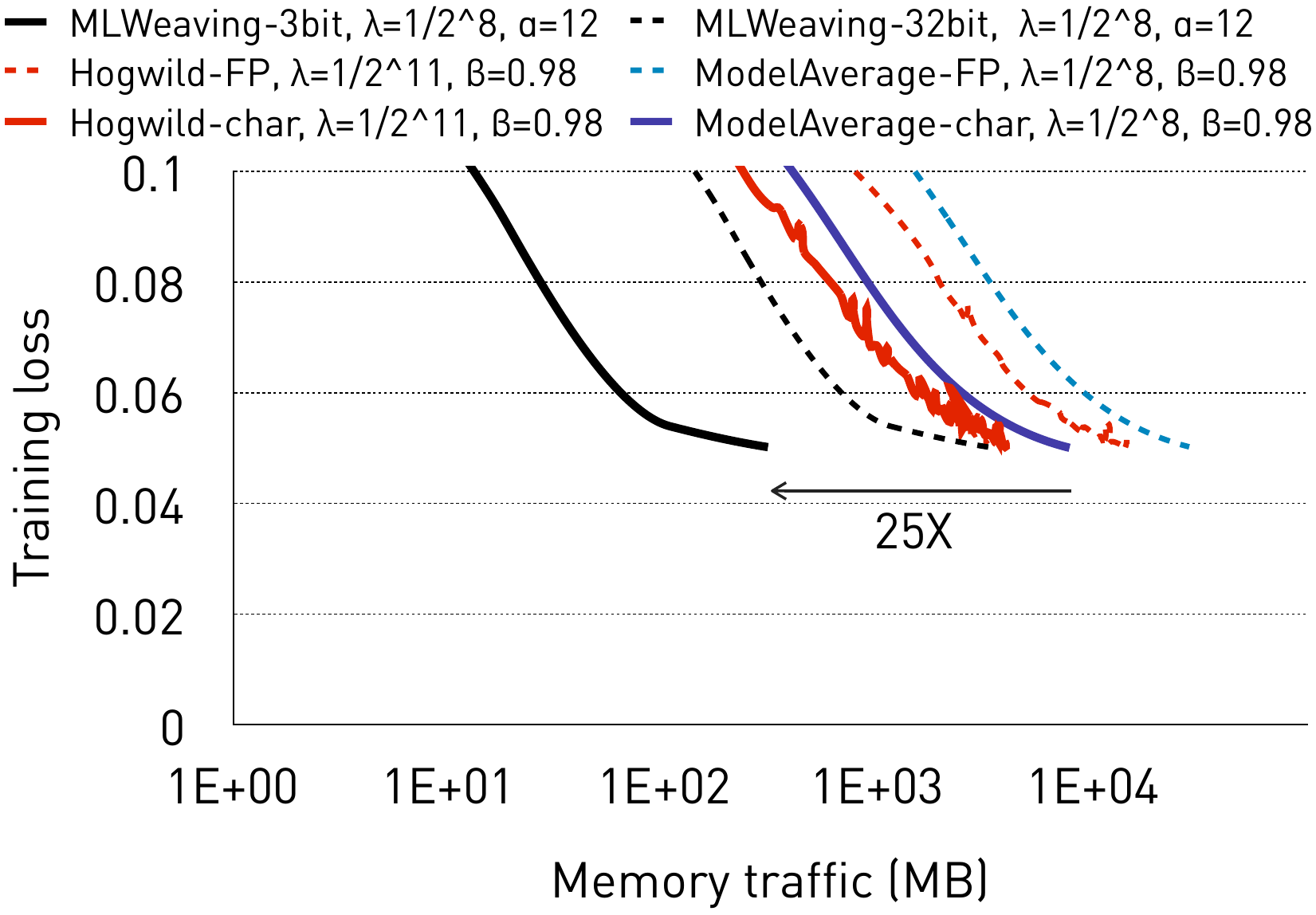} 
			\label{fig_loss_bytes_epsilon}} 
	\hfill
	\subfloat[KDD (loss vs. epoch)]{\includegraphics[width=2.05in]{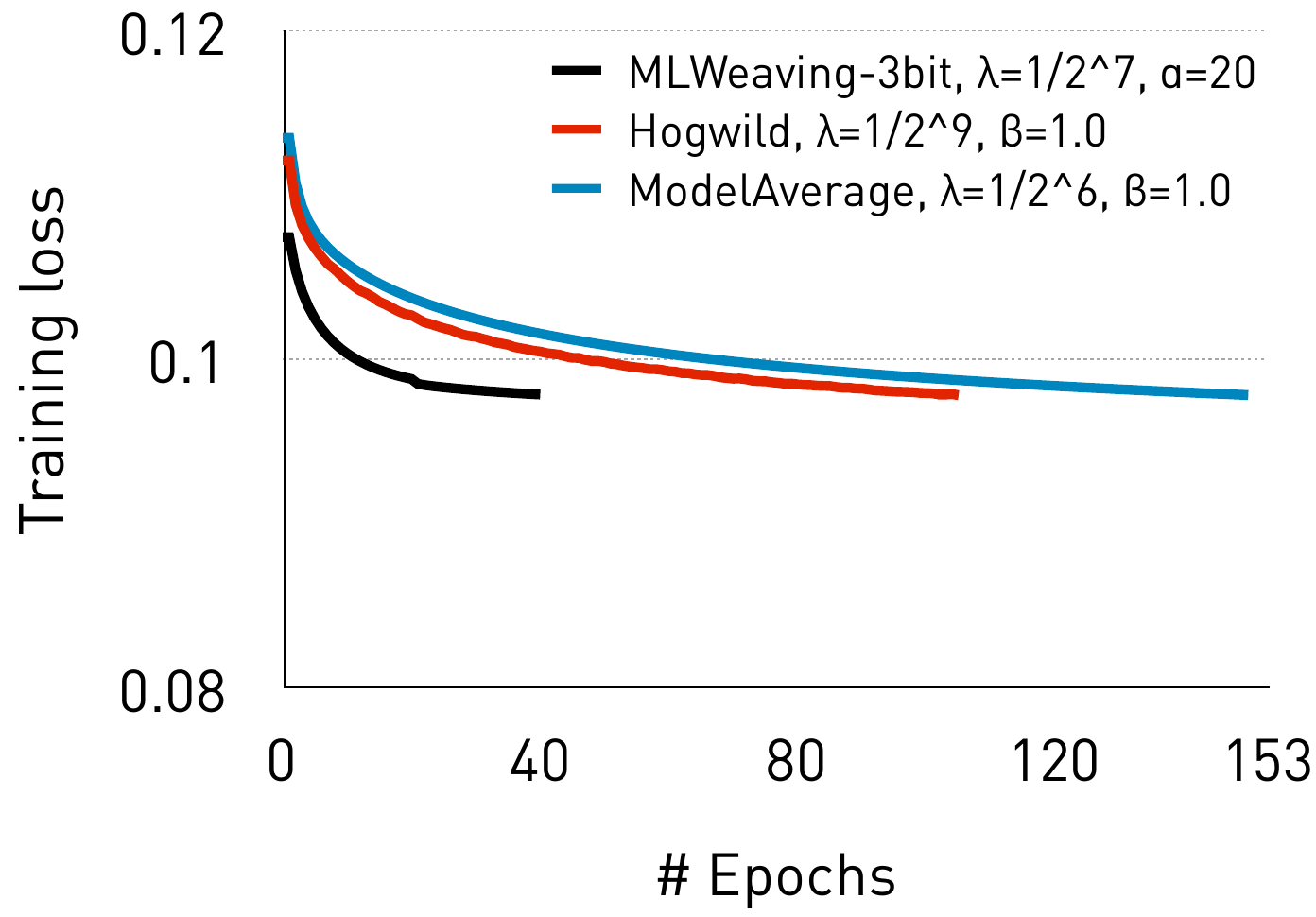} 
		\label{fig_loss_epoch_kdd}} 
	\subfloat[KDD (loss vs. time)]{\includegraphics[width=2.25in]{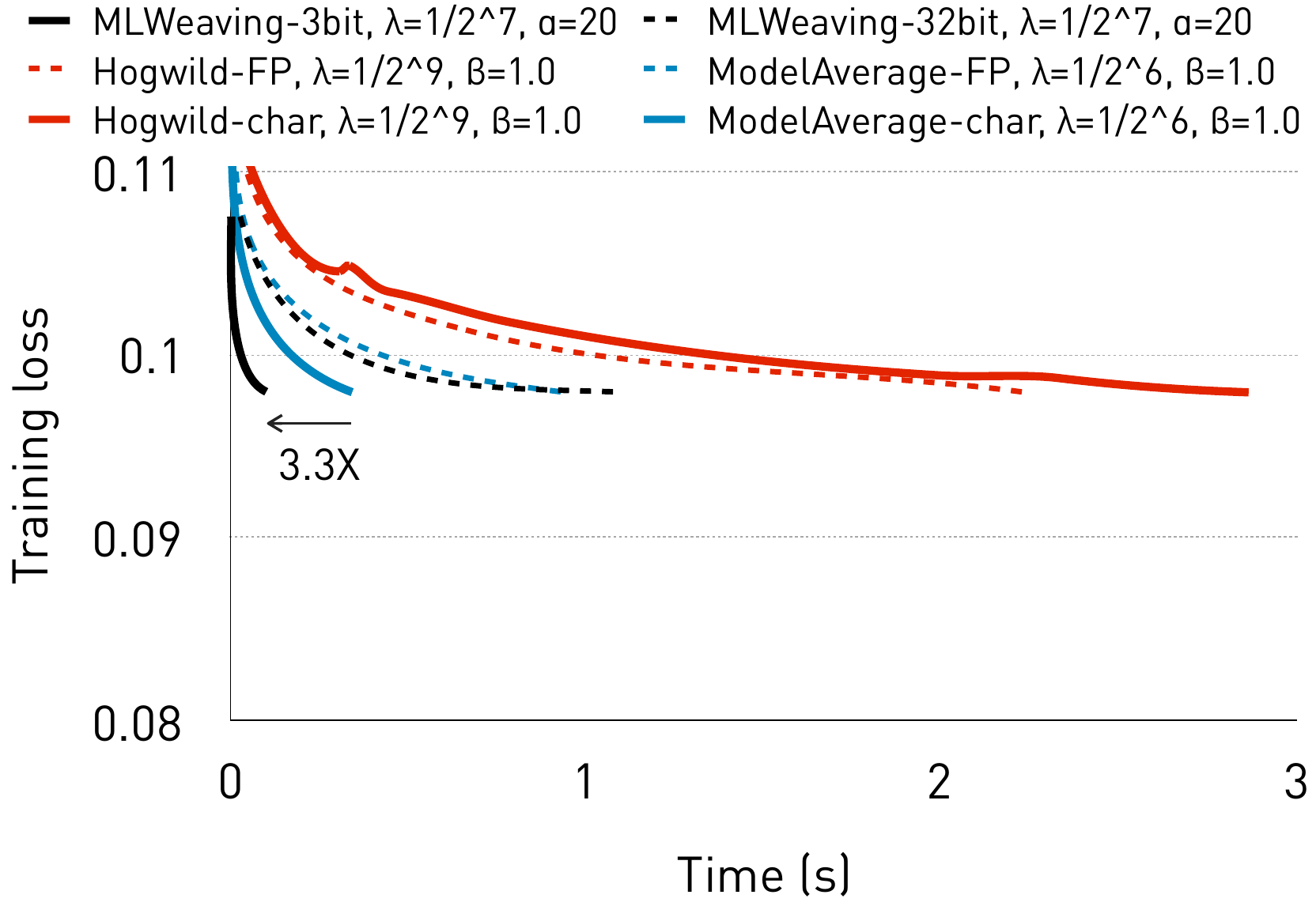} 
		\label{fig_loss_time_kdd}} 
	\subfloat[KDD (loss vs. memory traffic)]{\includegraphics[width=2.25in]{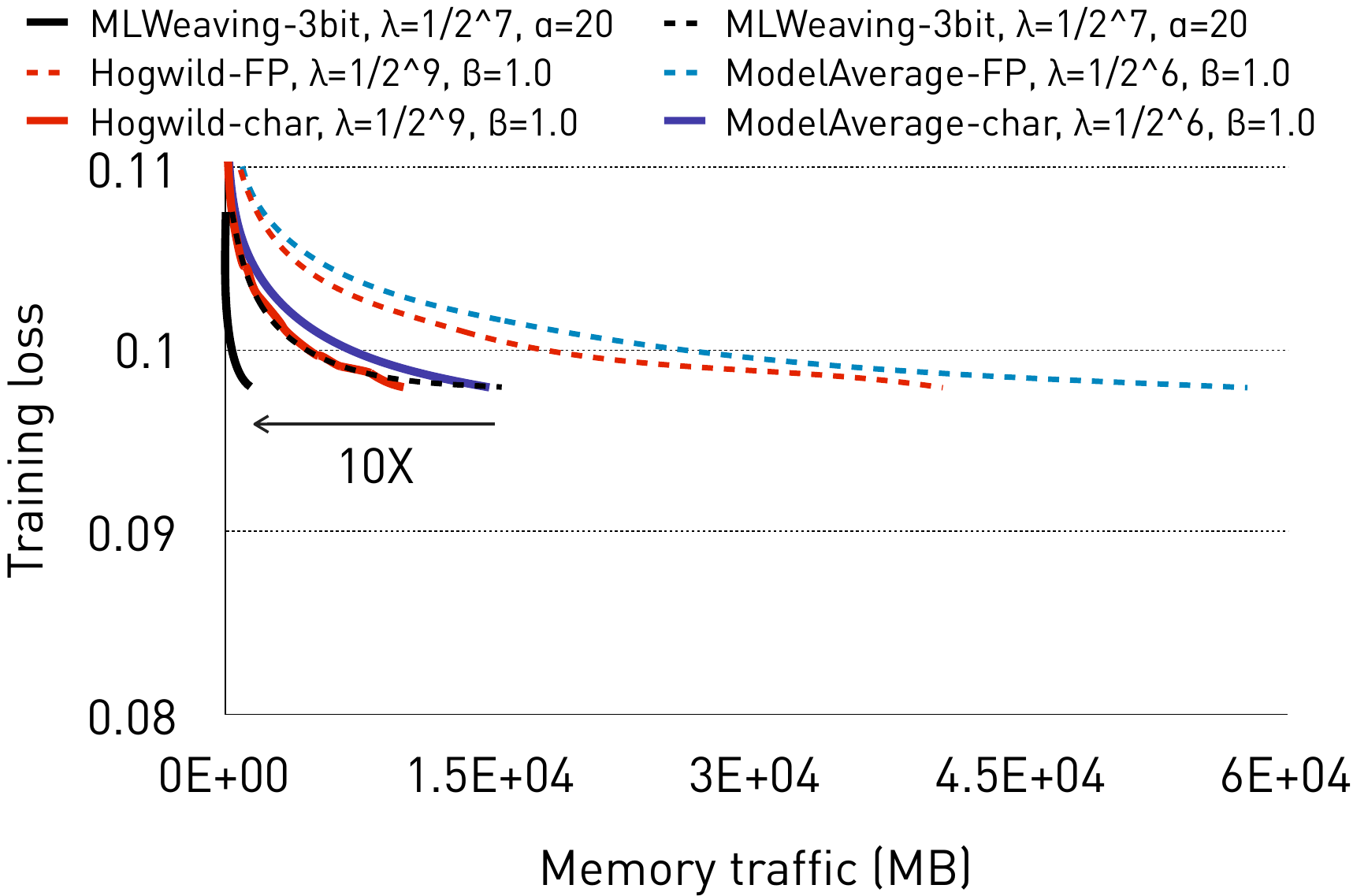} 
		\label{fig_loss_bytes_kdd}} 		
	\vspace{-2.5ex}	
	\caption{Convergence comparison: training loss vs. epoch/time/memory traffic. The batch size is 8. Speedup indicates MLWeaving versus the fastest 14-core AVX2-enhanced low-precision CPU approach, in terms of time and memory traffic. } %MLWeaving employ ``4-bit", ``3-bit" or ``6-bit" to train for the dataset Gisette, Epsilon or ImageNet. ``$\alpha = 40$'' indicates constant learning rate under MLWeaving, since at most 40 epochs are evaluated in our experiment.

	\label{fig_loss_time_epoch} 
	\vspace{-3ex}
\end{figure*}

\vspace{0.5ex}
\noindent
{\bf MLWeaving vs. CPUs. }We compare the statistical efficiency between MLWeaving and two full-precision CPU approaches for two datasets Epsilon and KDD\footnote{We do not examine lower-precision (i.e., less than 8 bits) CPU approaches, as a lower precision always leads to a slightly worse statistical efficiency. } in Figures~\ref{fig_loss_epoch_epsilon},\ref{fig_loss_epoch_kdd}.
We observe that MLWeaving requires a smaller number of epochs to converge than its CPU counterparts, even though MLWeaving uses 3-bit precision while CPU approaches use full precision. For example, MLWeaving requires only 40 epochs to converge for the dataset Epsilon, while ModelAverage (or Hogwild) needs 392 (or 199) epochs to converge to the same loss in Figure~\ref{fig_loss_epoch_epsilon}. The underlying reason is that MLWeaving is always working on the up-to-date model while ModelAverage and Hogwild are not. 

\begin{figure}
	\centering
	%\hfill
	\subfloat[Gisette, $\lambda = 1/2^7$]{\includegraphics[width=1.6in]{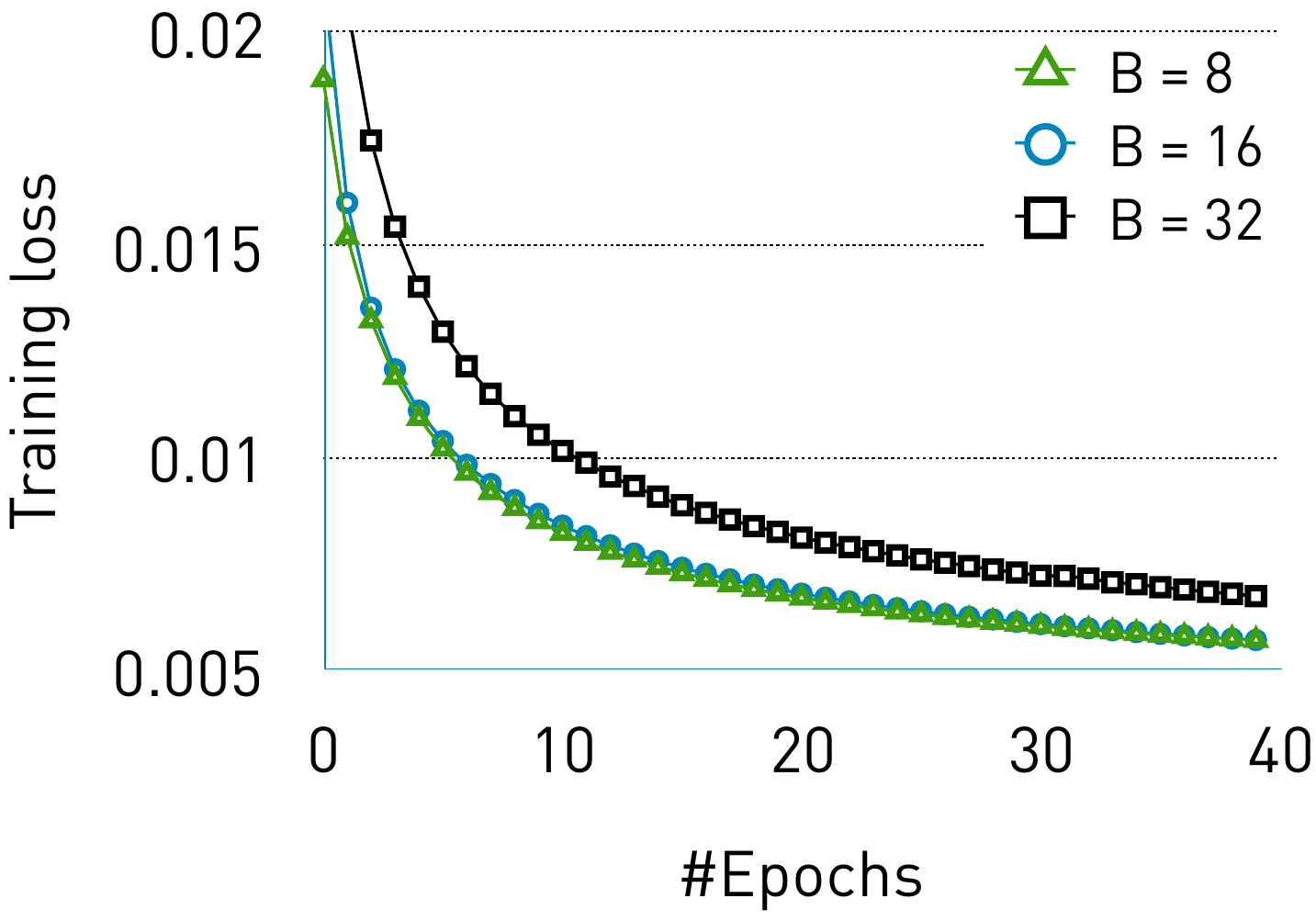} 
		\label{fig_impact_of_batch_size_gisette}} 
	\subfloat[KDD, $\lambda = 1/2^8$]{\includegraphics[width=1.56in]{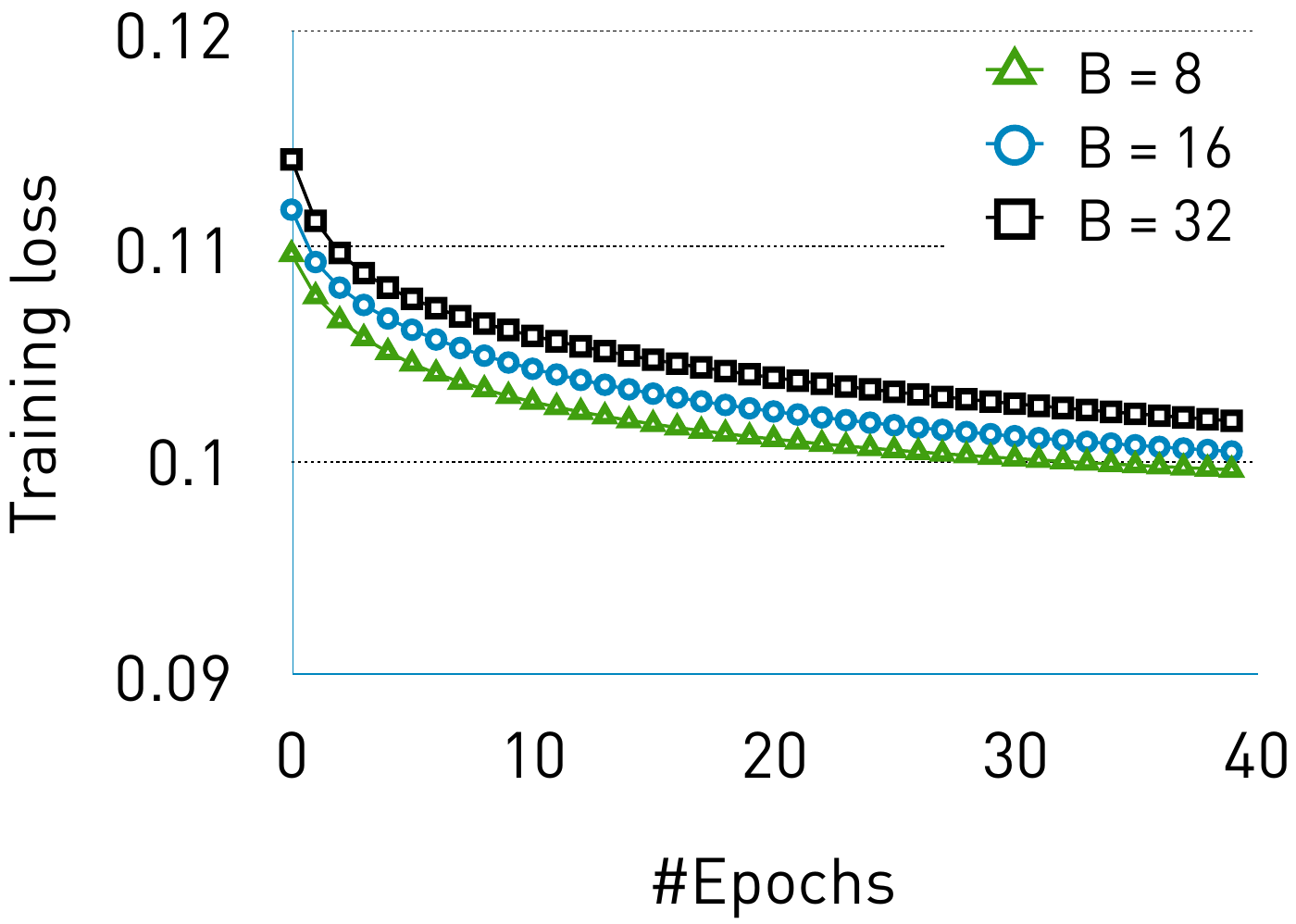} 
		\label{fig_impact_of_batch_size_kdd}} % \caption{}
	\vspace{-1.5ex}	
	\caption{Effect of batch size on statistical efficiency. $s$ is 8.} %: Hogwild and ModelAverage$\lambda$ is $1/2^{10}$ (``with RAW"), or $1/2^{13}$ (``without RAW"). 
	\label{fig_impact_of_batch_size_statistical_efficiency} 
	\vspace{1.5ex}
\end{figure}

\vspace{0.5ex}
%performance difference vs bits (application)  
\noindent
{\bf Impact of Mini Batch Size. }We examine the impact of batch size $B$ on statistical efficiency under MLWeaving. Figure~\ref{fig_impact_of_batch_size_statistical_efficiency} compares the convergence trend with different batch sizes for the datasets Gisette and KDD. We run 40 epochs and observe the training loss for each epoch. We find that a larger batch size leads to a slightly slower convergence speed. Thus, we prefer to use a small mini-batch size to train with MLWeaving. 
%	\vspace{-2ex}
%\newpage
%\vspace{-1ex}
\subsection{End-to-End Comparison: Loss vs. Time}
\label{subsec_experiment_loss_time}
In this subsection, we validate that MLWeaving outperforms its CPU rivals in terms of end-to-end performance (training loss vs. time), even though the evaluated CPU has 4 times more achievable memory bandwidth than the targeted FPGA. 
We employ two datasets Epsilon and KDD to demonstrate the comparison result, as shown in Figure~\ref{fig_loss_time_epoch}.\footnote{
We only put the results of the two datasets with the largest and smallest speedups in Figure~\ref{fig_loss_time_epoch}, 
\ifarxiv
while the results of all the other datasets are in Figure~\ref{fig_loss_time_epoch_appendix} of Appendix.
\else
while the results of all the other datasets are in Appendix of our technical report~\cite{mlweaving_tr}.
\fi
} 
We make three observations. 

First, MLWeaving uses a low-precision dataset to converge to the same training loss as ModelAverage or Hogwild, each of which works on the full-precision dataset. For example, 3-bit precision is good enough for MLWeaving to train the dataset Epsilon, indicating great potential for low-precision training. 
Second, low-precision Hogwild slightly slows down the training. Since Hogwild is bounded by cache coherence overhead, low precision, which potentially speeds up other parts of SGD, causes more cache coherence traffic among cores and then slows down the convergence speed, as shown in Figure~\ref{fig_loss_time_epsilon}. 
%First, low-precision training can converge much faster on both synchronous MLWeaving on FPGAs and asynchronous CPU counterparts.
Third, MLWeaving requires up to 25X less data movement to converge to the same loss, compared to its low-precision CPU counterpart, as illustrated in Figures~\ref{fig_loss_bytes_epsilon}. Thus, MLWeaving could provide 25X speedup if the FPGA had comparable memory bandwidth. It means that MLWeaving does significantly more energy-efficient training than its low-precision CPU counterparts. 
%MLWeaving have more advantages, in terms of memory traffic.ModelAverage is always faster than Hogwild on CPU, since it does not bring severe cache coherence overhead
We conclude that MLWeaving, with its low-precision and synchronous training on FPGAs, greatly outperforms the state-of-the-art low-precision and asynchronous first-order training on CPUs for training linear models. 
%\vspace{-1ex}
\subsection{Effect of Flexible Precision Schedule}
\label{subsec_experiment_precision_schedule}
In this subsection, we examine the effect of our flexible precision schedule (in Section~\ref{sec_per_epoch_tuning}) that increases the level of precision with a simple fixed schedule during training.
Table~\ref{table_of_per_epoch_tuning_precision} illustrates the speedup that MLWeaving with dynamic precision schedule achieves over MLWeaving with fixed precision schedule and over the fastest CPU method with fixed precision. The CPU approach is fully optimized. We make two observations. 

First, the dynamic precision schedule (i.e., ``adaptive" approach) reduces end-to-end training time on average by 1.19x, compared with the ``non-adaptive" approach that uses fixed precision under MLWeaving, indicating great potential of dynamic precision schedule.\footnote{
\ifarxiv
The exact per-epoch tuning process for the dataset Gisette is shown in Appendix. 
\else
The exact per-epoch tuning process for the dataset Gisette is shown in Appendix of our technical report~\cite{mlweaving_tr}.
\fi
 }
%This is in addition to the already provided improvements of MLWeaving-based hardware acceleration over the CPU implementations and emphasizes that the capability to dynamically change the precision is a valuable asset by itself. 
%Second, the dynamic precision approach removes the need to tune the hyperparameter (precision) instead of determining the precision in advance.
Second, the dynamic precision schedule sometimes can lead to slowdowns, for instance for Epsilon. The reason is that Epsilon needs only 40 epochs to train with a low-precision (i.e., 3-bit) dataset to converge to the targeted loss, while our dynamic precision schedule (in Section~\ref{sec_per_epoch_tuning}) uses the high precision (above 3 bits) after the first 8 epochs. The training that uses the high precision (above 3 bits) needs the same number of epochs to converge as using the 3-bit precision, while the throughput of using the 3-bit precision is much higher. Therefore, we need a sophisticated precision schedule to fully utilize MLWeaving.   
We conclude that even a simple dynamic precision schedule can significantly reduce end-to-end training time. % even with a simple precision schedule.%, which indicates a great potential for dynamic precision schedule. 
%\newline

\begin{table} %[!hbp]
	\centering
	\vspace{-1ex}
	\begin{scriptsize}
		\caption{End-to-end speedup due to flexible precision schedule.}
			\label{table_of_per_epoch_tuning_precision} 
			\vspace{-2ex}
\begin{tabular}{|l||r|r|r|r|r|}
	\hline
	\textbf{Adaptive Approach} & \textbf{TL} &  \textbf{Gisette}  &  \textbf{KDD} & \textbf{Madelon} &  \textbf{Epsilon} \\
	\hline
	%$B_{max}$ &        5 &    6 &     5 &  3 &     5         \\
	\hline
	 Vs. non-adaptive &       $1.26\times$ &   $1.5\times$ &      $0.92\times$ &      $1.35\times$ &   $0.9\times$         \\
	\hline
	Vs. fastest CPU & $10.7\times$ &   $16.5\times$ &      $3\times$ &      $6.1\times$ &   $9.9\times$         \\
	\hline
	% Vs. non-adaptive (mem) &       $1.28\times$ &   $1.57\times$ &      $1.25\times$ &      $1.39\times$ &   $0.91\times$         \\
	%\hline
	%Vs. fastest CPU (mem) & $12.4\times$ &   $25.1\times$ &      $8.1\times$ &      $5.3\times$ &   $22.7\times$         \\
	%\hline	
	%\wzk{Vs. fastest CPU method, pure} & $11.6\times$ &   $11\times$ &         $2.5\times$ & $4.5\times$ &     $11\times$         \\
	%\hline	
\end{tabular}
	\end{scriptsize}
		\vspace{2ex}
\end{table}

\section{Related Work}
\label{sec_related_work}
%\vspace{-2ex}
To our knowledge, MLWeaving is the first novel solution for data representation and hardware acceleration for ML in database engines. MLWeaving builds on previous work from multiple communities: databases, machine learning, and computer architecture.%, efficient and adaptive algorithms, 

\vspace{0.5ex}
\noindent
{\bf Bulk bitwise operations. } 
Bulk bitwise operations~\cite{variant_index_sigmod_97, bit_sliced_sigmod_01, ambit_micro17, bitweaving_sigmod13, widetable_vldb14, byteslice_sigmod16, compute_caches_hpca17, Pinatubo_dac16, bit_plane_compression_isca16, hebe_icde18, fast_bulk_cal15} have been used in a variety of applications, including 
database scans~\cite{bitweaving_sigmod13, widetable_vldb14, byteslice_sigmod16, ambit_micro17}.
Among them, MLWeaving is inspired by BitWeaving~\cite{bitweaving_sigmod13},
a transposed columnar storage layout designed for predicate evaluation on a per-column basis. BitWeaving is very efficient when answering a subfamily
of relational queries. The memory
layout used in MLWeaving is different
from BitWeaving to accommodate the access pattern
of batch SGD and the hardware implementation on an FPGA.

\vspace{0.5ex}
\noindent
{\bf FPGA-accelerated ML/DL. } 
MLWeaving builds on a growing line of research
accelerating machine learning with FPGAs~\cite{sgd_fpga_fccm17, asyn_lp_sgd_isca17, tabla_hpca16, svn_fccm09, lp_dsp_deep_fpl18, flexibilty_ispd18, dnn_fpga_micro16, snnap_hpca15, capult_micro16, configurable_dnn_isca18, brainwave_micro18, dana_vldb18, ernn_hpca_19}. Closest to MLWeaving is the work by De Sa et al.~\cite{asyn_lp_sgd_isca17} and Kara et al.~\cite{sgd_fpga_fccm17}
 using FPGAs to implement low-precision generalized
linear models, and they are asynchronous.
These previous methods achieve good performance using individual circuits for each level of precision. 
Microsoft Brainwave~\cite{brainwave_micro18} leverages FPGAs to implement low-precision programmable Neural Processing Unit for DNN inference at scale. It can support four low precision levels: 8/16-bit int and ms-fp9/8. In contrast, MLWeaving provides a single, flexible hardware design for all precision levels. We believe that MLWeaving has great potential to be used in Brainwave to enable any-precision inference.

\vspace{0.5ex}
\noindent
{\bf Low-Precision DNNs. }One specific application whose low-precision
implementation on hardware has been intensively
studied is Deep Neural Networks. 
Prior efforts~\cite{stripes_micro16, Flexpoint_NIPS2017, high_accuracy_low_precision_training_arxiv18, bit_pragmatic_micro17, bit_fusion_isca18, tartan_arxiv17, proteus_ics16, neural_cache_isca18} explore the fine-grained variation in bit-level precision for DNN inference, so their computation time is proportional to the bitwidth used.
Since the computation time dominates the overall performance for their DNN inference, the overall performance scales linearly with the bitwidth. 
Other research~\cite{SnaPEA_isca18, gist_isca18, nn_compiler_asplos18, MAERI_asplos18, vibnn_asplos18, minerva_isca16, dadiannao_micro14, low_precisin_dl_arxiv14, deep_learning_lp_icml15, Cnvlutin_isca16, tpu_isca17, eyeriss_isca16, eyeriss_ijssc17, pudianao_asplos15, shidiannao_isca15, eie_isca16, Scalpel_isca17, tpu_isca17} focuses on using a fixed-point, low-precision data representation and arithmetic to accelerate DNNs, instead of using full-precision. For example, Google's TPU~\cite{tpu_isca17} features 64K 8-bit fixed-point MACs to accelerate neural network inference. %, while MLWeaving supports the machine learning training with any precision
More information about DNNs can be found in the survey~\cite{deep_neural_network_survey_proceedingieee17}.
Compared with these efforts, MLWeaving focuses on a different workload, i.e., generalized linear models. We do not focus on the fixed quantization of the input data, but on flexible data retrieval.
As part of future work we would like to investigate whether the design in MLWeaving can also be used to reduce memory traffic for DNNs.

\vspace{0.5ex}
\noindent
{\bf Compression on DB/ML. }Previous work~\cite{dictory_encoding_sigmod_01, hyrise_vldb_10, chronons_sigmod_07, multi_byte_acm_commun_75, variant_index_sigmod_97, lightweight_compression_damon_15, bit_sliced_sigmod_01, integer_compression_damon_10, compressed_database_sigmod_rec_00, ibm_blink_icde_08, hana_ide_12, deep_compression_iclr16, compressed_page_micro13} employs compression techniques, e.g., dictionary encoding, to compress the data such that the further memory traffic can be significantly reduced at the cost of lightweight decompression overhead. Previous work~\cite{cla_vldb16, smvm_tpds13} directly performs compressed operations on sparse data representations to accelerate linear algebra. In contrast, MLWeaving exploits the low precision of dataset to accelerate machine learning training. 
\vspace{-1.5ex} 
\section{Conclusion} %and Future Work
\label{sec_Conclusion}
MLWeaving is an innovative solution for embedding machine learning in relational engines and taking advantage of modern hardware. It consists of an in-memory data storage layout that allows efficient retrieval of quantized data at \emph{any} level of precision, an efficient implementation of SGD on an FPGA, and an adaptive algorithm to learn a model using lower precision without having to determine the level of precision in advance. MLWeaving achieves linear speedup as precision level is decreased, and provides up to a 16X performance improvement compared to the state-of-the-art low-precision first-order CPU implementation. We make the MLWeaving design open-source.\footnote{Github: https://github.com/fpgasystems/MLWeaving}%MLWeaving opens a few exciting research directions. First, a variant of MLWeaving can be used to speed up other machine learning algorithms (e.g., k-means), which can benefit from low precision. Second, more sophisticated precision schedule can be explored to harvest the potential of MLWeaving. 

\vspace{0.6em}
\noindent
{\bf Future Directions and Limitations.} % in databases
The current prototype of MLWeaving has a number of limitations that will require additional work to make learning of generalized linear models possible in all cases. Some of these limitations are methodological and affect the ML algorithms used. Others are a question of exploring the design space enabled by MLWeaving in more detail, which cannot be done in this paper due to page limit. 
A first methodological limitation of MLWeaving is that, as is, it only supports dense,
numerical data. This is fine in {\em some} applications as it is not uncommon for sparse and categorical values (e.g., YouTube video IDs)
to be mapped to a dense embedding. However,
there are cases where training
directly on sparse or categorical data is still necessary.
To support the latter case, MLWeaving would need to be extended 
by potentially combining it not only with 
other lossless compression strategies but also with
ML techniques such as feature hashing~\cite{feature_hashing_arxiv09}
and weight sharing~\cite{deep_compression_iclr16}. A very interesting future research direction MLWeaving opens is the development of 
a unified data structure supporting dense, sparse, categorical,
or numerical data while providing similar level of
flexibility as MLWeaving.

Another methodological limitation of MLWeaving is that it currently
only supports generalized linear models. However,
as long as the loss
function is Lipschitz continuous over the input data and the data
access pattern is row-wise, it is likely that 
similar techniques could still be applied with some adjustments. Thus, another interesting future direction is to adapt MLWeaving to problems such as matrix
factorization~\cite{matrix_factorization_nips13} or clustering using K-Means~\cite{kmeans_pami02}.

An area where MLWeaving needs more work is
 dynamic precision scheduling. The current strategy is 
based on a simple intuition just to illustrate the
benefit provided by MLWeaving. More sophisticated dynamic
schedules will require a systematic analysis of the convergence properties with the dynamic precision changes. For example, one could use a technique similar to how AdaComm~\cite{adacomm_axiv18} dynamically
adjusts communication frequencies. In particular, we can determine the convergence upper bound with respect to the precision schedule and choose the schedule that minimizes the convergence upper bound.  We leave this direction to future work.

Limitations of MLWeaving caused by the current implementation and hardware can be solved by using different platforms or more complex designs.
For example, MLWeaving currently supports models up to 32K dimensions due to the available on-chip memory capacity. It is reasonable to expect that the capacity will increase in future FPGAs. Also, the learning rate supported by MLWeaving can only be $2^{-j}$, where $j$ is an integer, since MLWeaving uses a right-shift operator to control the learning rate. It is not clear how limiting this is in practice given the current results, but we will explore different hardware designs to provide more flexibility as part of future work. 

\vspace{0.6em}
\noindent
{\bf Acknowledgments.} Some experiments in the paper were obtained through the Intel Hardware Accelerator Research Program (HARP2) at the Paderborn Center for Parallel Computing (PC$^2$). We thank Intel for their donation of the HARP2 machine. 

%We will make the MLWeaving design open-source. 

%The optimization of conjunctive predicates is crucial to the modern database queries. %In the existing works, in order to produce the optimal evaluation order of predicates, the query optimizer needs to do the sampling which can be inaccurate. 
%State-of-the-art works perform the sampling to guess the optimal evaluation order of predicates, which can be inaccurate and incurs high cost.
%%We argue for an alternative approach to the optimization of conjunctive predicates, relying on memory-efficient storage layouts. Then, 
%Alternatively, we propose Hebe, a simplified execution scheme for the evaluation of conjunctive predicates. \NameOfModel is order-oblivious while maintaining high-performance. %The experimental result shows that \NameOfModel can significantly improve the overall performance over the state-of-the-art implementation with optimal evaluation order on TPC-H queries.

\newpage

\bibliographystyle{abbrv}
%\linespread{.2}
%	\scriptsize
\bibliography{myref}

\begin{thebibliography}{100}

\bibitem{compute_caches_hpca17}
S.~Aga, S.~Jeloka, A.~Subramaniyan, S.~Narayanasamy, D.~Blaauw, and R.~Das.
\newblock {Compute Caches}.
\newblock In {\em HPCA}, 2017.

\bibitem{SnaPEA_isca18}
V.~Akhlaghi, A.~Yazdanbakhsh, K.~Samadi, H.~Esmaeilzadeh, and R.~Gupta.
\newblock {SnaPEA: Predictive Early Activation for Reducing Computation in Deep
  Convolutional Neural Networks}.
\newblock In {\em ISCA}, 2018.

\bibitem{bit_pragmatic_micro17}
J.~Albericio, A.~Delm\'{a}s, P.~Judd, S.~Sharify, G.~O'Leary, R.~Genov, and
  A.~Moshovos.
\newblock {Bit-pragmatic Deep Neural Network Computing}.
\newblock In {\em MICRO}, 2017.

\bibitem{Cnvlutin_isca16}
J.~Albericio, P.~Judd, T.~Hetherington, T.~Aamodt, N.~E. Jerger, and
  A.~Moshovos.
\newblock {Cnvlutin: Ineffectual-neuron-free Deep Neural Network Computing}.
\newblock In {\em ISCA}, 2016.

\bibitem{lp_dsp_deep_fpl18}
A.~Boutros, S.~Yazdanshenas, and V.~Betz.
\newblock {Embracing Diversity: Enhanced DSP Blocks for Low-Precision Deep
  Learning on FPGAs}.
\newblock In {\em FPL}, 2018.

\bibitem{svn_fccm09}
S.~Cadambi, I.~Durdanovic, V.~Jakkula, M.~Sankaradass, E.~Cosatto,
  S.~Chakradhar, and H.~P. Graf.
\newblock {A Massively Parallel FPGA-Based Coprocessor for Support Vector
  Machines}.
\newblock In {\em FCCM}, 2009.

\bibitem{vibnn_asplos18}
R.~Cai, A.~Ren, N.~Liu, C.~Ding, L.~Wang, X.~Qian, M.~Pedram, and Y.~Wang.
\newblock {VIBNN: Hardware Acceleration of Bayesian Neural Networks}.
\newblock In {\em ASPLOS}, 2018.

\bibitem{capult_micro16}
A.~M. Caulfield, E.~S. Chung, A.~Putnam, H.~Angepat, J.~Fowers, M.~Haselman,
  S.~Heil, M.~Humphrey, P.~Kaur, J.~Kim, D.~Lo, T.~Massengill, K.~Ovtcharov,
  M.~Papamichael, L.~Woods, S.~Lanka, D.~Chiou, and D.~Burger.
\newblock {A Cloud-Scale Acceleration Architecture}.
\newblock In {\em MICRO}, 2016.

\bibitem{libsvm_tist_11}
C.~Chang and C.~Lin.
\newblock {LIBSVM: A Library for Support Vector Machines}.
\newblock {\em TIST}, 2(3):27:1--27:27, 2011.

\bibitem{eyeriss_isca16}
Y.~Chen, J.~Emer, and V.~Sze.
\newblock {Eyeriss: A Spatial Architecture for Energy-Efficient Dataflow for
  Convolutional Neural Networks}.
\newblock In {\em ISCA}, 2016.

\bibitem{eyeriss_ijssc17}
Y.~Chen, T.~Krishna, J.~S. Emer, and V.~Sze.
\newblock {Eyeriss: An Energy-Efficient Reconfigurable Accelerator for Deep
  Convolutional Neural Networks}.
\newblock {\em JSSC}, 2017.

\bibitem{dadiannao_micro14}
Y.~Chen, T.~Luo, S.~Liu, S.~Zhang, L.~He, J.~Wang, L.~Li, T.~Chen, Z.~Xu,
  N.~Sun, and O.~Temam.
\newblock {DaDianNao: A Machine-Learning Supercomputer}.
\newblock In {\em MICRO}, 2014.

\bibitem{dictory_encoding_sigmod_01}
Z.~Chen, J.~Gehrke, and F.~Korn.
\newblock {Query Optimization in Compressed Database Systems}.
\newblock In {\em SIGMOD}, 2001.

\bibitem{flexibilty_ispd18}
G.~R. Chiu, A.~C. Ling, D.~Capalija, A.~Bitar, and M.~S. Abdelfattah.
\newblock {Flexibility: FPGAs and CAD in Deep Learning Acceleration}.
\newblock In {\em ISPD}, 2018.

\bibitem{brainwave_micro18}
E.~Chung, J.~Fowers, K.~Ovtcharov, M.~Papamichael, A.~Caulfield, T.~Massengill,
  M.~Liu, D.~Lo, S.~Alkalay, M.~Haselman, M.~Abeydeera, L.~Adams, H.~Angepat,
  C.~Boehn, D.~Chiou, O.~Firestein, A.~Forin, K.~S. Gatlin, M.~Ghandi, S.~Heil,
  K.~Holohan, A.~E. Husseini, T.~Juhasz, K.~Kagi, R.~Kovvuri, S.~Lanka, F.~van
  Megen, D.~Mukhortov, P.~Patel, B.~Perez, A.~Rapsang, S.~Reinhardt,
  B.~Rouhani, A.~Sapek, R.~Seera, S.~Shekar, B.~Sridharan, G.~Weisz, L.~Woods,
  P.~Y. Xiao, D.~Zhang, R.~Zhao, and D.~Burger.
\newblock {Serving DNNs in Real Time at Datacenter Scale with Project
  Brainwave}.
\newblock {\em IEEE Micro}, 38(2):8--20, 2018.

\bibitem{low_precisin_dl_arxiv14}
M.~Courbariaux, Y.~Bengio, and J.~David.
\newblock {Low Precision Arithmetic For Deep Learning}.
\newblock {\em CoRR}, abs/1412.7024, 2014.

\bibitem{asyn_lp_sgd_isca17}
C.~De~Sa, M.~Feldman, C.~R{\'e}, and K.~Olukotun.
\newblock {Understanding and Optimizing Asynchronous Low-Precision Stochastic
  Gradient Descent}.
\newblock In {\em ISCA}, 2017.

\bibitem{high_accuracy_low_precision_training_arxiv18}
C.~{De Sa}, M.~{Leszczynski}, J.~{Zhang}, A.~{Marzoev}, C.~R. {Aberger},
  K.~{Olukotun}, and C.~{R{\'e}}.
\newblock {{High-Accuracy Low-Precision Training}}.
\newblock {\em ArXiv}, 2018.

\bibitem{tartan_arxiv17}
A.~Delmas, S.~Sharify, P.~Judd, and A.~Moshovos.
\newblock {Tartan: Accelerating Fully-Connected and Convolutional Layers in
  Deep Learning Networks by Exploiting Numerical Precision Variability}.
\newblock {\em CoRR}, abs/1707.09068, 2017.

\bibitem{shidiannao_isca15}
Z.~Du, R.~Fasthuber, T.~Chen, P.~Ienne, L.~Li, T.~Luo, X.~Feng, Y.~Chen, and
  O.~Temam.
\newblock {ShiDianNao: Shifting Vision Processing Closer to the Sensor}.
\newblock In {\em ISCA}, 2015.

\bibitem{neural_cache_isca18}
C.~Eckert, X.~Wang, J.~Wang, A.~Subramaniyan, R.~Iyer, D.~Sylvester, D.~Blaauw,
  and R.~Das.
\newblock {Neural Cache: Bit-Serial In-Cache Acceleration of Deep Neural
  Networks}.
\newblock In {\em ISCA}, 2018.

\bibitem{cla_vldb16}
A.~Elgohary, M.~Boehm, P.~J. Haas, F.~R. Reiss, and B.~Reinwald.
\newblock {Compressed Linear Algebra for Large-scale Machine Learning}.
\newblock {\em PVLDB}, 9(12):960--971, 2016.

\bibitem{Tarantula_isca02}
R.~Espasa, F.~Ardanaz, J.~Emer, S.~Felix, J.~Gago, R.~Gramunt, I.~Hernandez,
  T.~Juan, G.~Lowney, M.~Mattina, and A.~Seznec.
\newblock {Tarantula: a vector extension to the alpha architecture}.
\newblock In {\em ISCA}, 2002.

\bibitem{hana_ide_12}
F.~Farber, N.~May, W.~Lehner, I.~Muller, H.~Rauhe, J.~Dees, and S.~Ag.
\newblock {The SAP HANA Database: An Architecture Overview}.
\newblock In {\em IEEE Data Eng. Bull.}, 2012.

\bibitem{byteslice_sigmod16}
Z.~Feng, E.~Lo, B.~Kao, and W.~Xu.
\newblock {ByteSlice: Pushing the Envelop of Main Memory Data Processing with a
  New Storage Layout}.
\newblock In {\em SIGMOD}, 2015.

\bibitem{configurable_dnn_isca18}
J.~Fowers, K.~Ovtcharov, M.~Papamichael, T.~Massengill, M.~Liu, D.~Lo,
  S.~Alkalay, M.~Haselman, L.~Adams, M.~Ghandi, S.~Heil, P.~Patel, A.~Sapek,
  G.~Weisz, L.~Woods, S.~Lanka, S.~K. Reinhardt, A.~M. Caulfield, E.~S. Chung,
  and D.~Burger.
\newblock {A Configurable Cloud-Scale DNN Processor for Real-Time AI}.
\newblock In {\em ISCA}, 2018.

\bibitem{hyrise_vldb_10}
M.~Grund, J.~Kr\"{u}ger, H.~Plattner, A.~Zeier, P.~Cudre-Mauroux, and
  S.~Madden.
\newblock {HYRISE: A Main Memory Hybrid Storage Engine}.
\newblock {\em PVLDB}, 4(2):105--116, 2010.

\bibitem{harp2_fpl16}
P.~Gupta.
\newblock {Accelerating Datacenter Workloads}.
\newblock {\em FPL &}, 2016.

\bibitem{deep_learning_lp_icml15}
S.~Gupta, A.~Agrawal, K.~Gopalakrishnan, and P.~Narayanan.
\newblock {Deep Learning with Limited Numerical Precision}.
\newblock In {\em ICML}, 2015.

\bibitem{eie_isca16}
S.~Han, X.~Liu, H.~Mao, J.~Pu, A.~Pedram, M.~A. Horowitz, and W.~J. Dally.
\newblock {EIE: Efficient Inference Engine on Compressed Deep Neural Network}.
\newblock In {\em ISCA}, 2016.

\bibitem{deep_compression_iclr16}
S.~Han, H.~Mao, and W.~J. Dally.
\newblock {Deep Compression: Compressing Deep Neural Network with Pruning,
  Trained Quantization and Huffman Coding}.
\newblock In {\em ICLR}, 2015.

\bibitem{madlib_vldb12}
J.~M. Hellerstein, C.~R{\'e}, F.~Schoppmann, D.~Z. Wang, E.~Fratkin,
  A.~Gorajek, K.~S. Ng, C.~Welton, X.~Feng, K.~Li, and A.~Kumar.
\newblock {The MADlib Analytics Library: Or MAD Skills, the SQL}.
\newblock {\em PVLDB}, 5(12):1700--1711, 2012.

\bibitem{connection_machine_mit86}
W.~Hillis.
\newblock {\em The Connection Machine}.
\newblock MIT Press, 1986.

\bibitem{chronons_sigmod_07}
A.~L. Holloway, V.~Raman, G.~Swart, and D.~J. DeWitt.
\newblock {How to Barter Bits for Chronons: Compression and Bandwidth Trade
  Offs for Database Scans}.
\newblock In {\em SIGMOD}, 2007.

\bibitem{idreos2012monetdb}
S.~Idreos, F.~Groffen, N.~Nes, S.~Manegold, S.~Mullender, and M.~Kersten.
\newblock {MonetDB: Two Decades of Research in Column-oriented Database}.
\newblock {\em IEEE Data Engineering Bulletin}, 2012.

\bibitem{reg_fccm16}
Z.~Istvan, D.~Sidler, and G.~Alonso.
\newblock {Runtime Parameterizable Regular Expression Operators for Databases}.
\newblock In {\em FCCM}, 2016.

\bibitem{histogram_sigmod14}
Z.~Istvan, L.~Woods, and G.~Alonso.
\newblock {Histograms As a Side Effect of Data Movement for Big Data}.
\newblock In {\em SIGMOD}, 2014.

\bibitem{gist_isca18}
A.~Jain, A.~Phanishayee, J.~Mars, L.~Tang, and G.~Pekhimenko.
\newblock {Gist: Efficient Data Encoding for Deep Neural Network Training}.
\newblock In {\em ISCA}, 2018.

\bibitem{nn_compiler_asplos18}
Y.~Ji, Y.~Zhang, W.~Chen, and Y.~Xie.
\newblock {Bridge the Gap Between Neural Networks and Neuromorphic Hardware
  with a Neural Network Compiler}.
\newblock In {\em ASPLOS}, 2018.

\bibitem{bionicdb_cidr13}
R.~Johnson and I.~Pandis.
\newblock {The Bionic DBMS is Coming, but What Will It Look Like?}
\newblock In {\em CIDR}, 2013.

\bibitem{tpu_isca17}
N.~P. Jouppi, C.~Young, N.~Patil, D.~Patterson, G.~Agrawal, R.~Bajwa, S.~Bates,
  S.~Bhatia, N.~Boden, A.~Borchers, R.~Boyle, P.-l. Cantin, C.~Chao, C.~Clark,
  J.~Coriell, M.~Daley, M.~Dau, J.~Dean, B.~Gelb, T.~V. Ghaemmaghami,
  R.~Gottipati, W.~Gulland, R.~Hagmann, C.~R. Ho, D.~Hogberg, J.~Hu, R.~Hundt,
  D.~Hurt, J.~Ibarz, A.~Jaffey, A.~Jaworski, A.~Kaplan, H.~Khaitan,
  D.~Killebrew, A.~Koch, N.~Kumar, S.~Lacy, J.~Laudon, J.~Law, D.~Le, C.~Leary,
  Z.~Liu, K.~Lucke, A.~Lundin, G.~MacKean, A.~Maggiore, M.~Mahony, K.~Miller,
  R.~Nagarajan, R.~Narayanaswami, R.~Ni, K.~Nix, T.~Norrie, M.~Omernick,
  N.~Penukonda, A.~Phelps, J.~Ross, M.~Ross, A.~Salek, E.~Samadiani, C.~Severn,
  G.~Sizikov, M.~Snelham, J.~Souter, D.~Steinberg, A.~Swing, M.~Tan,
  G.~Thorson, B.~Tian, H.~Toma, E.~Tuttle, V.~Vasudevan, R.~Walter, W.~Wang,
  E.~Wilcox, and D.~H. Yoon.
\newblock {In-Datacenter Performance Analysis of a Tensor Processing Unit}.
\newblock In {\em ISCA}, 2017.

\bibitem{proteus_ics16}
P.~Judd, J.~Albericio, T.~Hetherington, T.~M. Aamodt, N.~E. Jerger, and
  A.~Moshovos.
\newblock {Proteus: Exploiting Numerical Precision Variability in Deep Neural
  Networks}.
\newblock In {\em ICS}, 2016.

\bibitem{stripes_micro16}
P.~Judd, J.~Albericio, T.~Hetherington, T.~M. Aamodt, and A.~Moshovos.
\newblock {Stripes: Bit-serial deep neural network computing}.
\newblock In {\em MICRO}, 2016.

\bibitem{kmeans_pami02}
T.~Kanungo, D.~M. Mount, N.~S. Netanyahu, C.~D. Piatko, R.~Silverman, and A.~Y.
  Wu.
\newblock {An Efficient k-Means Clustering Algorithm: Analysis and
  Implementation}.
\newblock {\em TPAMI}, 24:881--892, 2002.

\bibitem{sgd_fpga_fccm17}
K.~Kara, D.~Alistarh, G.~Alonso, O.~Mutlu, and C.~Zhang.
\newblock {FPGA-Accelerated Dense Linear Machine Learning: A
  Precision-Convergence Trade-Off}.
\newblock In {\em FCCM}, 2017.

\bibitem{columnml_vldb19}
K.~Kara, K.~Eguro, C.~Zhang, and G.~Alonso.
\newblock {ColumnML: Column-Store Machine Learning with On-the-Fly Data
  Transformation}.
\newblock {\em PVLDB}, 12(4):348--361, 2018.

\bibitem{kara2017fpga}
K.~Kara, J.~Giceva, and G.~Alonso.
\newblock {Fpga-Based Data Partitioning}.
\newblock In {\em SIGMOD}, 2017.

\bibitem{smvm_tpds13}
V.~Karakasis, T.~Gkountouvas, K.~Kourtis, G.~Goumas, and N.~Koziris.
\newblock {An Extended Compression Format for the Optimization of Sparse
  Matrix-Vector Multiplication}.
\newblock {\em TPDS}, 24(10):1930--1940, 2013.

\bibitem{bit_plane_compression_isca16}
J.~Kim, M.~Sullivan, E.~Choukse, and M.~Erez.
\newblock {Bit-Plane Compression: Transforming Data for Better Compression in
  Many-Core Architectures}.
\newblock In {\em ISCA}, 2016.

\bibitem{Flexpoint_NIPS2017}
U.~K\"{o}ster, T.~Webb, X.~Wang, M.~Nassar, A.~K. Bansal, W.~Constable,
  O.~Elibol, S.~Gray, S.~Hall, L.~Hornof, A.~Khosrowshahi, C.~Kloss, R.~J. Pai,
  and N.~Rao.
\newblock {Flexpoint: An Adaptive Numerical Format for Efficient Training of
  Deep Neural Networks}.
\newblock In {\em NIPS}. 2017.

\bibitem{vector_processor_micro02}
C.~Kozyrakis and D.~Patterson.
\newblock {Vector vs. Superscalar and VLIW architectures For Embedded
  Multimedia Benchmarks}.
\newblock In {\em MICRO}, 2002.

\bibitem{vector_magize_ieee_micro03}
C.~E. Kozyrakis and D.~A. Patterson.
\newblock {Scalable, Vector Processors For Embedded Systems}.
\newblock {\em IEEE Micro}, 2003.

\bibitem{imagenet_nips12}
A.~Krizhevsky, I.~Sutskever, and G.~Hinton.
\newblock {ImageNet Classification with Deep Convolutional Neural Networks}.
\newblock In {\em NIPS}. 2012.

\bibitem{MAERI_asplos18}
H.~Kwon, A.~Samajdar, and T.~Krishna.
\newblock {MAERI: Enabling Flexible Dataflow Mapping over DNN Accelerators via
  Reconfigurable Interconnects}.
\newblock In {\em ASPLOS}, 2018.

\bibitem{multi_byte_acm_commun_75}
L.~Lamport.
\newblock {Multiple Byte Processing with Full-word Instructions}.
\newblock {\em Commun. ACM}, 18(8):471--475, 1975.

\bibitem{matrix_factorization_nips13}
D.~Lee and S.~Sebastian.
\newblock {Algorithms for Non-negative Matrix Factorization}.
\newblock In {\em NIPS}. 2001.

\bibitem{mini_batch_sgd_kdd14}
M.~Li, T.~Zhang, Y.~Chen, and A.~J. Smola.
\newblock {Efficient Mini-batch Training for Stochastic Optimization}.
\newblock In {\em SIGKDD}, 2014.

\bibitem{Pinatubo_dac16}
S.~Li, C.~Xu, Q.~Zou, J.~Zhao, Y.~Lu, and Y.~Xie.
\newblock {Pinatubo: A Processing-in-Memory Architecture for Bulk Bitwise
  Operations in Emerging Non-Volatile Memories}.
\newblock In {\em DAC}, 2016.

\bibitem{bitweaving_sigmod13}
Y.~Li and J.~M. Patel.
\newblock {BitWeaving: Fast Scans for Main Memory Data Processing}.
\newblock In {\em SIGMOD}, 2013.

\bibitem{widetable_vldb14}
Y.~Li and J.~M. Patel.
\newblock {WideTable: An Accelerator for Analytical Data Processing}.
\newblock {\em PVLDB}, 7(10):907--918, 2014.

\bibitem{ernn_hpca_19}
Z.~Li, C.~Ding, S.~Wang, W.~Wen, Y.~Zhuo, C.~Liu, Q.~Qiu, W.~Xu, X.~Lin,
  X.~Qian, and Y.~Wang.
\newblock {E-RNN: Design Optimization for Efficient Recurrent Neural Networksin
  FPGAs}.
\newblock In {\em HPCA}, 2019.

\bibitem{pudianao_asplos15}
D.~Liu, T.~Chen, S.~Liu, J.~Zhou, S.~Zhou, O.~Teman, X.~Feng, X.~Zhou, and
  Y.~Chen.
\newblock {PuDianNao: A Polyvalent Machine Learning Accelerator}.
\newblock In {\em ASPLOS}, 2015.

\bibitem{mlbench_vldb18}
Y.~Liu, H.~Zhang, L.~Zeng, W.~Wu, and C.~Zhang.
\newblock {MLbench: Benchmarking Machine Learning Services Against Human
  Experts}.
\newblock {\em PVLDB}, 11(10):1220--1232, 2018.

\bibitem{dana_vldb18}
D.~Mahajan, J.~K. Kim, J.~Sacks, A.~Ardalan, A.~Kumar, and H.~Esmaeilzadeh.
\newblock {In-RDBMS Hardware Acceleration of Advanced Analytics}.
\newblock {\em PVLDB}, 11(11):1317--1331, 2018.

\bibitem{tabla_hpca16}
D.~Mahajan, J.~Park, E.~Amaro, H.~Sharma, A.~Yazdanbakhsh, J.~K. Kim, and
  H.~Esmaeilzadeh.
\newblock {TABLA: A Unified Template-based Framework For Accelerating
  Statistical Machine Learning}.
\newblock In {\em HPCA}, 2016.

\bibitem{convnets_vlsic16}
B.~Moons and M.~Verhelst.
\newblock {A 0.3 x2013;2.6 TOPS/W precision-scalable processor for real-time
  large-scale ConvNets}.
\newblock In {\em VLSI-Circuits}, 2016.

\bibitem{snnap_hpca15}
T.~Moreau, M.~Wyse, J.~Nelson, A.~Sampson, H.~Esmaeilzadeh, L.~Ceze, and
  M.~Oskin.
\newblock {SNNAP: Approximate Computing on Programmable SoCs via Neural
  Acceleration}.
\newblock In {\em HPCA}, 2015.

\bibitem{data_processing_vldb09}
R.~Mueller, J.~Teubner, and G.~Alonso.
\newblock {Data Processing on FPGAs}.
\newblock {\em PVLDB}, 2(1):910--921, 2009.

\bibitem{streams_vldb09}
R.~Mueller, J.~Teubner, and G.~Alonso.
\newblock {Streams on Wires: A Query Compiler for FPGAs}.
\newblock {\em PVLDB}, 2(1):229--240, 2009.

\bibitem{Glacier_sigmod10}
R.~Mueller, J.~Teubner, and G.~Alonso.
\newblock {Glacier: A Query-to-hardware Compiler}.
\newblock In {\em SIGMOD}, 2010.

\bibitem{hogwild_NIPS11}
F.~Niu, B.~Recht, C.~Re, and S.~Wright.
\newblock {Hogwild: A Lock-Free Approach to Parallelizing Stochastic Gradient
  Descent}.
\newblock In {\em NIPS}. 2011.

\bibitem{variant_index_sigmod_97}
P.~O'Neil and D.~Quass.
\newblock {Improved Query Performance with Variant Indexes}.
\newblock In {\em SIGMOD}, pages 38--49, 1997.

\bibitem{centaur_fccm17}
M.~Owaida, D.~Sidler, K.~Kara, and G.~Alonso.
\newblock {Centaur: A Framework for Hybrid CPU-FPGA Databases}.
\newblock In {\em FCCM}, 2017.

\bibitem{compressed_page_micro13}
G.~{Pekhimnko}, V.~{Seshadri}, Y.~{Kim}, H.~{Xin}, O.~{Mutlu}, P.~B. {Gibbons},
  M.~A. {Kozuch}, and T.~C. {Mowry}.
\newblock {Linearly Compressed Pages: A Low-Complexity, Low-Latency Main Memory
  Compression Framework}.
\newblock In {\em MICRO}, 2013.

\bibitem{lightweight_compression_damon_15}
O.~Polychroniou and K.~A. Ross.
\newblock {Efficient Lightweight Compression Alongside Fast Scans}.
\newblock In {\em DaMoN}, 2015.

\bibitem{ibm_blink_icde_08}
V.~Raman, G.~Swart, L.~Qiao, F.~Reiss, V.~Dialani, D.~Kossmann, I.~Narang, and
  R.~Sidle.
\newblock {Constant-Time Query Processing}.
\newblock In {\em ICDE}, 2008.

\bibitem{minerva_isca16}
B.~Reagen, P.~Whatmough, R.~Adolf, S.~Rama, H.~Lee, S.~K. Lee, J.~M.
  Hern\'{a}ndez-Lobato, G.-Y. Wei, and D.~Brooks.
\newblock {Minerva: Enabling Low-power, Highly-accurate Deep Neural Network
  Accelerators}.
\newblock In {\em ISCA}, 2016.

\bibitem{one_vs_all_jmlr04}
R.~Rifkin and A.~Klautau.
\newblock { In Defense of One-Vs-All Classification}.
\newblock In {\em JMLR}. 2004.

\bibitem{bit_sliced_sigmod_01}
D.~Rinfret, P.~O'Neil, and E.~O'Neil.
\newblock {Bit-Sliced Index Arithmetic}.
\newblock In {\em SIGMOD}, 2001.

\bibitem{chaining_cray_78}
R.~M. Russell.
\newblock {The CRAY-1 Computer System}.
\newblock {\em Commun. ACM}, 21(1):63--72, 1978.

\bibitem{integer_compression_damon_10}
B.~Schlegel, R.~Gemulla, and W.~Lehner.
\newblock {Fast Integer Compression Using SIMD Instructions}.
\newblock In {\em DaMoN}, 2010.

\bibitem{fast_bulk_cal15}
V.~{Seshadri}, K.~{Hsieh}, A.~{Boroum}, D.~{Lee}, M.~A. {Kozuch}, O.~{Mutlu},
  P.~B. {Gibbons}, and T.~C. {Mowry}.
\newblock {Fast Bulk Bitwise AND and OR in DRAM}.
\newblock {\em IEEE CAL}, 2015.

\bibitem{ambit_micro17}
V.~Seshadri, D.~Lee, T.~Mullins, H.~Hassan, A.~Boroumand, J.~Kim, M.~A. Kozuch,
  O.~Mutlu, P.~B. Gibbons, and T.~C. Mowry.
\newblock {Ambit: In-memory Accelerator for Bulk Bitwise Operations Using
  Commodity DRAM Technology}.
\newblock In {\em MICRO}, 2017.

\bibitem{dnn_fpga_micro16}
H.~Sharma, J.~Park, D.~Mahajan, E.~Amaro, J.~K. Kim, C.~Shao, A.~Mishra, and
  H.~Esmaeilzadeh.
\newblock {From High-Level Deep Neural Models to FPGAs}.
\newblock In {\em MICRO}, 2016.

\bibitem{bit_fusion_isca18}
H.~Sharma, J.~Park, N.~Suda, L.~Lai, B.~Chau, J.~Kim, V.~Chandra, and
  H.~Esmaeilzadeh.
\newblock {Bit Fusion: Bit-Level Dynamically Composable Architecture for
  Accelerating Deep Neural Networks}.
\newblock In {\em ISCA}, 2018.

\bibitem{pattern_matching_sigmod17}
D.~Sidler, Z.~Istv\'{a}n, M.~Owaida, and G.~Alonso.
\newblock {Accelerating Pattern Matching Queries in Hybrid CPU-FPGA
  Architectures}.
\newblock In {\em SIGMOD}, 2017.

\bibitem{sidler2017doppiodb}
D.~Sidler, Z.~Istv{\'a}n, M.~Owaida, K.~Kara, and G.~Alonso.
\newblock {doppioDB: A Hardware Accelerated Database}.
\newblock In {\em SIGMOD}, 2017.

\bibitem{filter_adaptive_precision_asicsoc99}
A.~Sinha and A.~P. Chandrakasan.
\newblock {Energy Efficient Filtering Using Adaptive Precision and Variable
  Voltage}.
\newblock In {\em IEEE International ASIC/SOC Conference}, 1999.

\bibitem{deep_neural_network_survey_proceedingieee17}
V.~Sze, Y.~H. Chen, T.~J. Yang, and J.~S. Emer.
\newblock {Efficient Processing of Deep Neural Networks: A Tutorial and
  Survey}.
\newblock {\em Proceedings of the IEEE}, 2017.

\bibitem{inception_v3_cvpr16}
C.~Szegedy, V.~Vanhoucke, S.~Ioffe, J.~Shlens, and Z.~Wojna.
\newblock { Rethinking the Inception Architecture For Computer Vision}.
\newblock In {\em CVPR}. 2016.

\bibitem{fpga_book_mc13}
J.~Teubner and L.~Woods.
\newblock {\em {Data Processing on FPGAs Synthesis Lectures on Data
  Management}}.
\newblock 2013.

\bibitem{umuroglu2017finn}
Y.~Umuroglu, N.~J. Fraser, G.~Gambardella, M.~Blott, P.~Leong, M.~Jahre, and
  K.~Vissers.
\newblock {Finn: A framework For Fast, Scalable Binarized Neural Network
  Inference}.
\newblock In {\em FPGA}, 2017.

\bibitem{bismo_fpl18}
Y.~Umuroglu, L.~Rasnayake, and M.~Själander.
\newblock {BISMO: A Scalable Bit-Serial Matrix Multiplication Overlay for
  Reconfigurable Computing}.
\newblock In {\em FPL}, 2018.

\bibitem{adacomm_axiv18}
J.~Wang and G.~Joshi.
\newblock {Adaptive Communication Strategies to Achieve the Best Error-Runtime
  Trade-off in Local-Update SGD}.
\newblock {\em CoRR}, abs/1810.08313, 2018.

\bibitem{partitioning_opencl_fpl15}
Z.~Wang, B.~He, and W.~Zhang.
\newblock {A Study of Data Partitioning on OpenCL-based FPGAs}.
\newblock In {\em FPL}, 2015.

\bibitem{query_opencl_fpga_fpl16}
Z.~{Wang}, J.~{Paul}, H.~Y. {Cheah}, B.~{He}, and W.~{Zhang}.
\newblock {Relational Query Processing on OpenCL-based FPGAs}.
\newblock In {\em FPL}, 2016.

\bibitem{multi_kernel_tvlsi17}
Z.~{Wang}, J.~{Paul}, B.~{He}, and W.~{Zhang}.
\newblock {Multikernel Data Partitioning With Channel on OpenCL-Based FPGAs}.
\newblock {\em TVLSI}, 25(6):1906--1918, 2017.

\bibitem{hebe_icde18}
Z.~{Wang}, K.~{Zhang}, H.~{Zhou}, X.~{Liu}, and B.~{He}.
\newblock {Hebe: An Order-Oblivious and High-Performance Execution Scheme for
  Conjunctive Predicates}.
\newblock In {\em ICDE}, 2018.

\bibitem{feature_hashing_arxiv09}
K.~Q. Weinberger, A.~Dasgupta, J.~Attenberg, J.~Langford, and A.~J. Smola.
\newblock {Feature Hashing for Large Scale Multitask Learning}.
\newblock {\em CoRR}, 2009.

\bibitem{compressed_database_sigmod_rec_00}
T.~Westmann, D.~Kossmann, S.~Helmer, and G.~Moerkotte.
\newblock {The Implementation and Performance of Compressed Databases}.
\newblock {\em SIGMOD Rec.}, 29(3):55--67, 2000.

\bibitem{bit_serial_assp89}
S.~White.
\newblock {Applications of Distributed Arithmetic to Digital Signal Processing:
  A Tutorial Review}.
\newblock {\em IEEE ASSP Magazine}, 6(3):4--19, 1989.

\bibitem{Intel_PCM}
T.~Willhalm, R.~Dementiev, and P.~Fay.
\newblock {Intel Performance Counter Monitor - A better way to measure CPU
  utilization},
  https://software.intel.com/en-us/articles/intel-performance-counter-monitor,2016.

\bibitem{Ibex_vldb14}
L.~Woods, Z.~Istv\'{a}n, and G.~Alonso.
\newblock {Ibex: An Intelligent Storage Engine with Support for Advanced SQL
  Offloading}.
\newblock {\em VLDB}, 7(11):963--974, 2014.

\bibitem{dct_vlsi99}
T.~Xanthopoulos and A.~Chandrakasan.
\newblock {A Low-power DCT Core Using Adaptive Bitwidth and Arithmetic Activity
  Exploiting Signal Correlations and Quantization}.
\newblock In {\em Symposium on VLSI Circuits}, 1999.

\bibitem{Scalpel_isca17}
J.~Yu, A.~Lukefahr, D.~Palframan, G.~Dasika, R.~Das, and S.~Mahlke.
\newblock {Scalpel: Customizing DNN Pruning to the Underlying Hardware
  Parallelism}.
\newblock In {\em ISCA}, 2017.

\bibitem{dimmwitted_vldb14}
C.~Zhang and C.~R{\'e}.
\newblock {DimmWitted: A Study of Main-memory Statistical Analytics}.
\newblock {\em PVLDB}, 7(12):1283--1294, 2014.

\bibitem{zipml_icml17}
H.~Zhang, J.~Li, K.~Kara, D.~Alistarh, J.~Liu, and C.~Zhang.
\newblock {{Z}ip{ML}: Training Linear Models with End-to-End Low Precision, and
  a Little Bit of Deep Learning}.
\newblock In {\em ICML}, volume~70, pages 4035--4043, 2017.

\bibitem{modelAveraging_NIPS2010}
M.~Zinkevich, M.~Weimer, L.~Li, and A.~J. Smola.
\newblock {Parallelized Stochastic Gradient Descent}.
\newblock In {\em NIPS}. 2010.

\end{thebibliography}

\ifarxiv
\newpage
\section{Appendix}

\subsection{Cost Model}
In this section, we propose a hybrid analytical/empirical cost model to predict the performance of MLWeaving on FPGAs. We predict the throughput $Th$ to be the minimum of the computing throughput $Th_{comp}$ and memory throughput $TH_{mem}$, as shown in Equation~\ref{E_throughput_overall}.
\begin{equation} \begin{scriptsize}
\label{E_throughput_overall}
%\vspace{-1ex}
Th = Min (Th_{comp},Th_{mem})
%\vspace{-1ex}
\end{scriptsize} \end{equation}

\subsubsection{Evaluating Computing Throughput}
In this subsection, we present an analytical model to evaluate the computing throughput for ``chaining" and ``no chaining", under the assumption that the memory subsystem can provide the data as soon as the computing logic requires. The MLWeaving hardware design can consume 512 bits per cycle, so its theoretical computing throughput is 512 bits*400MHz=25.6GB/s. However, the practical computing throughput is lower, since MLWeaving has to guarantee the RAW dependency in the SGD model. One unavoidable factor is the pipeline latency $L$, which is the latency between the dot product module and the model update module. The value of $L$ is $40+2s$ cycles, where $s$ is the precision level. 

%\noindent
{\bf ``Chaining". }We evaluate the computing throughput of ``chaining", as shown in
Equation~\ref{E_comp_with_chaining}.
\begin{equation} \begin{scriptsize}
\label{E_comp_with_chaining}
%\vspace{-1ex}
Th_{comp} = \cfrac{(B/8) \times \lceil M/64\rceil \times s}{(B/8) \times \lceil M/64\rceil \times s + L} \times 25.6GB/s, 
\vspace{-1ex}
\end{scriptsize} \end{equation}
where the right part is the theoretical computing throughput and the left part is the utilization of the computing logic (i.e., the dot product module). Since chaining is enabled, the overhead of updating the model can be removed. However, the pipeline latency $L$ cannot be ignored, especially when $M$ and $s$ are small. 

%\noindent
{\bf ``No chaining". }We evaluate the computing throughput of ``no chaining", as shown in
Equation~\ref{E_comp_wo_chaining}.
\begin{equation} \begin{scriptsize}
\label{E_comp_wo_chaining}
%\vspace{-1ex}
Th_{comp} = \cfrac{(B/8) \times \lceil M/64\rceil \times s}{(1+B/8) \times \lceil M/64\rceil \times s + L} \times 25.6GB/s, 
%\vspace{-1ex}
\end{scriptsize} \end{equation}
where the overhead of updating the model is $\lceil M/64\rceil \times s$. The main difference from ``chaining" is that the dot product module is idle when the model update module is active. 

\subsubsection{Evaluating Memory Throughput}
MLWeaving accesses the host memory via two PCIe and one QPI links, so the memory throughput is affected by both the memory subsystem and the PCIe/QPI links. This makes the explicit prediction (in terms of the analytical cost model) of memory throughput extremely difficult, especially when the memory-related IP core on the FPGA side is encrypted. Therefore, we present an empirical cost model to evaluate the memory throughput $Th_{mem}$. 
In order to predict the memory throughput of MLWeaving, we analyze the memory access pattern of MLWeaving and measure the memory throughput. % according to the memory access pattern 
Note, both ``chaining" and ``no chaining" use the same memory subsystem. 

{\bf Memory Access Pattern. }The memory access pattern of MLWeaving is quite fixed. Essentially, for the precision level $s$, MLWeaving sequentially fetches $s$ cache lines (512 bits each) every 32 cache lines, regardless of the number of features. We conclude that MLWeaving's memory bandwidth is quite fixed. %Thanks to the fixed memory access pattern, it is quite easy to benchmark the memory bandwidth 

{\bf Benchmarking Memory Throughput. }Based on the memory access pattern, we can benchmark the memory throughput for each precision level $s$ under our framework Centaur~\cite{centaur_fccm17}. We can get the empirical memory throughput, as illustrated in Equation~\ref{E_mem_bandwidth}. We can observe that when $s$ is small (e.g., $< 4$), the memory throughput becomes noticeably lower due to its low utilization of row buffer contents. When $s$ is larger than 4, the throughput of the PCIe/QPI links becomes the new bottleneck, so the achievable throughput stays constant.
\begin{equation} \begin{scriptsize}
\label{E_mem_bandwidth}
%\vspace{-1ex}
Th_{mem} = \begin{cases}
10.2GB/s &s = 1\\
13.3GB/s &s=2\\
13.8GB/s  & s=3 \\
14.8GB/s  & s\ge  4 \\
\end{cases}
%\vspace{-1ex}
\end{scriptsize} \end{equation}

\subsubsection{Predicting MLWeaving Performance via the Cost Model}
In this subsection, we leverage our cost model to predict the speedup of “chaining” over “no chaining” to demonstrate the effect of chaining. Actually, the intuition behind this speedup is not straightforward. In order to understand this speedup, we leverage our cost model to predict the throughput of “chaining” and “no chaining” individually.  Then, we can calculate the speedup to be the throughput (Th) of “chaining” divided by the throughput (Th) of “no chaining” with different numbers of features ($M$) and different precision level ($s$). Figure~\ref{fig_cost_model} shows the different peak speed up for different datasets. ``Actual" indicates the real speedup measured via our experiments, while ``predicted" shows the estimated speedup using our cost model. We make two observations. 

First, our cost model can roughly predict the speedup for different precision levels. Second, our cost model can predict the peak speedup for different numbers of features ($M$) and different precision levels ($s$). In particular, the peak speedup is for $s$=4 for Epsilon (2K features), while the peak is at $s$=2 for Gisette (5k features). We conclude that our cost model can guide us to find the peak speedup with different precision levels and different numbers of features.  

\begin{figure}
	\centering
	%\hfill
	\subfloat[Epsilon (2000 features)]{\includegraphics[width=2.2in]{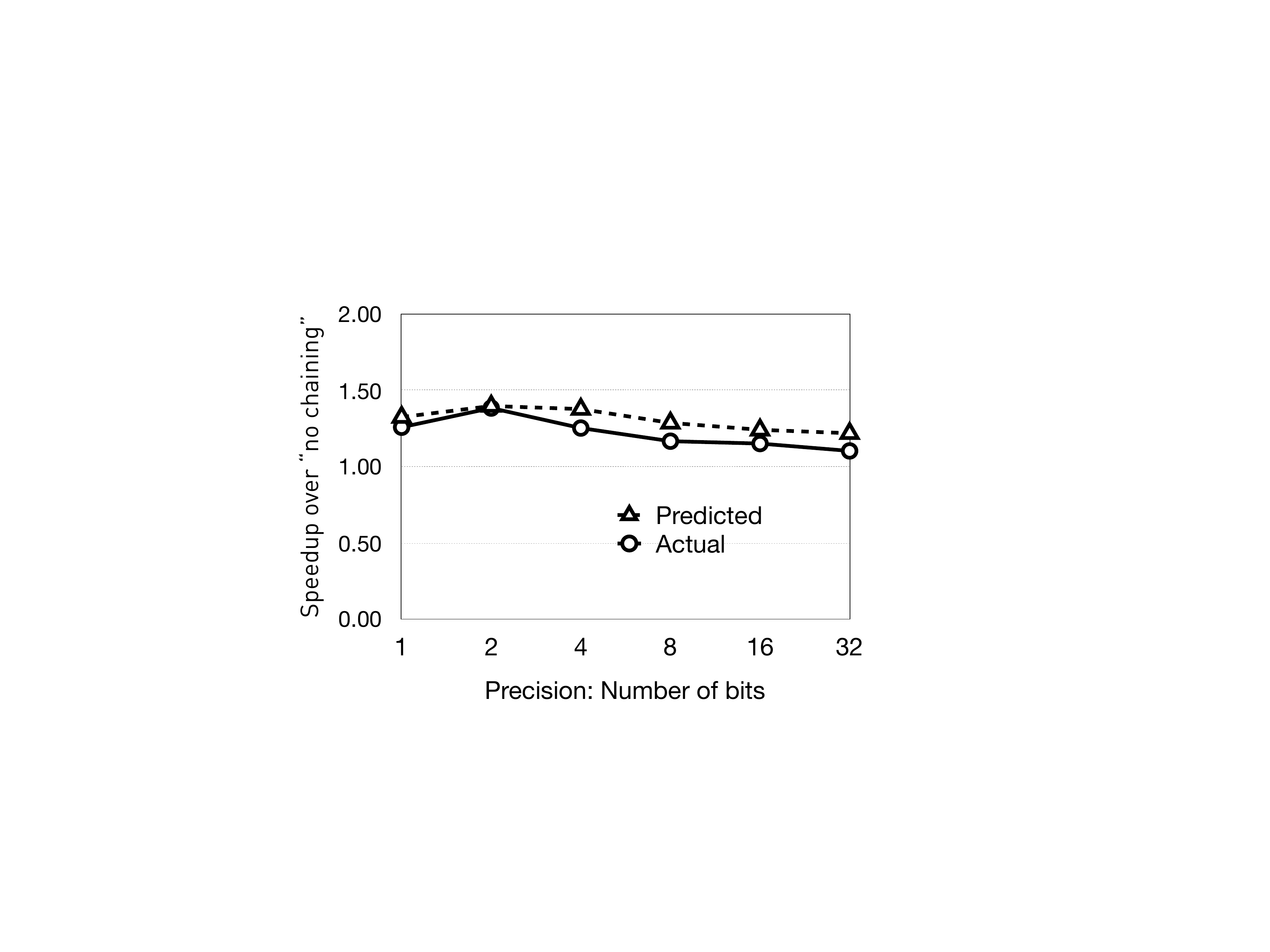} 
		\label{fig_mlweaving_on_cpu_modelaverage}} % \caption{}
		\hfill
	\subfloat[Gisette (5000 features)]{\includegraphics[width=2.2in]{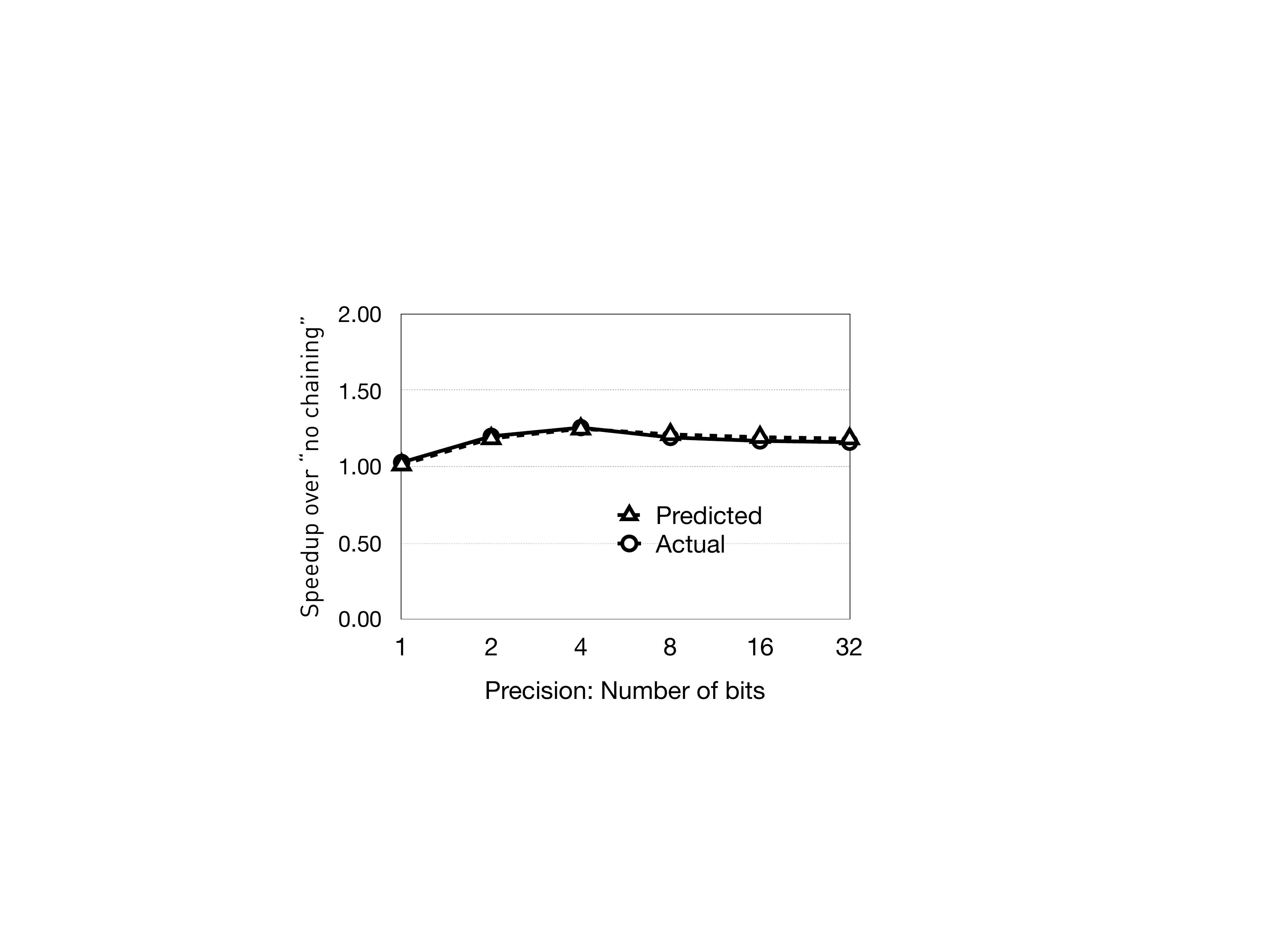} 
		\label{fig_mlweaving_on_cpu_hogwild}} 
%	\vspace{-1.5ex}	
	\caption{Speedup of ``chaining" over ``no chaining.} %: Hogwild and ModelAverage$\lambda$ is $1/2^{10}$ (``with RAW"), or $1/2^{13}$ (``without RAW"). 
	\label{fig_cost_model} 
%	\vspace{-3ex}
\end{figure}

\subsection{MLWeaving on Modern CPUs}
We examine the performance of MLWeaving on CPUs. Since modern CPUs do not yet have any custom instructions to efficiently consume the bit stream from MLWeaving layout, we have to employ regular instructions to retrieve each data element for further computing. For instance, an 8-bit element contains bits from eight different memory locations, leading to a very significant memory lookup overhead. We make our best effort that our implementation is able to retrieve 32 elements (i.e., 32-way parallelism) in a single AVX2 instruction. Figure~\ref{fig_mlweaving_on_cpu} illustrates the performance of MLWeaving on CPUs. We make three observations. 

First, MLWeaving can converge faster than the floating-point implementation under ModelAverage, when the precision level is less than 8 bits, as shown in Figure~\ref{fig_mlweaving_on_cpu_modelaverage}. This means that MLWeaving can also bring reasonable performance benefits on CPUs. Second, MLWeaving makes the low-precision implementation slower under ModelAverage. In particular, ``ModelAverage-MLWeaving-8-bit" is much slower than ``ModelAverage-char", even though both have the same precision of 8 bits. It means that the memory lookup overhead of MLWeaving cannot be fully amortized when deploying it on CPUs. 
Third, the low-precision implementation makes HogWild slower, as HogWild is always bounded by the cache coherence overhead. This means that the low-precision implementation, which can generate more shared model update operations in a fixed period, incurs more pressure to cache coherence module inside a CPU. 
\begin{figure}
	\centering
	%\hfill
	\subfloat[ModelAverage, $\lambda = 1/2^8, \beta = 0.98$]{\includegraphics[width=2.8in]{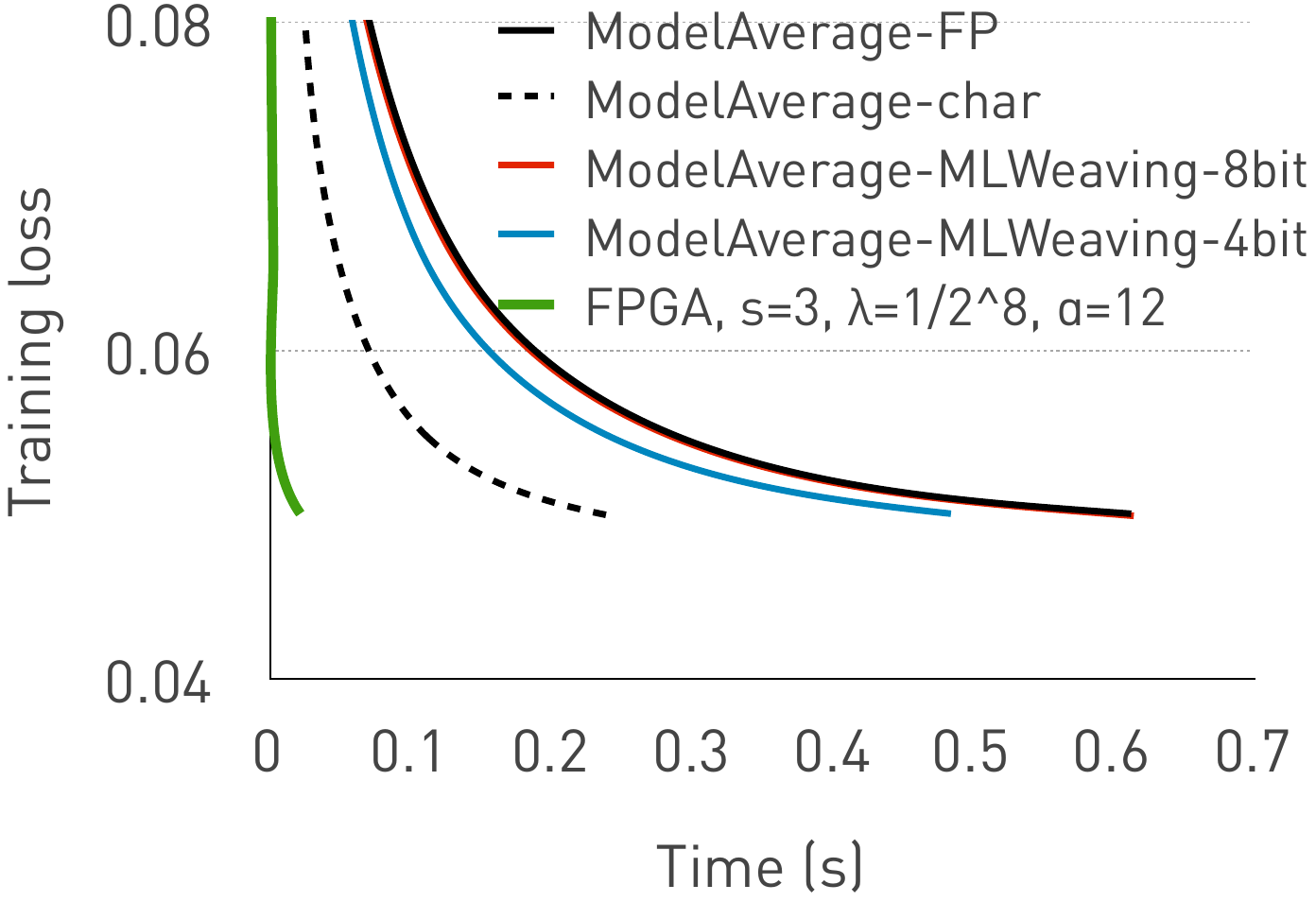} 
		\label{fig_mlweaving_on_cpu_modelaverage}} % \caption{}
		\hfill
	\subfloat[Hogwild, $\lambda = 1/2^{11}, 
	\alpha = 12$]{\includegraphics[width=2.8in]{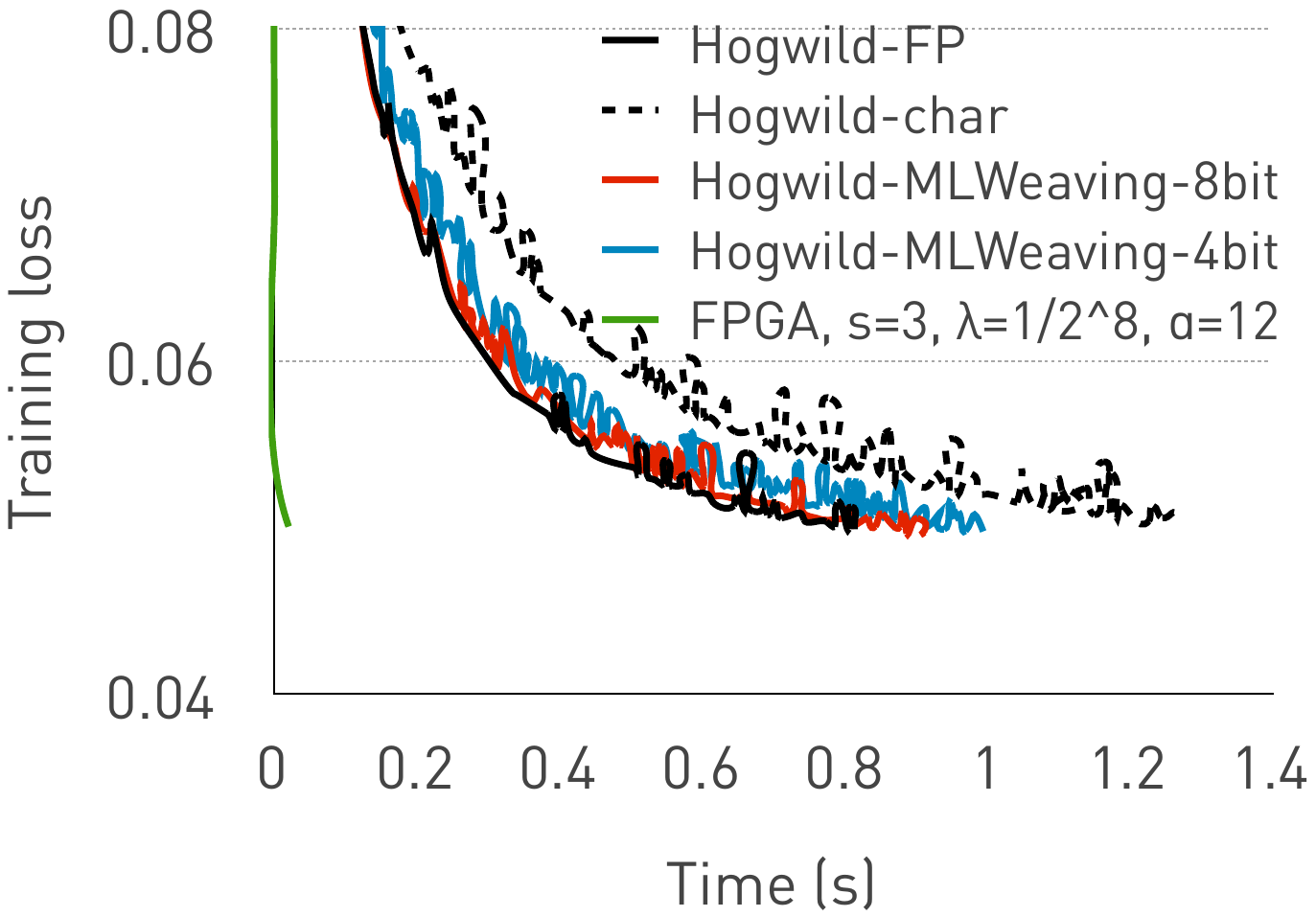} 
		\label{fig_mlweaving_on_cpu_hogwild}} 
%	\vspace{-1.5ex}	
	\caption{MLWeaving on CPU with the dataset of Epsilon.} %: Hogwild and ModelAverage$\lambda$ is $1/2^{10}$ (``with RAW"), or $1/2^{13}$ (``without RAW"). 
	\label{fig_mlweaving_on_cpu} 
%	\vspace{-3ex}
\end{figure}

\subsection{Flexible Precision Schedule}
Figure~\ref{fig_adaptive_schedule_gisette} shows the per-epoch tuning process for the Gisette dataset. 
The x-axis is the elapsed time and the y-axis is the training loss. 
The baseline here is the 4-bit precision, denoted by ``Non-adaptive, 4-bit", which can converge to the same training loss as full precision does. 
The proposed adaptive approach increases the level of precision during the training. In particular, it starts with 2-bit precision (denoted by ``2-bit") for the first four epochs, followed by 3-bit precision (denoted by ``3-bit") for the next four epochs. Then, 4-bit precision is used for the next eight epochs. Finally, ten epochs are employed under 5-bit precision to reach the target loss. We make two observations. 
%Now we experimentally show the impact of per-epoch precision tuning. The 

First, the proposed flexible precision schedule can reach the same training loss, while achieving 1.5X performance improvement over the baseline (``Non-adaptive, 4-bit"), even though the schedule is preliminary, indicating a great potential for using per-epoch precision schedules.  
Second, each precision level transition (e.g., 2-bit to 3-bit) brings a significant reduction in loss, coinciding with our theory that tuning the precision level higher (e.g., 2-bit to 3-bit) at runtime can converge to the same loss as a fixed $3$-bit precision for all the epochs.

\begin{figure}[t]
	\centering
	\includegraphics[width=8.0cm]{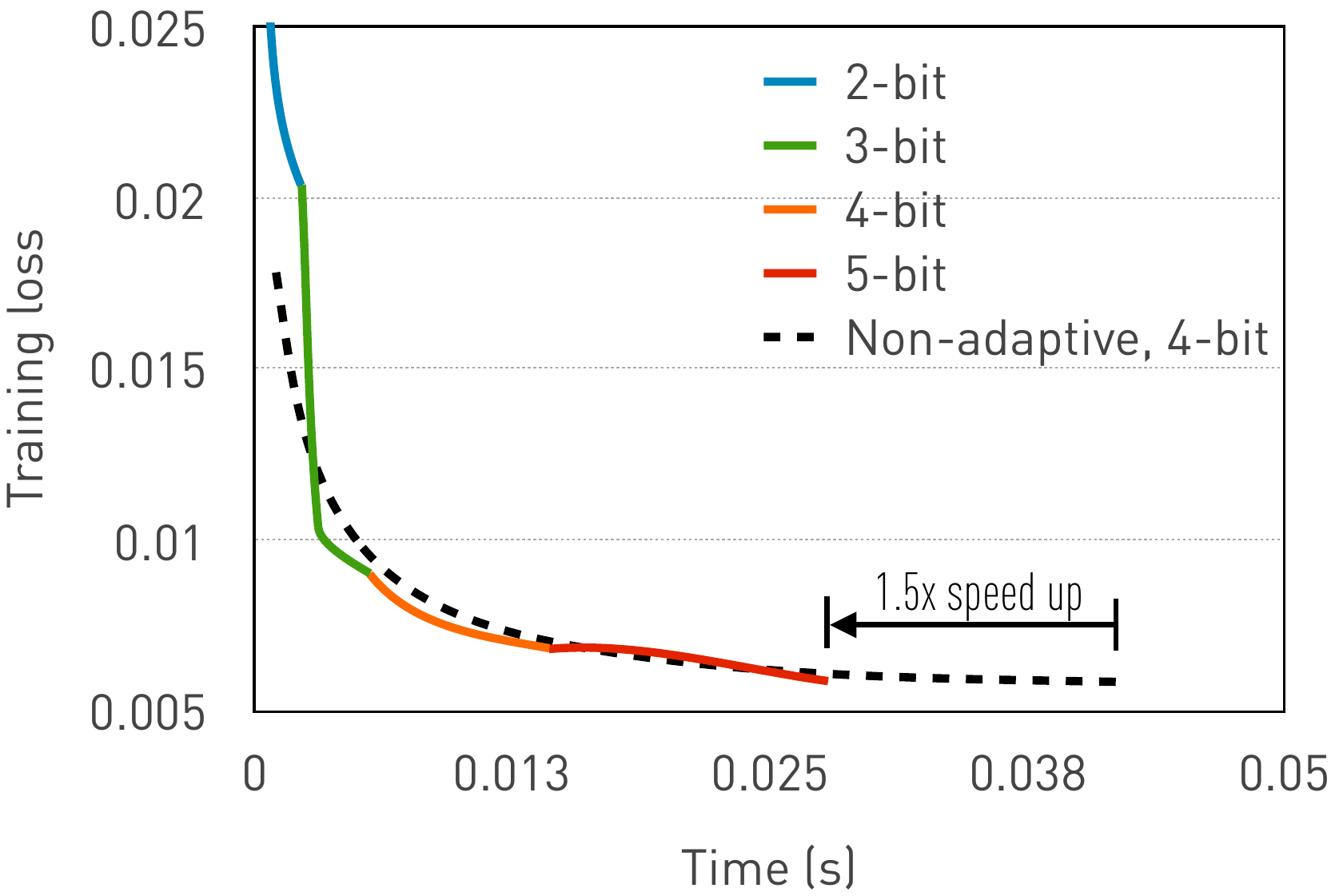}
	\caption{Effect of flexible precision schedule.  }
	\label{fig_adaptive_schedule_gisette}
\end{figure}

\subsection{Convergence Analysis}
We add the experimental results for the remaining three datasets, as shown in Figure~\ref{fig_loss_time_epoch_appendix}. 
%Direct connection + analytical model. 
\begin{figure*}[t!]
	\centering
	\subfloat[Gisette (loss vs. epoch)]{\includegraphics[width=2.05in]{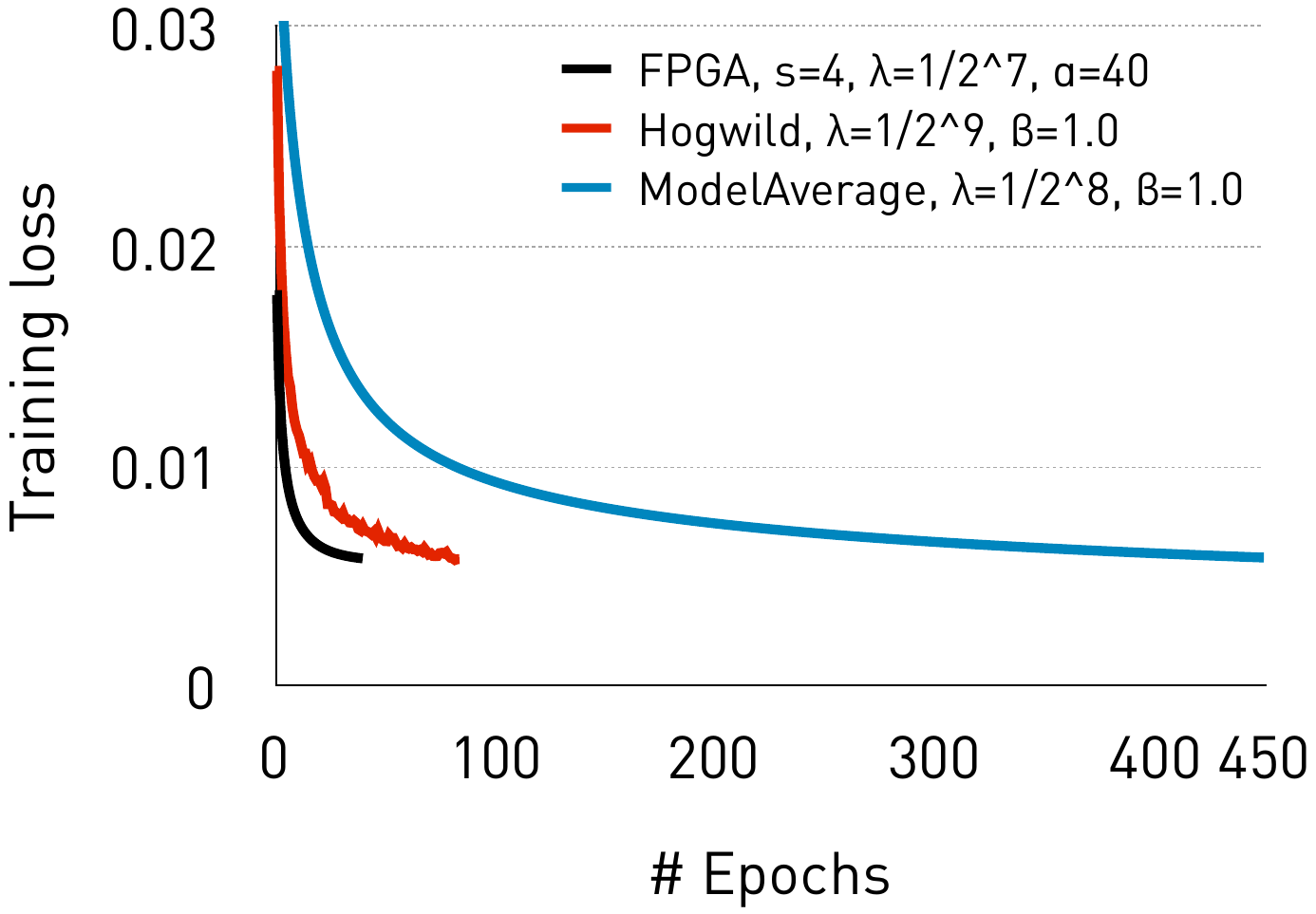} 
		\label{fig_loss_epoch_gisette}} 
	\subfloat[Gisette (loss vs. time)]{\includegraphics[width=2.25in]{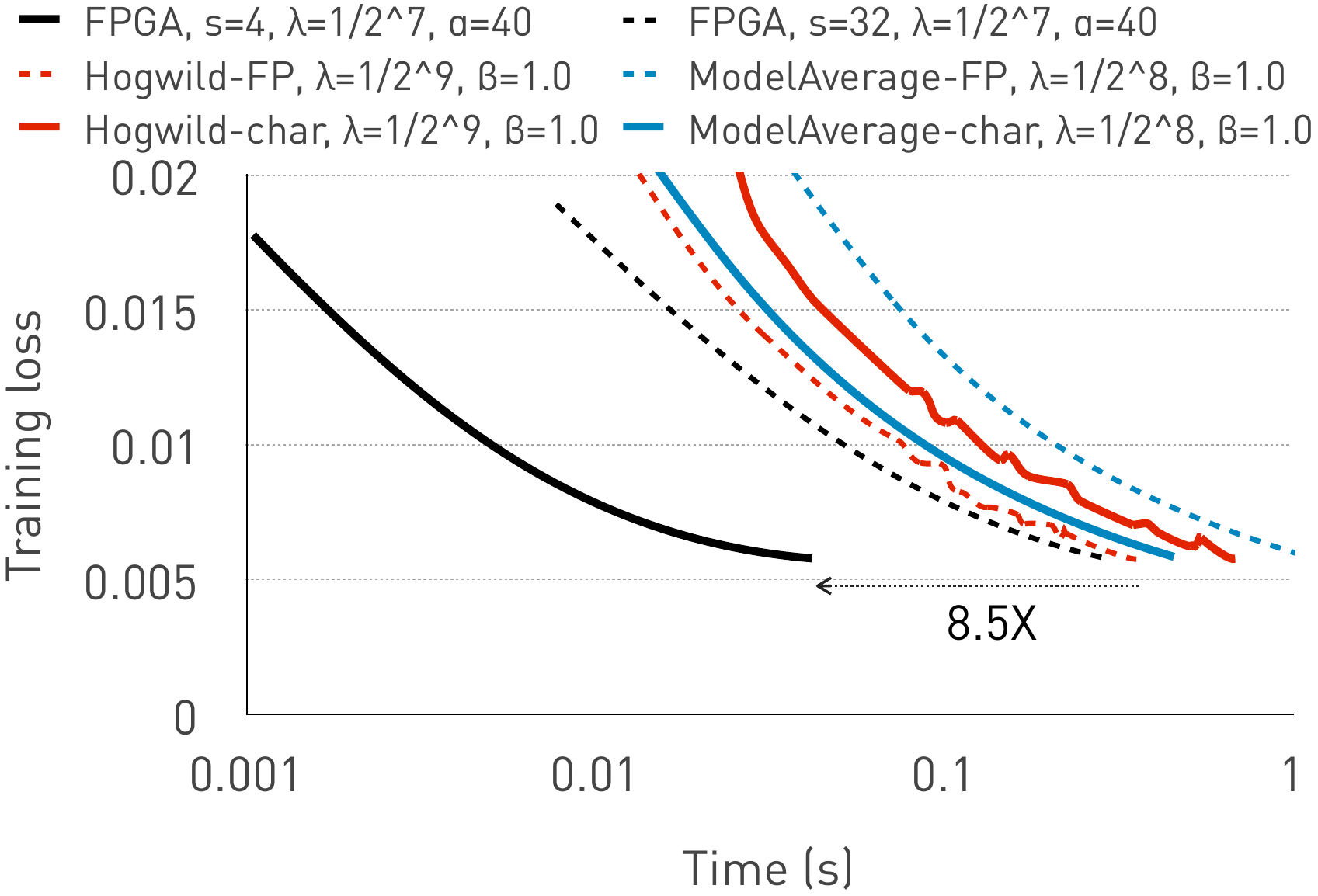} 
		\label{fig_loss_time_gisette}} 
	\subfloat[Gisette (loss vs. memory traffic)]{\includegraphics[width=2.25in]{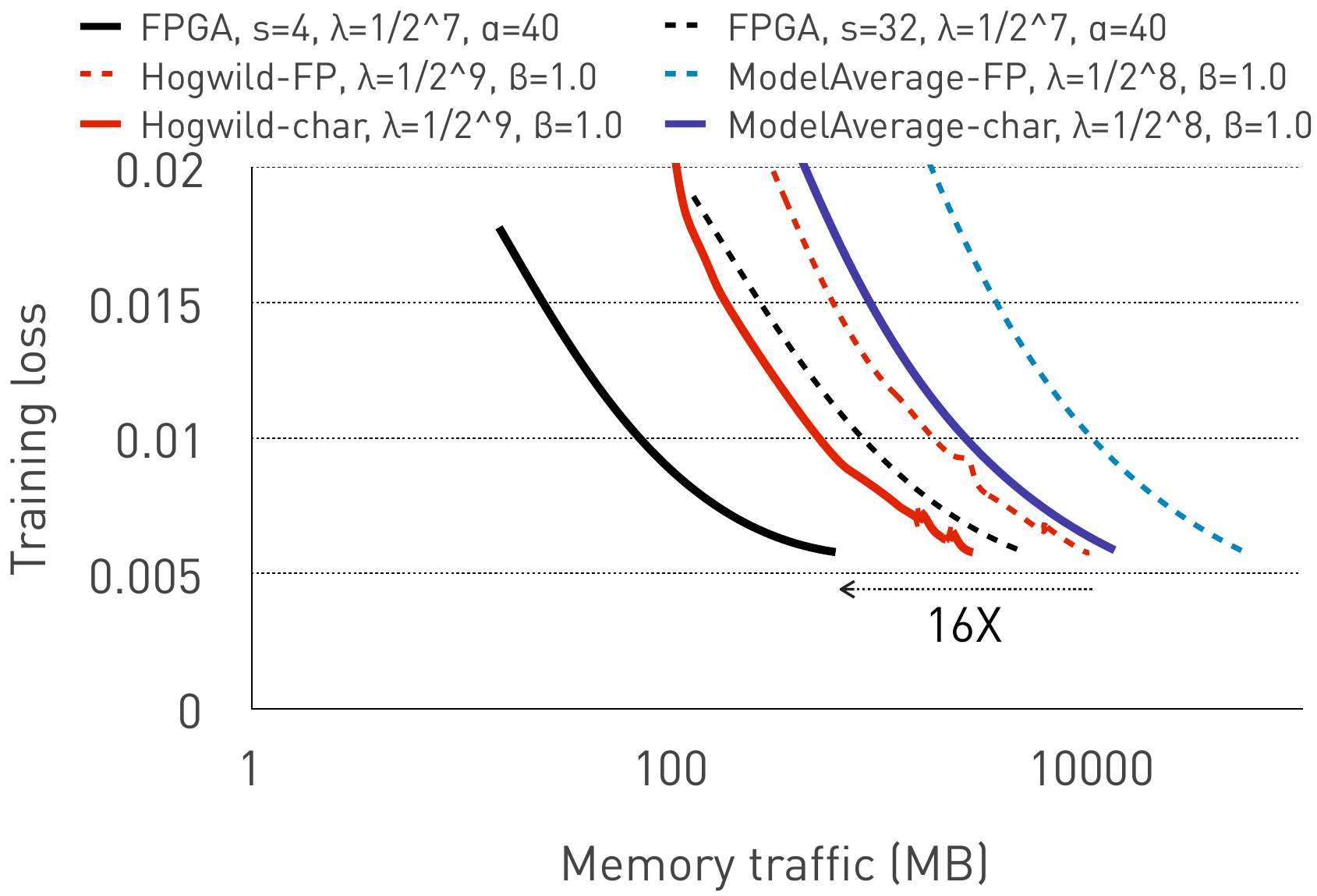} 
		\label{fig_loss_bytes_gisette}} % \caption{}	
	\hfill
	\subfloat[TL (loss vs. epoch)]{\includegraphics[width=2.05in]{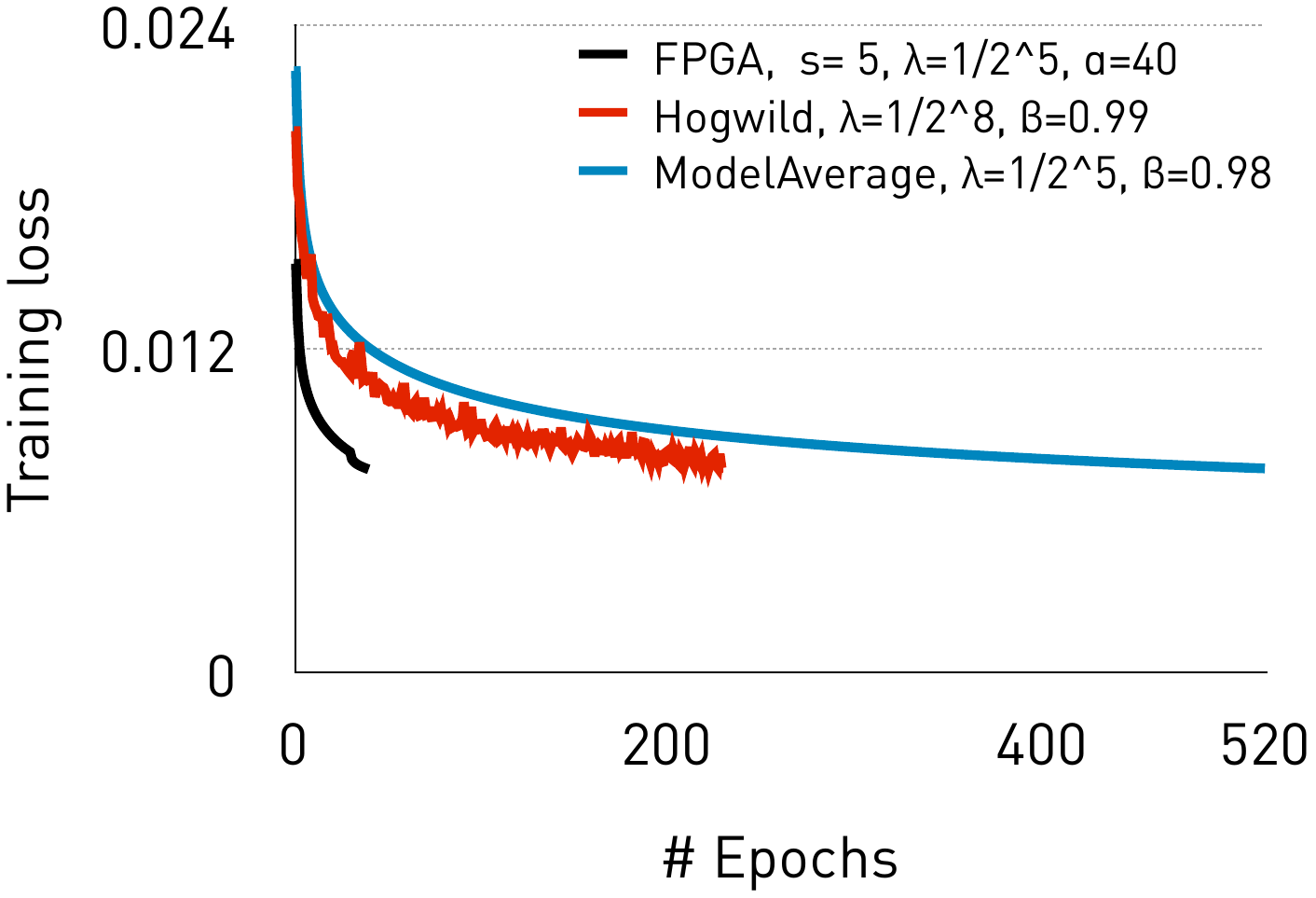} 
		\label{fig_loss_epoch_imagenet}} % \caption{}
	\subfloat[TL (loss vs. time)]{\includegraphics[width=2.25in]{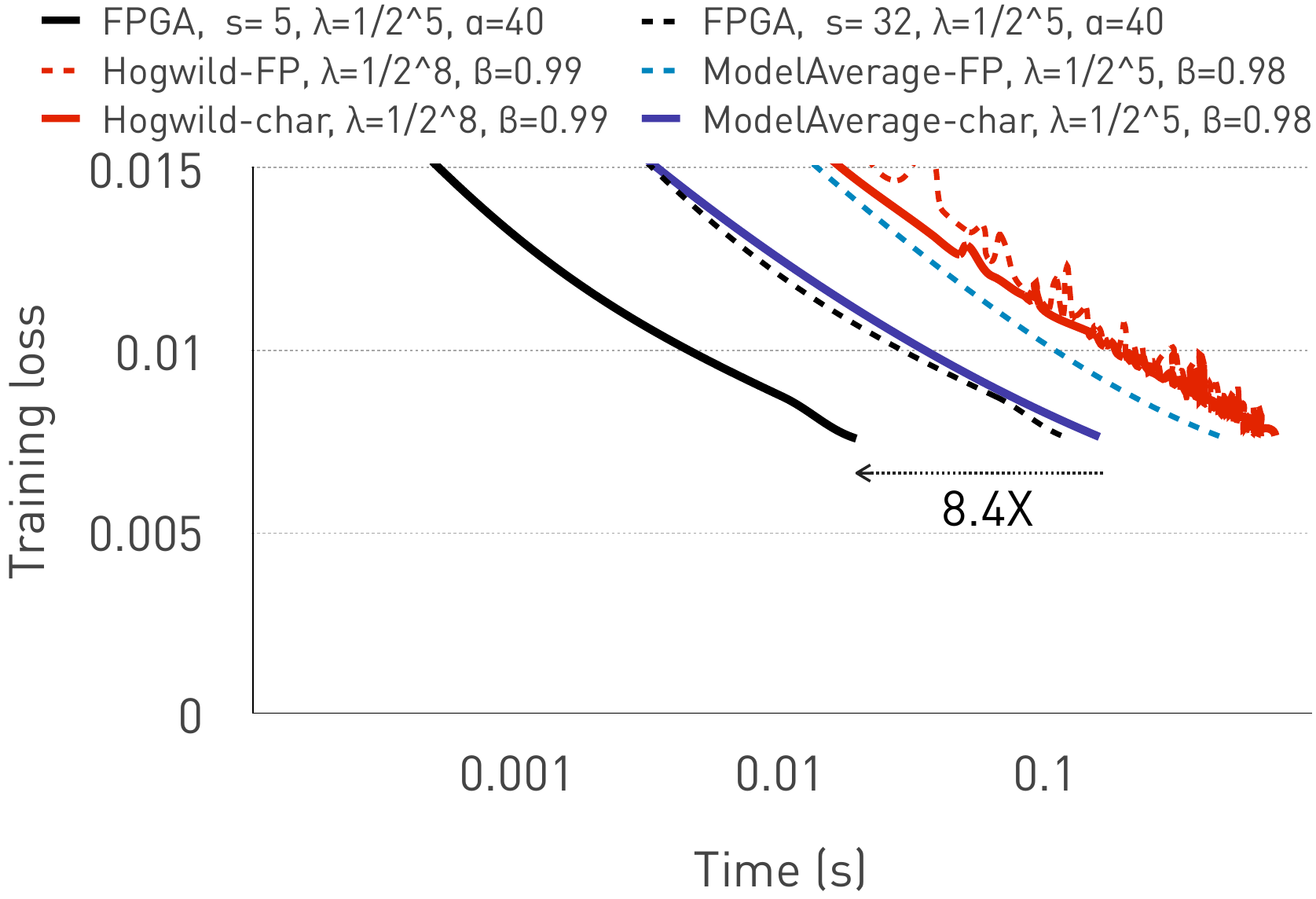} 
	\label{fig_loss_time_imagenet}} % \caption{}
	\subfloat[TL (loss vs. memory traffic)]{\includegraphics[width=2.25in]{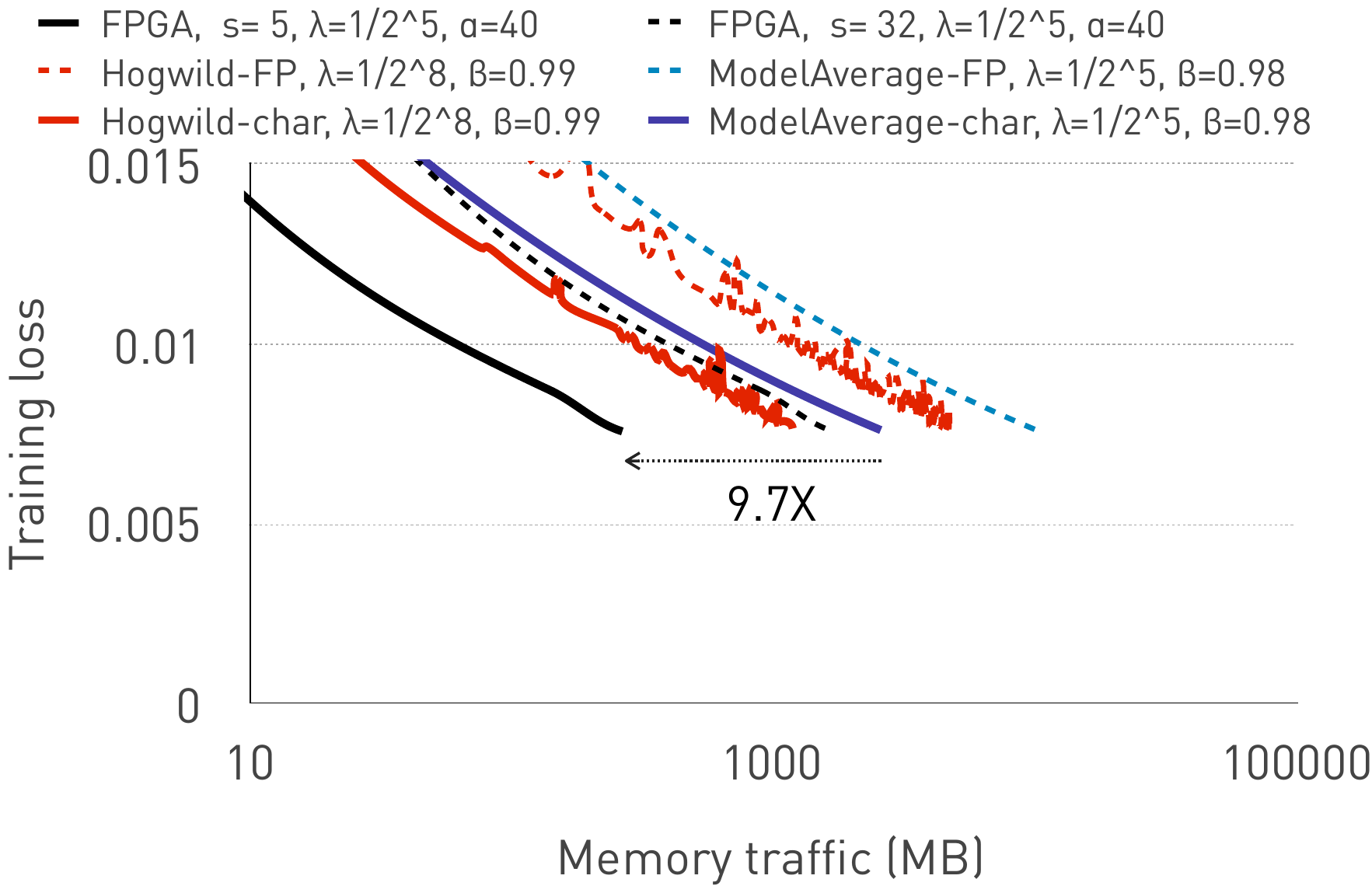} 
		\label{fig_loss_bytes_imagenet}} 
	\hfill
	\subfloat[Madelon (loss vs. epoch)]{\includegraphics[width=2.05in]{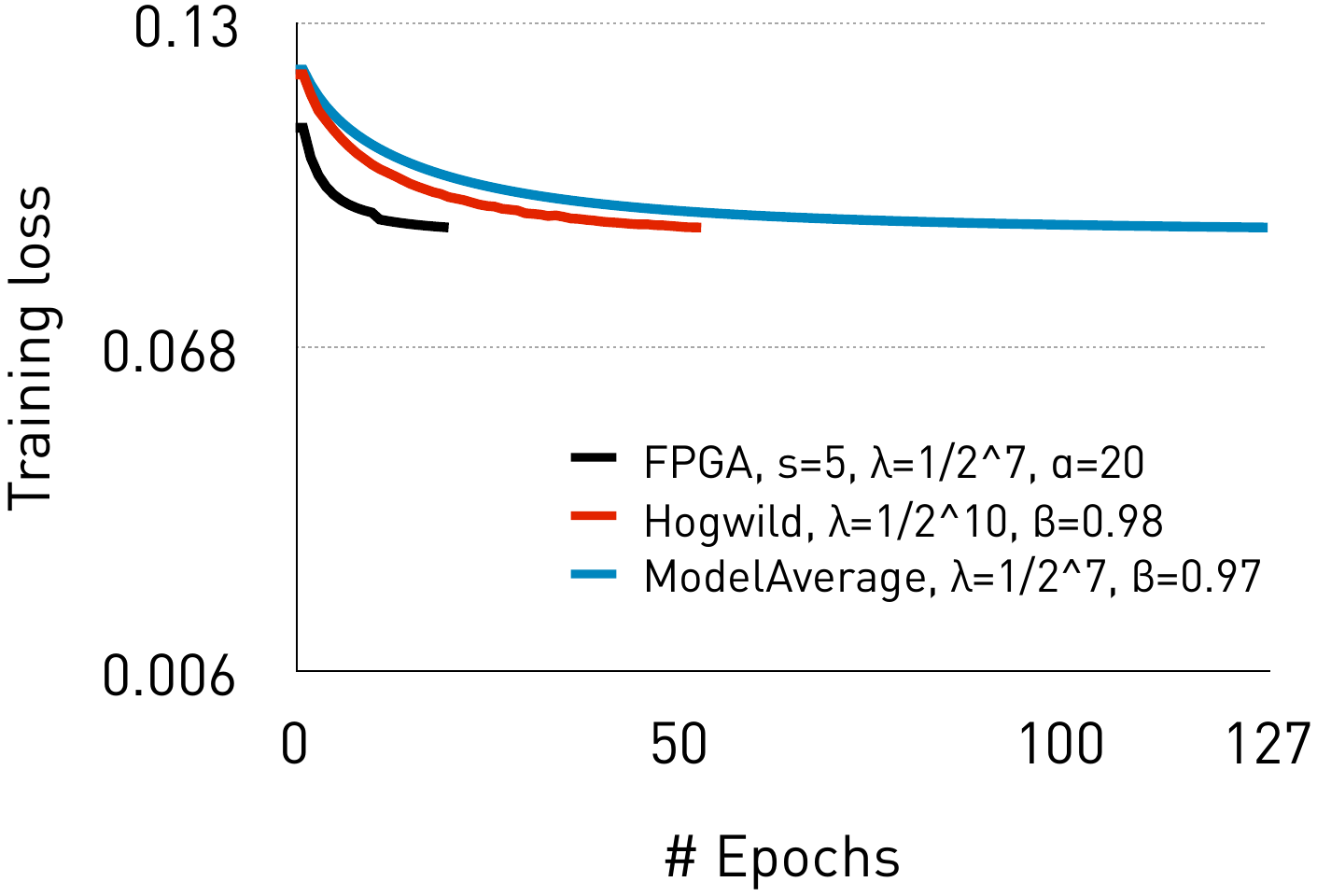} 
		\label{fig_loss_epoch_madelon}} % \caption{
	\subfloat[Madelon (loss vs. time)]{\includegraphics[width=2.25in]{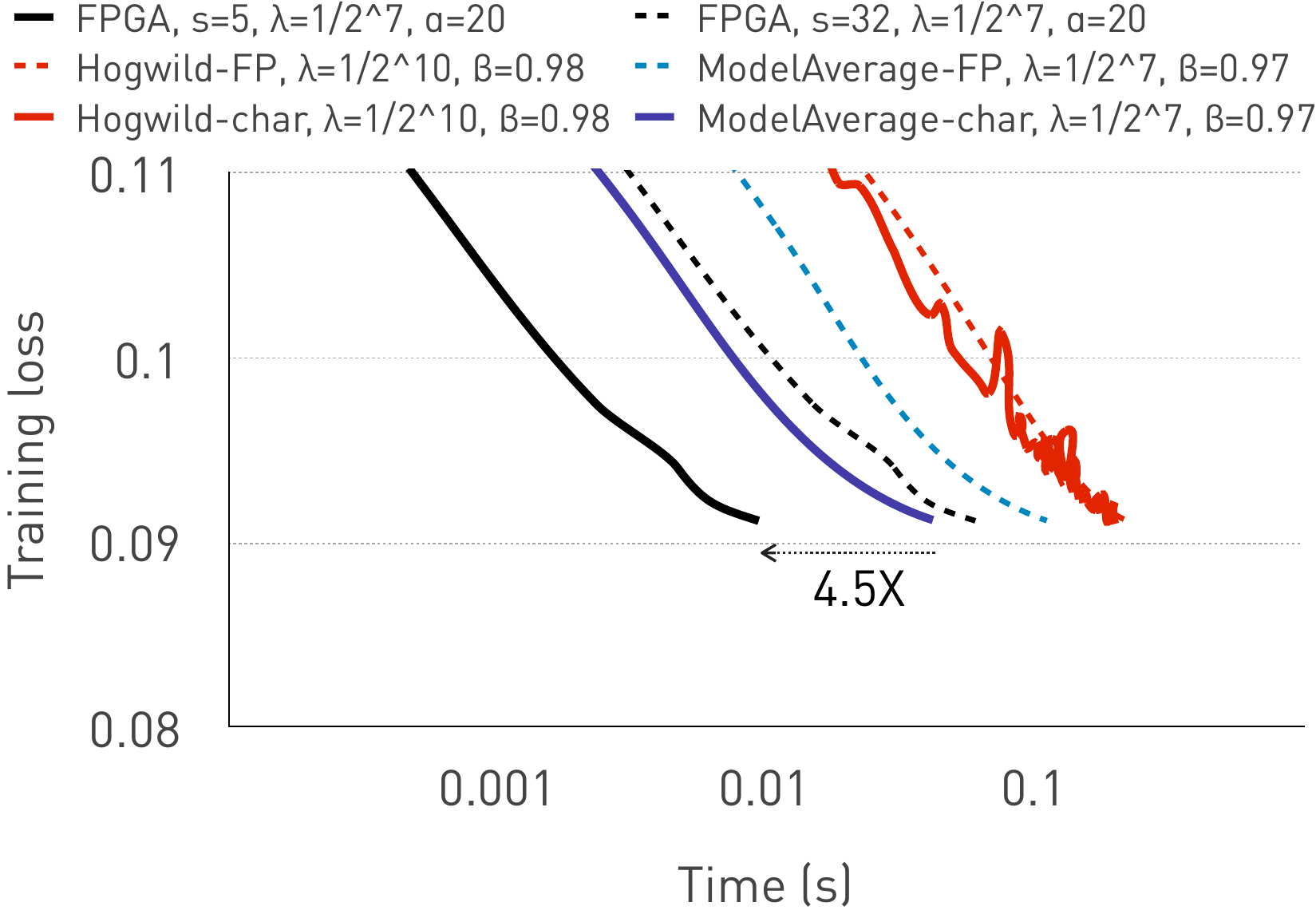} 
		\label{fig_loss_time_madelon}} % \caption{}
	\subfloat[Madelon (loss vs. memory traffic)]{\includegraphics[width=2.25in]{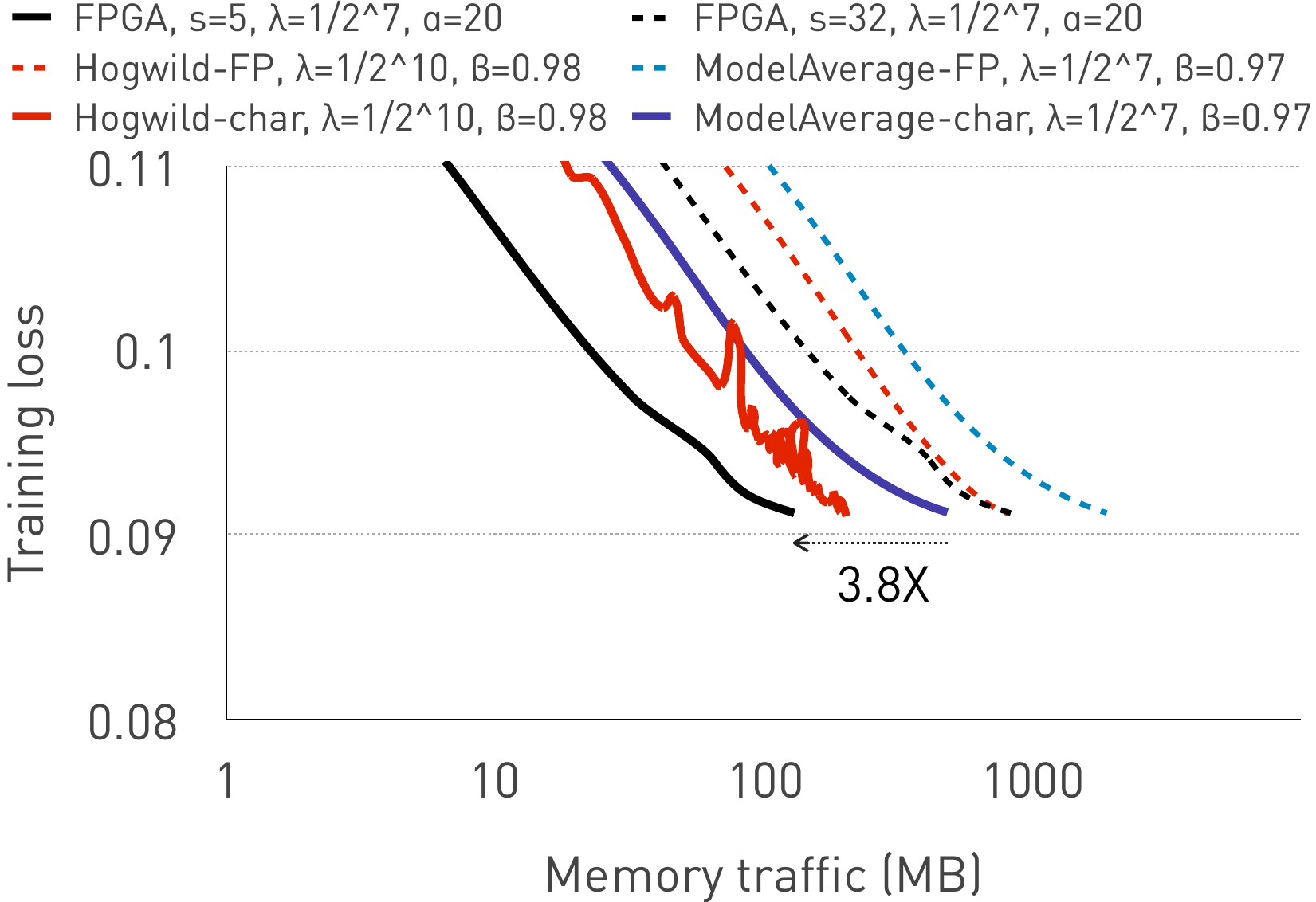} 
		\label{fig_loss_bytes_madelon}} % \caption{}	
	\caption{Convergence comparison: loss vs. epoch/time/memory traffic. The batch size is 8. Speedup indicates MLWeaving vs. fastest 14-core AVX2-enhanced low-precision CPU approach, in terms of time and memory traffic.} %MLWeaving employ ``4-bit", ``3-bit" or ``6-bit" to train for the dataset Gisette, Epsilon or ImageNet. 
	\vspace{-1ex}
	\label{fig_loss_time_epoch_appendix} 
	\vspace{-2ex}
\end{figure*}

\fi

\end{document}